\documentclass[prd,amsmath,amsfonts,floatfix,nofootinbib,preprintnumbers,superscriptaddress]{revtex4}
\usepackage{graphicx}
\allowdisplaybreaks[4]

\def\la{\langle}
\def\ra{\rangle}
\def\op{{\mathcal{O}}}
\def\calI{{\mathcal{I}}}
\def\calIdp{{\mathcal{I}}_{\eta^{\prime}}}
\def\calF{{\mathcal{F}}}

\def\calH{{\mathcal{H}}}
\def\calHdp{{\mathcal{H}}_{\eta^{\prime}}}
\def\calW{{\mathcal{W}}}
\def\calK{{\mathcal{K}}}
\def\calKdp{{\mathcal{K}}_{\eta^{\prime}}}
\def\frakA{{\mathfrak{A}}}
\def\frakB{{\mathfrak{B}}}
\def\frakC{{\mathfrak{C}}}
\def\frakD{{\mathfrak{D}}}
\def\frakE{{\mathfrak{E}}}
\def\frakF{{\mathfrak{F}}}
\def\quu{Q_{uu}}
\def\qss{Q_{ss}}
\def\ruu{R_{uu}}
\def\rss{R_{ss}}
\def\rus{R_{us}}
\def\suu{S_{uu}}
\def\sss{S_{ss}}
\def\sus{S_{us}}
\def\tuu{T_{uu}}
\def\tss{T_{ss}}
\def\tus{T_{us}}
\def\hqone{\hat{y}_{1}}
\def\hqtwo{\hat{y}_{2}}
\def\hqthree{\hat{y}_{3}}

\def\exponential{{\mathrm{e}}}
\newcommand{\xpt}[0]{{$\chi$PT}}
\newcommand{\qxpt}[0]{{Q$\chi$PT}}
\newcommand{\pqxpt}[0]{{PQ$\chi$PT}}
\newcommand{\hbxpt}[0]{{HB$\chi$PT}}
\newcommand{\pbar}[0]{{\bar{p}}}

\newcommand{\str}[0]{\rm str}

\begin{document}

\preprint{NT@UW 04-024, INT-PUB-04-28, UW/PT 04-21}

\title{Twist-two matrix elements at finite and infinite volume}

\author{William Detmold} \affiliation{Department of Physics,
  University of Washington, Box 351560, Seattle, WA 98195, U.S.A.}

\author{C.-J. David Lin} \affiliation{Department of Physics,
  University of Washington, Box 351560, Seattle, WA 98195, U.S.A.}
\affiliation{Institute for Nuclear Theory, University of Washington,
  Box 351550, Seattle, WA 98195, U.S.A.}

\begin{abstract}
  We present one-loop results for the forward twist-two matrix
  elements relevant to the unpolarised, helicity and transversity
  baryon structure functions, in partially-quenched ($N_{f}=2$ and
  $N_{f}=2+1$) heavy baryon chiral perturbation theory.  The full-QCD
  limit can be straightforwardly obtained from these results and we
  also consider SU(2$|$2) quenched QCD. Our calculations are performed
  in finite volume as well as in infinite volume.  We discuss
  features of lattice simulations and investigate finite volume
  effects in detail. We find that volume effects are not negligible,
  typically around 5--10\% in current partially-quenched and full QCD
  calculations, and are possibly larger in quenched QCD.  Extensions
  to the off-forward matrix elements and potential difficulties that
  occur there are also discussed.
\end{abstract}

\date\today \maketitle

\section{Introduction}

The quark and gluon sub-structure of hadrons has been probed for many
years in high energy scattering experiments. Much of the information
that has been gleaned is encoded in the parton distribution functions
(PDFs) that describe the longitudinal momentum distributions of quarks
and gluons within hadrons. Cross-sections for deep inelastic
scattering, for example, have been shown to factorise into short and
long distance contributions \cite{Collins:1985ue,Brock:1993sz}. The
short distance pieces (Wilson coefficients) are perturbatively
calculable whilst the long range effects are expressed in terms of the
PDFs. The utility of such PDFs is that they are universal; the same
set of PDFs appear in deep-inelastic scattering, Drell-Yan processes
and heavy vector boson production. Whilst PDFs are scale dependent,
once they are known at one scale they can be calculated at higher
scales via the DGLAP \cite{DGLAP} evolution equations. A number of
groups \cite{unpolPDFs,polPDFs} have exploited the universality of
PDFs and their known scale-dependence by performing global analyses of
experimental data, thereby providing convenient parameterisations of
the PDFs.  Such parameterisations have proven very useful in testing
perturbative QCD in high energy processes and constraining new
physics, but nothing is learnt about the non-perturbative origins of
the PDFs.

Whilst experiments continually increase our knowledge of PDFs, there
is much that is still unknown. Recent results
\cite{Amaudruz:1991at,Hawker:1998ty,Peng:1998pa} have shown that $\bar
u(x) \ne \bar d(x)$, however other simple qualitative questions such
as whether $\Delta \bar u(x)=\Delta \bar d(x)$ or $s(x)=\bar s(x)$
remain unanswered.  Even for the unpolarised valence quark
distributions, information is scarce at large $x$ and there is no
experimental information about the transversity distributions.
Consequently, any insight that can be gained directly from QCD would
be very useful.  Since the PDFs encode the soft, hadronic scale
physics of QCD bound states, perturbative QCD is of little use.  One
can turn to models to suggest the qualitatively important features of
PDFs (for example $\bar u(x) \ne \bar d(x)$ was predicted on the basis
of the pion cloud \cite{Thomas:1983fh}), but to make concrete
predictions with systematically improvable errors one must solve QCD
non-perturbatively. Currently this means one must use lattice QCD.

In lattice QCD, one discretises space-time and uses Monte-Carlo
techniques to evaluate the functional integrals over the quark and
gluons fields, necessarily making a Wick rotation to Euclidean space
in the process.  However, deep-inelastic scattering and related
processes are dominated by distances that are light-like, and as such
are inaccessible in Euclidean space calculations.  The way around this
difficulty is provided by the operator product expansion (OPE) which
relates matrix elements of certain local operators to Mellin moments
of the various quark and gluon distributions (defined below).  For
quark distributions, the twist-two (twist = dimension $-$ spin)
operators that arise from the OPE of the bilocal light-cone operators
in $N_f=2$ QCD are
\begin{eqnarray}
\label{eq:twist2op}
^{\rm QCD}{\cal O}^{(A)}_{\mu_0\ldots\mu_n} 
&=& i^{n}\left[\bar\psi
  \gamma_{\{\mu_0}\tensor{D}_{\mu_1}\ldots 
  \tensor{D}_{\mu_n\}}\tau_A\psi - {\rm traces}\right], \\
\label{eq:twist25op}
^{\rm QCD}\widetilde{\cal O}^{(A)}_{\mu_0\ldots\mu_n} 
&=& i^{n}\left[\bar\psi 
  \gamma_{\{\mu_0}\gamma_5\tensor{D}_{\mu_1}\ldots
  \tensor{D}_{\mu_n\}}\tau_A \psi - {\rm traces}\right], \\
\label{eq:twist2sigmaop}
^{\rm QCD}\widetilde{\cal O}^{T;(A)}_{\mu_0\ldots\mu_{n+1}} 
&=& i^{n}\left[\bar\psi
  \sigma_{\mu_0\{\mu_1}\gamma_5\tensor{D}_{\mu_2}\ldots
  \tensor{D}_{\mu_{n+1}\}}\tau_A\psi - {\rm traces}\right],
\end{eqnarray}
where $\tau_A$ is an isospin matrix ($\tau_0=\mathbf{1}$,
$\tau_{1,2,3}$ are the Pauli matrices), $\{\ldots\}$ indicates
symmetrisation of indices, the gauge covariant derivative
$\tensor{D}^\mu =\frac{1}{2}( \roarrow{D}^\mu-\loarrow{D}^\mu)$, and
'traces' are subtracted in order that the operator transforms
irreducibly under the Lorentz group. Additional twist-two operators
can be built exclusively from gluon fields and towers of higher twist
operators can also be constructed, but we shall not consider these
here.

The above operators are related to the spin averaged, longitudinally
polarised, and transversely polarised quark distributions. We first
define the Mellin moments of the quark distributions
$q=q^\uparrow+q^\downarrow$, $\Delta q=q^\uparrow-q^\downarrow$ and
$\delta q=q^\top-q^\perp$ (where $q^{\uparrow(\downarrow)}$
corresponds to quarks with helicity aligned (anti-aligned) with that
of a longitudinally polarized target, and $q^{\top(\perp)}$
corresponds to quarks with spin aligned (anti-aligned) with that of a
transversely polarized target) for flavour $q$ as
\begin{eqnarray}
  \langle x^n\rangle_q &=& \int_{0}^1 dx\; x^n
  \left[q(x)-(-1)^n\overline{q}(x)\right] \,,
\nonumber \\ 
  \label{eq:26}
  \langle x^n\rangle_{\Delta q} &=& \int_{0}^1 dx\; x^n 
  \left[\Delta q(x)+(-1)^n\Delta\overline{q}(x)\right]  \,,\\
  \langle x^n\rangle_{\delta q} &=& \int_{0}^1 dx\; x^n 
\left[ \delta q(x) -(-1)^n\delta\overline{q}(x)\right] \,.
\nonumber
\end{eqnarray}
These moments are then related to the forward hadron matrix elements
of the operators in Eqs.~(\ref{eq:twist2op})--(\ref{eq:twist2sigmaop})
through
\begin{eqnarray}
  \mbox{$\frac{1}{2}\sum_{S}$}\langle p,S | ^{\rm QCD}{\cal O}_{\mu_0\ldots\mu_{n}}^{(0,3)}|p,S
  \rangle &=&
  2 \langle x^n\rangle_{u\pm d} p_{\mu_0}\ldots p_{\mu_n}\,,
\nonumber \\
\nonumber \\
  \label{eq:27}
  \langle p,S | ^{\rm QCD}\widetilde{\cal
    O}_{\mu_0\ldots\mu_{n}}^{(0,3)}|p,S \rangle 
  &=& 
  2\langle x^n\rangle_{\Delta u\pm \Delta d}
  S_{\{\mu_0}p_{\mu_1}\ldots p_{\mu_n\}}\,,
\\
\nonumber \\
  \langle p,S | ^{\rm QCD} \widetilde{\cal
    O}_{\mu_0\ldots\mu_{n+1}}^{T;(0,3)}|p,S \rangle 
  &=&
  \mbox{$\frac{2}{M_H}$} \langle x^n\rangle_{\delta u\pm \delta d}
  S_{[\mu_{0}}p_{\{\mu_1]}\ldots 
  p_{\mu_{n+1}\}} \,,
\nonumber
\end{eqnarray}
where $p$ is the momentum of the hadron, $M_H$ is its mass and $S$ is its spin. The plus
or minus signs in Eq.~(\ref{eq:27}) correspond to choosing isospin
index $0$ or $3$ respectively. The corresponding off-forward matrix
elements are similarly related to Bjorken-$x$ moments of generalised
parton distributions (GPDs) which shall be discussed briefly below
(see \cite{Diehl:2003ny} for a comprehensive review).

The hadronic matrix elements of the twist-two operators in
Eq.~(\ref{eq:27}) can be calculated using standard lattice techniques.
Although a parametric form must be assumed in order to invert
\cite{Gockeler:1997xy,Detmold:2001dv} the Mellin transforms,
Eqs.~(\ref{eq:26}), such calculations will then lead to information
about the parton distributions directly from QCD\footnote{The reduced
  symmetry of the hyper-cubic lattice leads to lower dimensional
  operator mixing for twist-two operators with $n>3$, and consequently
  calculations are only currently available for $n=1,\,2,\,3$.}.
However, all lattice calculations are necessarily performed on finite
volumes and at finite lattice spacings.  Additionally, with current
computational resources, statistically meaningful simulations can only
be performed at quark masses, $m_q$, considerably larger than those
found in nature. These three restrictions have significant effects on
calculations of twist-two matrix elements which must be taken into
account if realistic predictions are to be made.

Conveniently, the low energy QCD dynamics that these matrix elements
characterize can be described using effective field theory. Standard
chiral perturbation theory (\xpt) as formulated in the infinite volume
continuum allows systematic exploration of the quark mass dependence
of low energy hadronic observables in the region where $m_\pi,\,|{\vec
  p}| <\Lambda_\chi$ where $\vec p$ is a typical momentum and
$\Lambda_\chi\sim1$~GeV is the chiral symmetry breaking scale.
Extensions to include finite volume (FV) and finite lattice spacing
effects are also well developed (see Refs.~\cite{Colangelo:2004sc} and
\cite{Bar:2004xp} respectively for recent reviews) as are the
modifications necessary to treat valence and sea quark masses
independently -- quenched and partially-quenched \xpt\ (\qxpt\ and
\pqxpt)
\cite{Morel:1987xk,Sharpe:1992ft,Bernard:1992mk,Bernard:1993sv}. In
our study, we shall ignore the effects of the discretisation of
space-time\footnote{The additional, Lorentz non-invariant
  contributions to unpolarised twist-two operators that must be
  included when the lattice spacing is non-zero have been considered
  in Ref.~\cite{Beane:2003xv}.} (whereby our results will only be
strictly applicable to lattice calculations in which a continuum
extrapolation has been performed) and consider continuum
partially-quenched chiral perturbation theory in a finite spatial
volume of dimension $L^3$.  If the size of the box is large compared
to the inverse pion mass (the lightest asymptotic state), $M_\pi L \gg
1$, the power counting of infinite volume \xpt\ ($p$-counting) applies
and the necessary modifications are easily made, replacing momentum
integrals by sums over allowed momentum modes (see
Refs.~\cite{Bernard:1995ez,GoltermanLeung,SPQR,Arndt:2004bg,Becirevic:2003wk,
  Detmold:2004qn,Beane:2003da,Beane:2003yx,
  AliKhan:2003cu,Beane:2004tw,Beane:2004rf,Leinweber:2001ac,Young:2004tb}
for recent examples).  On the other hand if $M_\pi L\sim1$, one needs
to treat pion zero modes (components of the pion field with zero
momentum) carefully since they correspond to vacuum fluctuations of
order unity. In such a regime, modified power countings are required
\cite{eregime,Detmold:2004ap,Bedaque:2004dt}. In this analysis, we
will restrict ourselves to the region $M_\pi L \gg 1$.

Using the low energy effective theory, it is possible to compute the
quark mass and volume dependence of hadronic observables such as the
matrix elements of twist-two operators. For the most part, the quark
mass dependence of the various twist-two matrix elements has been
studied extensively
\cite{Detmold:2001jb,Arndt:2001ye,Chen:2001eg,Chen:2001gr,Chen:2001et,
  Chen:2001yi,Beane:2002vq} and lattice data have been analysed
assuming an infinite volume
\cite{Detmold:2001jb,Detmold:2002nf,Detmold:2003tm}. However, the
volume dependence of these observables has been ignored (with the
exception of the matrix element of the $n=0$ helicity-dependent
twist-two operator, the isovector axial coupling $g_A$
\cite{Beane:2004rf,Detmold:2004ap}) in such analyses.  Nonetheless,
finite volume effects have been found to be important in many
observables; here we investigate the effect they have on nucleon, and
other octet baryon matrix elements of twist-two operators.

In Section \ref{S2}, we introduce aspects of heavy baryon chiral
perturbation theory relevant for the analysis of twist-two matrix
elements and define our notation.  In Section \ref{S3}, we discuss the
twist-two operators in QCD and their matching in the low-energy
effective theory and present examples of results for the quark-mass
dependence of the nucleon matrix elements using two degenerate
flavours of quarks.  Full results in the two flavour
partially-quenched case and results including the strange quark are
relegated to Appendices \ref{A2} and \ref{A3} respectively. In Appendix
\ref{A4}, the quenched theory is discussed. In Section \ref{S4}, we
discuss the general form of finite volume corrections to these matrix
elements and make comparisons with available data.  Section \ref{S5}
discusses the complications that arise when non-forward matrix
elements are considered and Section \ref{S6} presents our conclusions.


\section{Heavy baryon chiral perturbation theory}
\label{S2}

Heavy baryon chiral perturbation theory (HB$\chi$PT) was first
constructed in
Refs.~\cite{Jenkins:1990jv,Jenkins:1991ne,Jenkins:1991es,Bernard:1993nj}.
In current lattice calculations, valence and sea quarks are often
treated differently, with sea quarks either absent (quenched QCD) or
having different masses than the valence quarks (partially-quenched
QCD)\footnote{At finite lattice spacing, different actions can even be
  used for the different quark sectors (e.g., staggered sea quarks and
  domain wall (DW) valence quarks), but we do not consider this
  complication here.}. The extensions of \hbxpt\ to quenched \hbxpt\ 
\cite{Labrenz:1996jy} and partially quenched \hbxpt\ 
\cite{Chen:2001yi,Beane:2002vq} to accommodate these modifications are
also well established and have been used to calculate many baryon
properties
\cite{Labrenz:1996jy,Savage:2001dy,Chen:2001yi,Chen:2001gr,Beane:2002vq,Arndt:2003vd,Walker-Loud:2004hf,Tiburzi:2004kd}.
In this and the next sections, we will primarily focus on the two
flavour partially-quenched theory; here we briefly introduce the
relevant details following the conventions set out in
Ref.~\cite{Beane:2002vq}.  We leave the three flavour and quenched
cases to Appendices~\ref{A3} and \ref{A4}.

\subsection{Pseudo-Goldstone mesons}

We consider a partially-quenched theory of valence ($u$, $d$), sea
($j$, $l$) and ghost ($\tilde u,\,\tilde d$) quarks with masses
corresponding to the matrix
\begin{equation}
\label{eq:Mq_def}
m_Q = {\rm  diag}(m_u,m_d,m_j,m_l,m_{\tilde u},m_{\tilde d})\,,
\end{equation}
where $m_{\tilde u,\tilde d}=m_{u,d}$ such that the QCD path-integral
determinants corresponding to the valence and ghost sectors exactly
cancel.

The corresponding low-energy meson dynamics are described by the
SU(4$|$2) \pqxpt\ Lagrangian. At leading order
\begin{eqnarray}
{\cal L }_\Phi & = & 
{f^2\over 8} 
{\rm str}\left[ \partial^\mu\Sigma^\dagger\partial_\mu\Sigma \right] + 
\lambda {f^2\over 4} 
{\rm str}\left[ m_Q\Sigma^\dagger + m_Q\Sigma \right]
 + 
\alpha_\Phi\partial^\mu\Phi_0\partial_\mu\Phi_0 - m_0^2\Phi_0^2
\, ,
\label{eq:PGBlagrangian}
\end{eqnarray}
where the pseudo-Goldstone mesons are embedded non-linearly in
\begin{equation}
\label{eq:Sigma_def}
\Sigma = \xi^2= \exp{\left({2\ i\ \Phi\over f}\right)} ,
\end{equation}
with the matrix $\Phi$ given by
\begin{equation}
  \label{eq:Phi_def}
\Phi = \begin{pmatrix} M &\chi^\dagger \cr \chi &\tilde{M} \end{pmatrix}
\,,
\end{equation}
where
\begin{eqnarray}
  \label{eq:Mchi_def}
  M=\begin{pmatrix} \eta_u    & \pi^+     & J^0      & L^+      \\
                    \pi^-     & \eta_d    & J^-      & L^0      \\
                    \bar{J}^0 & J^+       & \eta_j   & Y_{jl}^+ \\
                    L^-       & \bar{L}^0 & Y_{jl}^- & \eta_l   
                    \end{pmatrix}\,, 
& 
\hspace*{15mm}
\tilde{M}=\begin{pmatrix} \tilde\eta_u    & \tilde\pi^+        \\
                          \tilde\pi^-     & \tilde\eta_d       
                          \end{pmatrix}\,, 
 \hspace*{15mm}
&
\chi=\begin{pmatrix} \chi_{\eta_u} & \chi_{\pi^+}  & \chi_{J^0} & \chi_{L^+}      \\
                     \chi_{\pi^-}  & \chi_{\eta_d} & \chi_{J^-} & \chi_{L^0}      
                     \end{pmatrix}\,.
\end{eqnarray}
The upper left $2\times2$ block of $M$ corresponds to the usual
valence--valence mesons, the lower right to sea--sea mesons and the
remaining entries of $M$ to valence--sea mesons. Mesons in $\tilde{M}$
are composed of ghost quarks and anti-quarks and those in $\chi$ of
ghost--valence or ghost--sea quark--anti-quark pairs. Because of the
graded symmetry of the partially-quenched theory, the mesons in $\chi$
are fermionic. In terms of the quark masses, the tree-level meson
mases are given by
\begin{eqnarray}
  \label{eq:mesonmass_def}
  M^2_{\Phi_{ij}}=M_{{\cal Q}_i {\cal Q}_j}^2 = 
\lambda\left[ \left(m_Q\right)_{ii}
    + \left(m_Q\right)_{jj} \right] \,,
\end{eqnarray}
where ${\cal Q}=(u,\,d,\,j,\,l,\,\tilde{u},\,\tilde{d})$.

The singlet field $\Phi_0=\str\left(\Phi\right)/\sqrt{2}$ has mass
$m_0$ at tree level. The terms proportional to $\alpha_\Phi$ and $m_0$
in Eq.~(\ref{eq:PGBlagrangian}) are only relevant in the quenched
theory (see Appendix~\ref{A4}); in \pqxpt\ and \xpt\, the singlet
mesons acquire large masses and can be integrated out. Furthermore,
$\alpha_\Phi$ is suppressed by $1/N_c$ and we set it to zero
throughout.

\subsection{Baryons}

In SU(4$|$2) \hbxpt, the physical nucleons (those composed of three
valence quarks) enter as part of a {\bf 70}-dimensional
representation.  This is described by a three index flavour-tensor,
${\cal B}$ \cite{Labrenz:1996jy,Beane:2002vq,Chen:2001yi}.  The
embedding of the physical nucleon fields into ${\cal B}$ and the
symmetry properties of ${\cal B}$ are described in
Ref.~\cite{Beane:2002vq}.  The $\Delta$-isobar must also be included
in the theory since the mass-splitting, $\Delta$, between the nucleon
and $\Delta$-isobar is $\sim300$~MeV, comparable to the physical pion
mass (and less than pion masses used in current lattice simulations).
The parameter $\Delta$ is assumed to be small compared to the chiral
symmetry breaking scale.  These fields are represented in a three
index flavour-tensor ${\cal T}^\mu$ (a Rarita-Schwinger field)
transforming in the {\bf 44}-dimensional representation of SU(4$|$2).

The relevant part of leading-order Lagrangian describing these baryons
and their interactions with Goldstone mesons is
\begin{eqnarray}
{\cal L}_B & = & 
i\left(\overline{\cal B} v\cdot {\cal D} {\cal B}\right)
 - i \left(\overline{\cal T}^\mu v\cdot {\cal D} {\cal T}_\mu\right)
+ \Delta\ \left(\overline{\cal T}^\mu {\cal T}_\mu\right)
\nonumber\\
 && + 2\alpha \left(\overline{\cal B} S^\mu {\cal B} A_\mu\right)
 +  2\beta \left(\overline{\cal B} S^\mu A_\mu {\cal B} \right)
 +  2{\cal H} \left(\overline{\cal T}^\nu S^\mu A_\mu {\cal T}_\nu \right)
+ \sqrt{3\over 2}{\cal C} 
\left[
\left( \overline{\cal T}^\nu A_\nu {\cal B}\right) + 
\left(\overline{\cal B} A_\nu {\cal T}^\nu\right) \right]
\,,
\label{eq:free_lagrangian}
\end{eqnarray}
where $v^\mu$ is the baryon velocity, $S^\mu$ is the covariant
spin-vector \cite{Jenkins:1990jv,Jenkins:1991es} and ${\cal D}^\mu$ is
the usual covariant derivative
\begin{equation}
\label{eq:Cov_deriv_def}
  {\cal D}^{\mu} {\cal B} = \partial^{\mu} {\cal B}
  + \left [ V^{\mu}, {\cal B} \right ] 
\hspace*{0.3cm}{\mathrm{and}}\hspace*{0.3cm}
  {\cal D}^{\nu} {\cal T}_{\mu} = \partial^{\nu} {\cal T}_{\mu}
  + \left [ V^{\nu}, {\cal T}_{\mu} \right ] .
\end{equation}
The vector and axial-vector currents appearing in the above
expressions are given by
\begin{eqnarray}
V^\mu  =  {1\over 2}\left(\ \xi\partial^\mu\xi^\dagger
\ + \ 
\xi^\dagger\partial^\mu\xi \ \right)
\,,
& \hspace*{1cm} &
A^\mu \ =\  {i\over 2}\left(\ \xi\partial^\mu\xi^\dagger
\ - \ 
\xi^\dagger\partial^\mu\xi \ \right)
\, ,
\label{eq:VmuAmu}
\end{eqnarray}
where $\xi$ is defined in Eq.~(\ref{eq:Sigma_def}).  The various
Lorentz and flavour contractions (indicated by the parentheses) are
defined in Ref.~\cite{Beane:2002vq}.  In order that ${\cal T}^\mu$
correctly describes the spin-$3/2$ sector, $v\cdot {\cal T}= S \cdot
{\cal T}=0$.

In what follows, we will substitute $\alpha = \frac{4}{3}g_A +
\frac{1}{3}g_1$, $\beta = \frac{2}{3}g_1 - \frac{1}{3}g_A$, ${\cal
  C}=-g_{\Delta N}$ and ${\cal H}=g_{\Delta\Delta}$ since these
correspond to the usual \xpt\ couplings when the QCD limit, where
$m_{j}=m_{u}$ and $m_{l}=m_{d}$, of the theory is taken.


\section{Twist-two operators and matrix elements}
\label{S3}

\subsection{Twist-two operators in (PQ)\xpt}

In the low energy effective theory, the twist-two quark bilinear
operators in Eqs.~(\ref{eq:twist2op})--(\ref{eq:twist2sigmaop}) match
onto hadronic analogues constructed to obey the same symmetry
transformation properties. In two flavour QCD, the unpolarised and
helicity operators transform as either $({\bf 3},{\bf 1})\oplus({\bf
  1},{\bf 3})$ (isovector) or $({\bf 1},{\bf 1})$ (isoscalar) of
SU(2)$_L\times$SU(2)$_R$.  When one considers
SU(4$|$2)$_L\times$SU(4$|$2)$_R$ partially quenched QCD, there is more
than one way to extend these operators
\cite{Golterman:2000fw,Chen:2001yi,Beane:2002vq}.  Imposing
super-tracelessness and the correct QCD limit in the valence sector,
the most general extension of $\tau_3$ (isovector) to the adjoint
representation of SU(4$|$2)$_{L,R}$ is
\begin{eqnarray}
  \label{eq:1}
  \bar\tau_3={\rm diag}\left( 1, -1, q_j, q_l, q_k, q_j + q_l - q_k
  \right) \,.
\end{eqnarray}
The freedom in choosing the values of the $q_i$'s can be advantageous
in lattice simulations; certain choices of the $q_i$'s eliminate
disconnected contributions (diagrams in which the operator is on a
quark line connected to the external states only through gluons which
are notoriously hard to compute \cite{Detmold:2004kw}) even away from
the isospin limit.  The non-uniqueness of the extension of Gell-Mann
flavour matrices to PQQCD has additional consequences in that are
discussed in Appendix~\ref{A3}.

For the isosinglet operator, the most convenient choice is
\begin{eqnarray}
  \label{eq:2455}
  \bar\tau_0={\rm diag}\left( 1, 1, 1, 1, 1, 1 \right)\, ,
\end{eqnarray}
because it is purely in the singlet representation of SU(4$|$2). Any
other choice, such as ${\rm diag}\left( 1, 1, 0, 0, 1, 1 \right)$
[which one might choose as disconnected diagrams would be absent],
will contain contributions from other representations, and hence
introduce additional low energy constants.

The transversity operators in QCD are chiral-odd and belong to the
representation $(\overline{\bf 2},{\bf 2})\oplus({\bf 2},\overline{\bf
  2})$. The most general choice for their extension to SU(4$|$2) PQQCD
is
\begin{eqnarray}
  \label{eq:2}
  \bar\tau_T={\rm diag}\left( 1,y_i,y_j,y_k,y_l,y_m\right)\,.
\end{eqnarray}
For this operator disconnected contributions vanish as the matrix
element involves helicity flip. Thus clean calculations of $\langle
x^n \rangle_{\delta u}$ and $\langle x^n \rangle_{\delta d}$ are
possible.

Based on these symmetry properties, at leading order in PQ$\chi$PT the
hadronic operators that match onto those of PQQCD are
\begin{eqnarray}
  \label{eq:hadron_op}
  {\cal O}_{\mu_0\ldots\mu_n}^{(A)} &\equiv& 
  a^{(r_A)}_n \frac{i^{n+1}}{\Lambda_\chi^{n}} \frac{f^2}{4}
  \str\left[\Sigma^\dagger \bar\tau_A
    \roarrow{\partial}_{\mu_0}\ldots \roarrow{\partial}_{\mu_n}\Sigma +
    \Sigma \bar\tau_A
    \roarrow{\partial}_{\mu_0}\ldots \roarrow{\partial}_{\mu_n}\Sigma^\dagger
  \right] 
  \\
  &&+ \alpha^{(r_A)}_n v_{\mu_0}\ldots v_{\mu_n} \left(\overline{\cal
      B}{\cal B}\bar\tau^{\xi^+}_{A}\right) 
  + \beta^{(r_A)}_n v_{\mu_0}\ldots v_{\mu_n} \left(\overline{\cal
      B}\bar\tau^{\xi^+}_{A}{\cal B} \right) 
  \nonumber \\
  &&+ \gamma^{(r_A)}_n v_{\mu_0}\ldots v_{\mu_n} \left(\overline{\cal
      T}^\rho\bar\tau^{\xi^+}_{A}{\cal T}_\rho \right) 
  + \sigma^{(r_A)}_n v_{\{\mu_0}\ldots v_{\mu_{n-2}} \left(\overline{\cal
      T}_{\mu_{n-1}}\bar\tau^{\xi^+}_{A}{\cal T}_{\mu_n\}} \right) 
  - {\rm traces}\,, 
  \nonumber 
\end{eqnarray}
\begin{eqnarray}
  \label{eq:hadron_5op}
  \widetilde{\cal O}^{(A)}_{\mu_0\ldots\mu_n} &\equiv& 
  \Delta\alpha^{(r_A)}_n v_{\{\mu_0}\ldots v_{\mu_{n-1}} \left(\overline{\cal
      B}S_{\mu_n\}}{\cal B}\bar\tau^{\xi^+}_{A}\right) 
  + \Delta\beta^{(r_A)}_n v_{\{\mu_0}\ldots v_{\mu_{n-1}} \left(\overline{\cal
      B}S_{\mu_n\}}\bar\tau^{\xi^+}_{A}{\cal B} \right) 
   \\
  &&  + \Delta\gamma^{(r_A)}_n v_{\{\mu_0}\ldots v_{\mu_{n-1}} \left(\overline{\cal
      T}^\rho S_{\mu_n\}}\bar\tau^{\xi^+}_{A}{\cal T}_\rho \right) 
  + \Delta\sigma^{(r_A)}_n v_{\{\mu_0}\ldots v_{\mu_{n-3}} \left(\overline{\cal
      T}_{\mu_{n-2}}S_{\mu_{n-1}}\bar \tau^{\xi^+}_{A}{\cal T}_{\mu_n\}} \right) 
  \nonumber \\
  &&    + (1-\delta_{A0})\Delta c^{(r_A)}_n v_{\{\mu_0}\ldots v_{\mu_{n-1}} \left[ \left(\overline{\cal
        T}_{\mu_n\}}\bar\tau^{\xi^+}_{A}{\cal B} \right) + \left(\overline{\cal
        B}\bar\tau^{\xi^+}_{A}{\cal T}_{\mu_n\}} \right)\right]
  - {\rm traces}\,, 
  \nonumber 
\end{eqnarray}
\begin{eqnarray}
  \label{eq:hadron_sigmaop}
  \widetilde{\cal O}^{T}_{\mu_0\ldots\mu_n\alpha} & \equiv & 
  \delta\alpha_n v_{\{\mu_0}\ldots v_{[\mu_{n}\}} \left(\overline{\cal
      B}S_{\alpha]}{\cal B}\bar\tau^{\xi^+_T}\right) 
  + \delta\beta_n v_{\{\mu_0}\ldots v_{[\mu_{n}\}} \left(\overline{\cal
      B}S_{\alpha]}\bar\tau^{\xi^+_T}{\cal B} \right) 
  \\
  &&  + \delta\gamma_n v_{\{\mu_0}\ldots v_{[\mu_{n}\}} \left(\overline{\cal
      T}^\rho S_{\alpha]}\bar\tau^{\xi^+_T}{\cal T}_\rho \right) 
  + \delta\sigma_n v_{\{\mu_0}\ldots v_{\mu_{n-2}} \left(\overline{\cal
      T}_{\mu_{n-1}} S_{[\alpha}\bar\tau^{\xi^+_T}{\cal T}_{\mu_n]\}} \right) 
  \nonumber \\
  &&  + \delta c_n v_{\{\mu_0}\ldots v_{[\mu_{n}\}} \left[ \left(\overline{\cal
        T}_{\alpha]}\bar\tau^{\xi^+_T}{\cal B} \right) + \left(\overline{\cal
        B}\bar\tau^{\xi^+_T}{\cal T}_{\alpha]} \right)\right]
  - {\rm traces}\,, 
  \nonumber 
\end{eqnarray}
where $\bar \tau_A^{\xi^\pm}=\frac{1}{2}\left( \xi^\dagger \bar\tau_A
  \xi \pm \xi \bar\tau_A \xi^\dagger \right)$ and $\bar
\tau^{\xi^\pm_T}=\frac{1}{2}\left( \xi^\dagger \bar\tau_T \xi^\dagger
  \pm \xi \bar\tau_T \xi \right)$, and the different Lorentz and
flavour contractions (indicated by the parentheses) are given in
Ref.~\cite{Beane:2002vq}. The super-script on the low energy constants
(LECs; $\alpha^{(r_A)}$, $\Delta\sigma^{(r_A)}$, etc.) in the
unpolarised and helicity operators labels the chiral representation to
which they belong; for $A=0$, $r_0=s$ (singlet) otherwise $r_A=a$
(adjoint).  In what follows, we take $A$ to be either $0$ or $3$. In
QCD, the two different flavour contractions of the operators
proportional to $\alpha_n^{(r_A)}$ and $\beta_n^{(r_A)}$ (and their spin
dependent analogues) are identical.

There are additional classes of operators that formally enter these
expressions at the same order but do not contribute to the
next-to-leading order (NLO) matrix elements, {\it i.e.}, their
contributions to one-loop diagrams vanish; for example,
\begin{equation}
\label{eq:omitted_ops}
v_{\{\mu_0}\ldots v_{\mu_{n-2}} \left(\overline{\cal
    T}_{\mu_{n-1}}S_{\mu_n\}}\bar\tau_A^{\xi^+}{\cal B}\right)\,,
\ \ \
{\rm and}
\ \ \ 
v_{\{\mu_0}\ldots v_{\mu_{n-1}} \left(\overline{\cal
    B}S_{\mu_n\}}\tau^{\xi^-}_{A} {\cal B}\right)\,.
\end{equation}
Such operators are omitted in
Eqs.~(\ref{eq:hadron_op})--(\ref{eq:hadron_sigmaop}).  Also, NLO
counter terms, such as
\begin{equation}
\left ( \overline{\cal B} \left \{ \tau^{\xi^{+}}_{A} , 
 {\cal M}_{+} \right \} {\cal B} \right ), \hspace*{0.3cm}
 {\mathrm{where}}\hspace*{0.2cm} {\cal M}_{+} = \frac{1}{2}
 \left (\xi^{\dagger} m_{Q} \xi^{\dagger} + \xi m_{Q} \xi \right ) ,
\end{equation}
are neglected in this work, since we are focusing on finite volume
effects arising from one-loop diagrams at NLO. For the unpolarised
isovector operators, these counterterms are explicitly displayed in
Ref.~\cite{Beane:2002vq}; for the other operators, they are simple
generalisations.

Additional, higher-order operators arise when powers of the baryon
velocity are replaced by derivatives, such as $v_{\{\mu_0}\ldots
(\frac{i}{M} \partial)_{\mu_{n-1}} \left(\overline{\cal
    B}S_{\mu_n\}}\tau^{\xi^-}_{A} {\cal B}\right)$.  In the forward
limit, these operators only appear in loop diagrams, so their
contributions to matrix elements start at next-to-next-to-leading
order (NNLO), therefore we do not include them in this work.

\subsection{Nucleon matrix elements}

\begin{figure}[!t]
  \centering
  \includegraphics[width=4.5cm]{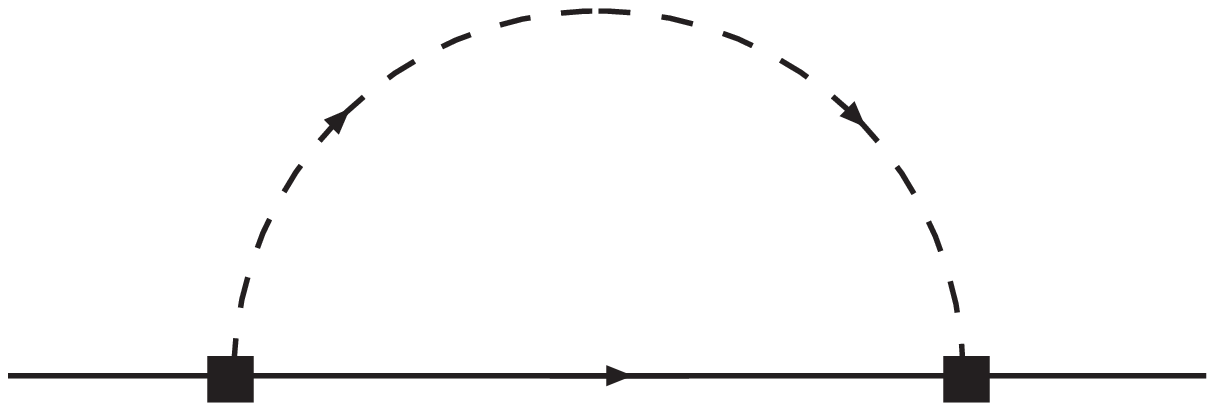}
  \includegraphics[width=4.5cm]{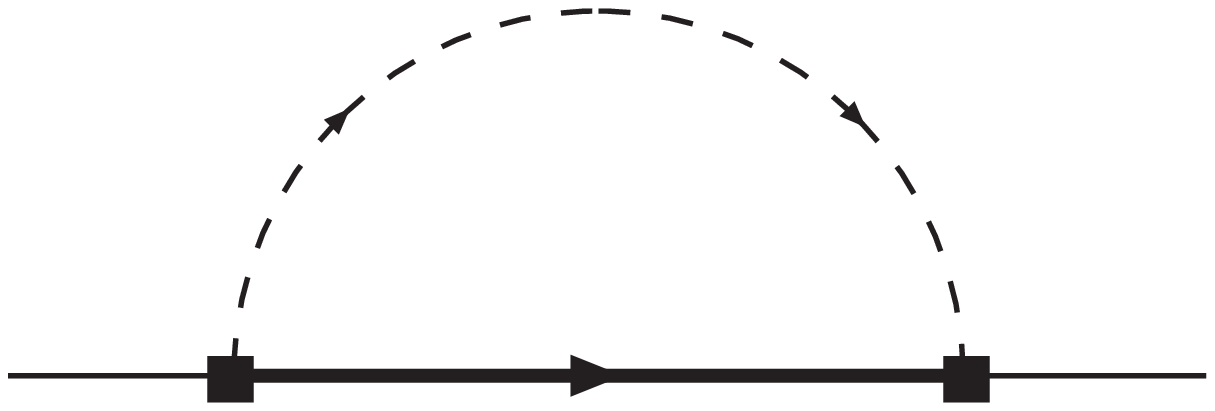}\\
  (a) \hspace{4.cm} (b) \\
  \vspace*{5mm}
  \includegraphics[width=4.5cm]{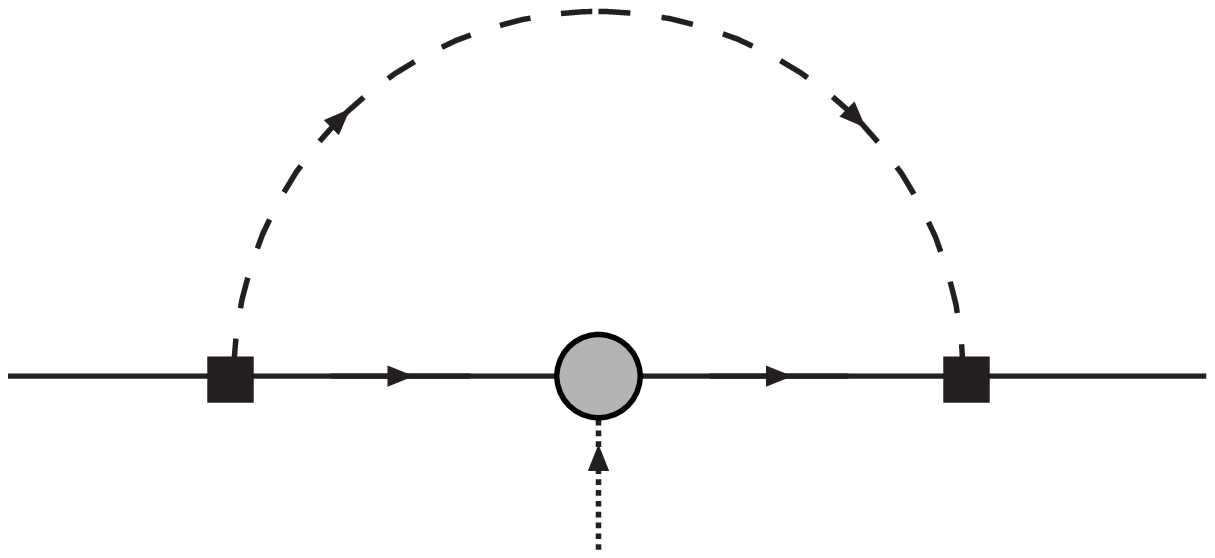}
  \includegraphics[width=4.5cm]{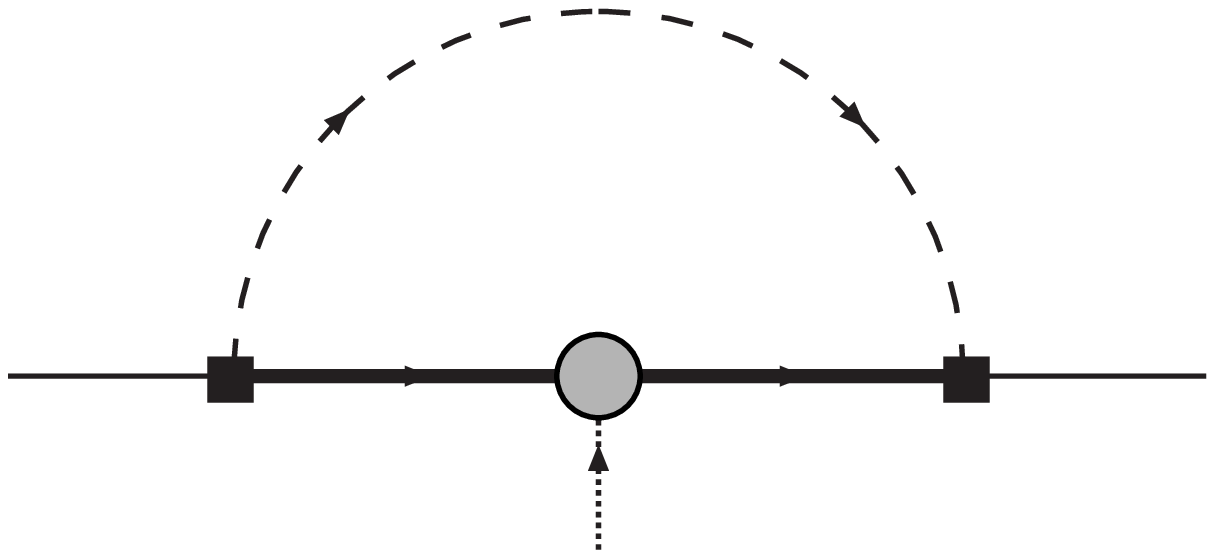}
  \includegraphics[width=4.5cm]{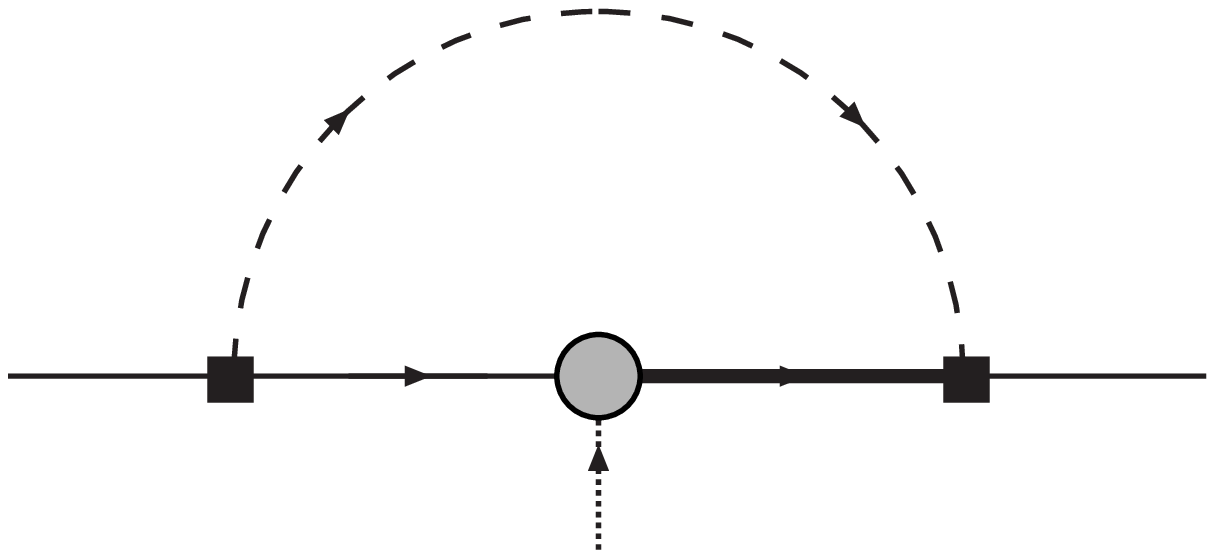}\\
  (c) \hspace{4.cm} (d)  \hspace{4.cm} (e)\\
  \vspace*{5mm}
  \includegraphics[width=4.5cm]{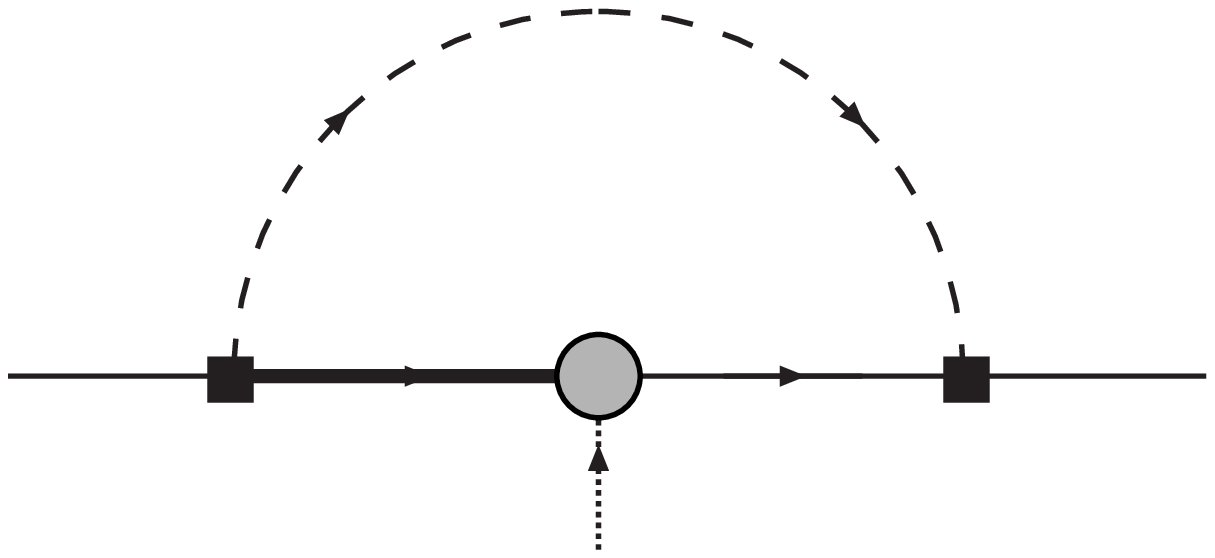}
  \includegraphics[width=4.5cm]{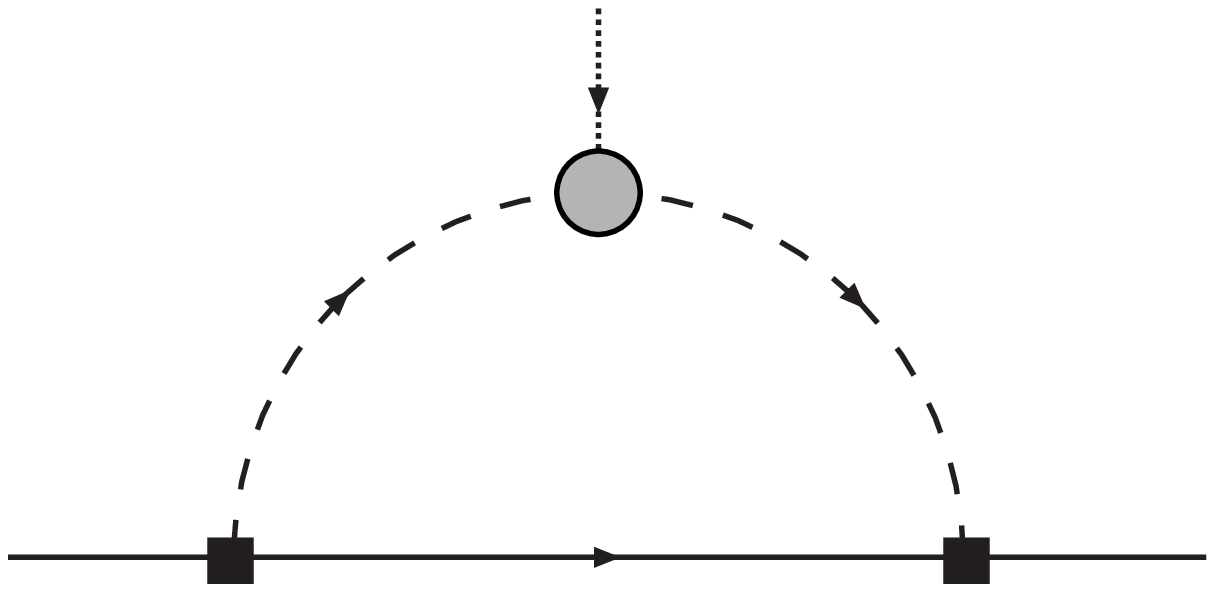}
  \includegraphics[width=4.5cm]{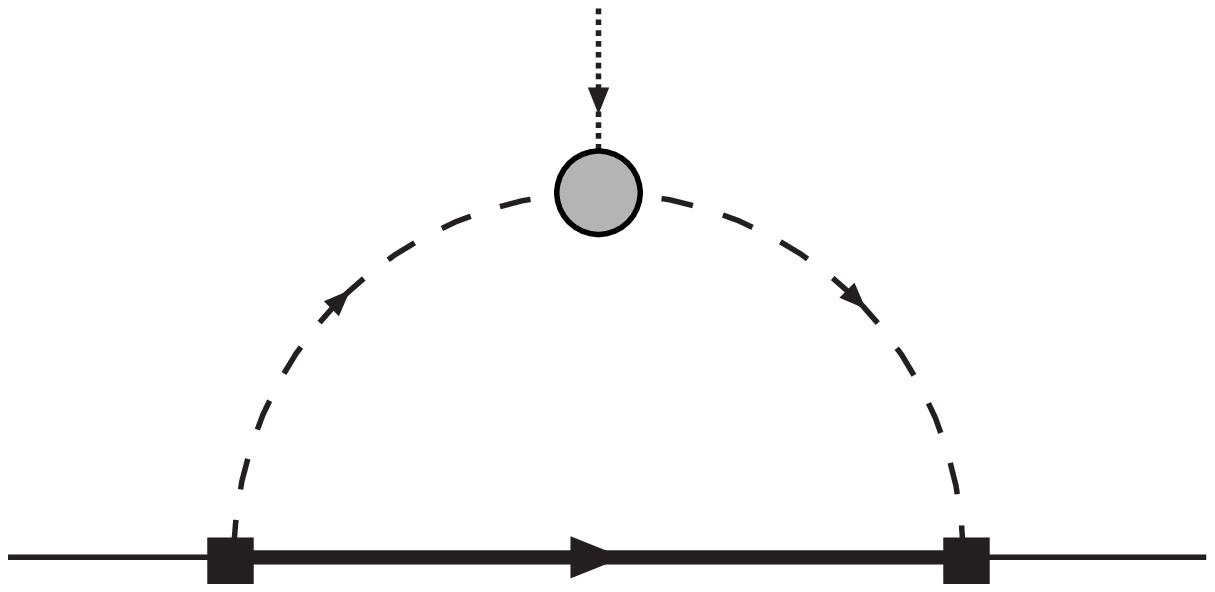}\\
  (f)\hspace{4cm} (g) \hspace{4cm} (h) \\
  \vspace*{5mm}
  \includegraphics[width=3.5cm]{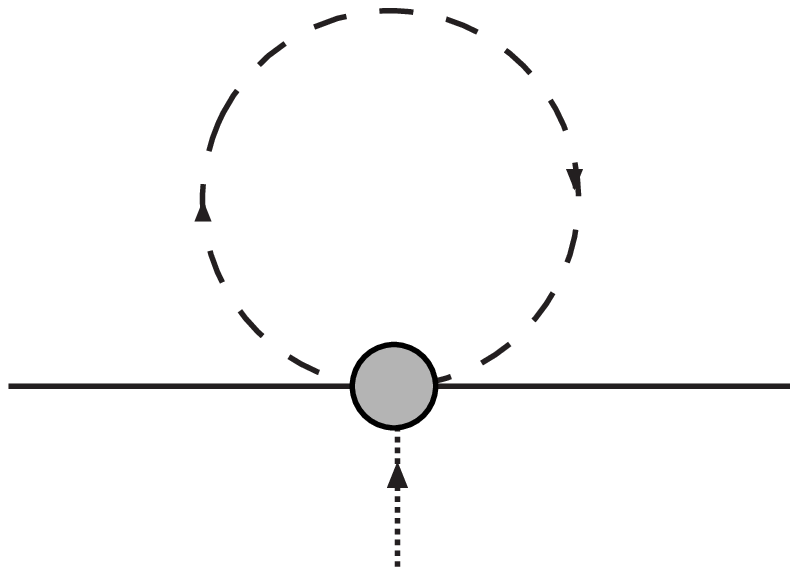}
  \hspace*{10mm}
  \includegraphics[width=3.5cm]{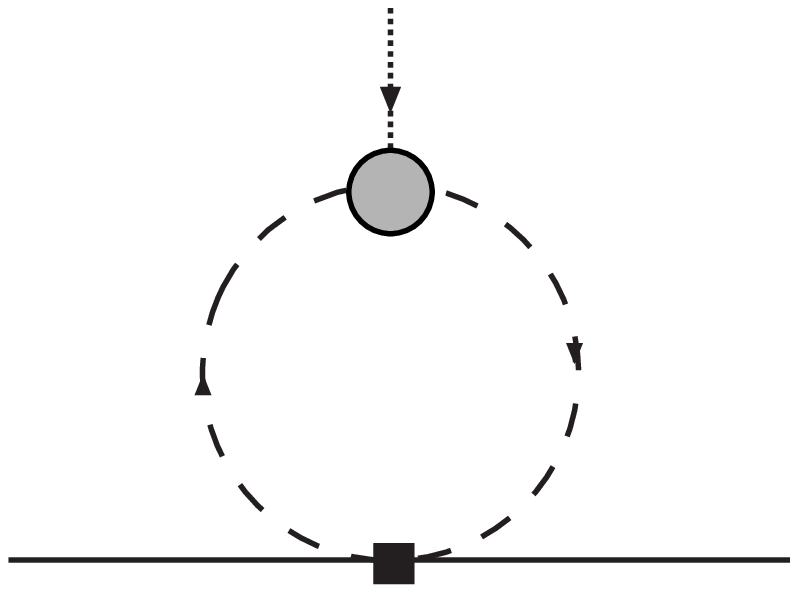}\\
  (i) \hspace{4.2cm} (j) \\
  \caption{Diagrams contributing to nucleon matrix elements of the
    twist-two operators. The black square corresponds to an
    interaction from the strong Lagrangian and the gray circle
    represents an insertion of the twist-two operators in
    Eqs.~\protect(\ref{eq:hadron_op})--\protect(\ref{eq:hadron_sigmaop}).
    The thin, thick and dashed lines are {\bf 70}--plet baryons, {\bf
      44}--plet baryons and mesons respectively. The first two
    diagrams represent the wave-function renormalisation and the
    remainder are operator renormalisations. Diagrams (e) and (f) are
    absent for the unpolarised, and isoscalar operator matrix
    elements, and diagrams in which the twist-two operator is inserted
    on a meson line are only present in the unpolarised case.}
  \label{fig:nucleondiagrams}
\end{figure}

The one-loop diagrams that contribute to nucleon twist-two matrix
elements at NLO are shown in Fig.~\ref{fig:nucleondiagrams}. The first
two diagrams, (a) and (b), represent the wave-function renormalisation
whilst the other diagrams are operator renormalisations. Diagrams (e)
and (f) are absent for the unpolarised or isoscalar operator matrix
elements as the transition between {\bf 70}--plet and {\bf 44}--plet
baryon states changes spin and isospin. Finally, diagrams in which the
twist-two operator is inserted on a meson line [(g), (h) and (j)] are
only present in the spin-averaged cases.

In Appendices~\ref{A2}, \ref{A3} and \ref{A4}, we give the results for
the independent matrix elements in SU(4$|$2) \pqxpt, for SU(6$|$3)
\pqxpt\ and for SU(2$|$2) quenched \xpt\ in the isospin limit.  As an
example, here we present the SU(4$|$2) isospin limit ($m_u=m_d,\,
m_j=m_l$), $q_j=q_l$ result for the nucleon matrix element of the
isovector, unpolarised operators ${\cal O}_{\mu_0\ldots\mu_n}^{(3)}$:
\begin{eqnarray}
 \langle N|{\cal O}^{(3)}_{\mu_0\ldots\mu_n}|N\rangle &=&  
\frac{1}{3}{\overline{U}_p v_{\mu_0}\ldots
  v_{\mu_n}U_p}(2\alpha_n^{(a)}-\beta_n^{(a)})\times(1+(1-\delta_{n0}){\cal W}_{SU(4|2)})
\nonumber \\&&+\frac{i(1-\delta_{n0}) }{6f^2}\,{v_{\mu_0}\ldots v_{\mu_n}}\,
    \Bigg\{ \frac{8}{3}\,g_{\Delta N}^2\,
         \left( {{\gamma }_n^{(a)}} - \frac{{{\sigma }_n^{(a)}}}{3} \right)
         \Big[ 3\,{\cal H}({M_{\pi }},\Delta )\, + 
           2\,{\cal H}({M_{uj}},\Delta )\,  \Big] \,
\nonumber \\&& \hspace*{1cm}
       + {{\alpha }_n^{(a)}}\,\Bigg[ 8\,i \,{\cal I}({M_{uj}})\, 
          -3\,{\cal H}({M_{uj}},0)\,
               \Big[ 2\,{g_1}\,{g_A}\, + 3\,g_1^2\,  \Big]
           + 3{\cal H}({M_{\pi }},0)\,\Big[ 2\,g_A^2 +  
               2\,{g_1}\,{g_A}  + 3\,g_1^2\,  \Big]  
\nonumber \\&& \hspace*{3cm}
              + 12\,{\delta }^2\,{\left( {g_1} + {g_A} \right)
               }^2\,{{\cal H}_{\eta^\prime}}({M_{\pi }},0)
         \Bigg]  
\nonumber \\&& \hspace*{1cm}
 -  {{\beta }_n^{(a)}}\,\Bigg[ 4\,i \,{\cal I}({M_{uj}})\,
         -12{\cal H}({M_{uj}},0)\,{g_1}{g_A}  + 
            3{\cal H}({M_{\pi }},0)\,\left[ 4\,{g_1}\,{g_A} + g_A^2\right] 
\nonumber \\&& \hspace*{3cm}
             +  6 \,{\delta }^2\,{\left( {g_1} + {g_A} \right)
               }^2\,{{\cal H}_{\eta^\prime}}({M_{\pi }},0)
            \Bigg]  
\Bigg\} ,
\label{eq:unpol_isovec_su42_maintext}
\end{eqnarray}
where $\calW_{\rm SU(4|2)}$ is the nucleon wave-function
renormalisation given in Eq.~(\ref{eq:waverenorm_su42}) and
\begin{equation}
\label{eq:delta_def}
 \delta^{2} = M_{\pi}^{2} - M_{uj}^{2} = \lambda (m_{u} - m_{j}) ,
\end{equation}
is proportional to the difference between sea and valence quark
masses. The functions $\calI(M)$, $\calH(M,\Delta)$ and
$\calHdp(M,\Delta)$ are defined in Eqs.~(\ref{eq:def_tadint}),
(\ref{eq:def_calH}) and (\ref{eq:def_calHdp}), below. Finally, $U_B$
corresponds to the type $B$ baryon spinor. To take the QCD limit, we
would set $\delta\to0$ and $j\to u$. For equivalent choices of the
$\bar\tau_A$ our results reproduce those found previously for the
unpolarised isovector operator \cite{Beane:2002vq,Chen:2001yi}.  One
can also calculate the matrix elements of these operators in the
$\Delta$-isobar (and the N--$\Delta$ transition in the spin dependent
cases).  However since these are not stable particles in much of the
region where \xpt\ converges, we do not present the expressions.

In these results, the only effect of the diagrams in which the
twist-two operator couples to a meson (diagrams (g), (h) and (j) in
Fig.~\ref{fig:nucleondiagrams}) is to satisfy the number sum rule for
the $n=0$ matrix elements, producing the $\delta_{n0}$ factors in the
above expression.  For $n>0$, these diagrams give sub-leading
contributions, entering at ${\cal O}(p^{n+2})$. The number sum-rule
also fixes
\begin{eqnarray}
&2\alpha_0^{(a)}-\beta_0^{(a)}=3,\quad\quad
\gamma_0^{(a)}=3,\quad \quad \sigma_0^{(a)}=0 \,,
& \nonumber \\
&\alpha_0^{(s)}+\beta_0^{(s)}=3, \quad\quad \gamma_0^{(s)}=3, 
\quad \quad \sigma_0^{(s)}=0 \,,
&
\label{eq:n0number}
\end{eqnarray}
and the $n=0$ low energy constants of the spin-dependent operators can
be fixed in terms of the usual axial couplings
\begin{eqnarray}
&2\Delta \alpha_0^{(a)}-\Delta \beta_0^{(a)}=6g_A, \quad\quad
\Delta\gamma_0^{(a)}=2g_{\Delta\Delta}, \quad \quad 
\Delta\sigma_0^{(a)}=0, \quad\quad
\Delta c_0^{(a)}= -\sqrt{\frac{3}{2}} g_{\Delta N} \,,
&\nonumber\\
&\Delta \alpha_0^{(s)}+\Delta \beta_0^{(s)}=2\left(g_A+g_1\right), \quad\quad
\Delta\gamma_0^{(s)}=0, \quad \quad 
\Delta\sigma_0^{(s)}=0\,.
&
\label{eq:n0spin}
\end{eqnarray}


\section{Finite volume corrections}
\label{S4}

\subsection{General discussion}

In momentum space, the finite volume of a lattice simulation restricts
the available momentum modes. Here we shall consider a hyper-cubic box
of dimensions $L^3\times T$ with $T\gg L$. Imposing periodic boundary
conditions on mesonic fields leads to quantised momenta $k=(k_0,{\vec
  k})$, ${\vec k}=\frac{2\pi}{L} {\vec j}=\frac{2\pi}{L}
(j_1,j_2,j_3)$ with $j_i\in \mathbb{Z}$, but $k_0$ treated as
continuous.  On such a finite volume, spatial momentum integrals are
replaced by sums over the available momentum modes. This leads to
modifications of the infinite volume results presented in the previous
section; the various functions arising from loop integrals are
replaced by their FV counterparts. In a system where $M_\pi L\gg 1$,
finite volume effects are predominantly from Goldstone mesons
propagating to large distances where they are sensitive to boundary
conditions and can even ``wrap around the world''.  Since the lowest
momentum mode of the Goldstone propagator is $\sim \exp(-M_\pi L)$ in
position space, finite volume effects will behave as a polynomial in
$1/L$ times this exponential if no multi-particle thresholds are
reached in the loop.

To investigate this behaviour, we consider the various finite volume
sums occurring in the twist-two matrix elements. We first define
\begin{eqnarray}
 & &\frac{1}{L^{3}} \sum_{\vec{k}} \int dk_{0} \frac{k_{\mu} k_{\nu}}
 {(k^{2}-m^{2}+i\epsilon)(k\cdot v - \Delta + i\epsilon)}
 -\frac{i g_{\mu\nu}}{16\pi^{2}} 
  \bar{\lambda} \left ( \frac{2\Delta^{2}}{3}-m^{2} 
   \right )\Delta\nonumber\\&&\nonumber\\
\label{eq:def_calF}
 && = g_{\mu\nu} \calF (m,\Delta) + v_{\mu} v_{\nu} {\mathcal{G}} (m,\Delta)\,,
\end{eqnarray}
where the ultra-violet divergence has been subtracted in dimensional
regularisation\footnote{It is important to note that because of the
  separation of scales, FV effects (infrared) are essentially
  independent of method chosen to regulate the divergent integrals
  (ultraviolet).  Also, the results presented in this work are derived
  in Minkowski space. We are free to work in Minkowski space since the
  sicknesses of quenched and partially quenched theories discussed in
  Refs.~\cite{Bernard:1995ez,Colangelo:1997ch} do not occur in our
  calculations.} with $\bar{\lambda} = \frac{2}{4-d} - \gamma_{E} +
{\mathrm{log}}(4\pi) + 1$ ($d$ is the number of dimensions).  All
finite volume sums that occur in the baryon wave function and operator
renormalisations involving baryon propagators (diagrams (a)--(h) in
Fig.~\ref{fig:nucleondiagrams}, sunset-type diagrams) can be expressed
in terms of $\calF(m,\Delta)$ and its derivatives. The tadpole
diagrams in Fig.~\ref{fig:nucleondiagrams} are discussed in
Appendix~\ref{A1}.  In the baryon rest frame where $v = (1,0,0,0)$,
Poisson's summation formula allows us to decompose $\calF$ into its
infinite-volume limit and a volume-dependent part,
\begin{equation}
\label{eq:decompose_calF}
\calF(m,\Delta) = F(m,\Delta) + F^{\mathrm{FV}}(m,\Delta)\,.
\end{equation}
It is straightforward to show that the infinite volume piece is
\begin{equation}
\label{eq:InfV_F}
 F(m,\Delta) = \frac{i}{16\pi^{2}} \bigg \{
   \bigg [ m^{2} - \frac{2 \Delta^{2}}{3} \bigg ]\,\Delta\,
    {\mathrm{log}}\left (\frac{m^{2}}{\mu^{2}} \right )
    + \bigg [ \frac{10\Delta^{2}}{9} - \frac{4 m^{2}}{3} \bigg ]\,
    \Delta + \frac{2}{3} (\Delta^{2} - m^{2})^{3/2} 
    \log\left(\frac{\Delta - \sqrt{\Delta^{2}-m^2+i\epsilon}}{\Delta +
        \sqrt{\Delta^{2}-m^2 +i\epsilon}}\right)
 \bigg \}\,
\end{equation}
($\mu$ is the renormalisation scale), and the finite volume
corrections are given by
\begin{eqnarray}
 F^{\mathrm{FV}}(m,\Delta) &=& \frac{i}{12\pi^{2}} \sum_{\vec{u}\not=
   \vec{0}} \frac{1}{u\, L}
 \int_{0}^{\infty} d |\vec{k}| \,
 \frac{|\vec{k}|\,{\mathrm{sin}}(u |\vec{k}| L)}{
     \sqrt{|{\vec k}|^2+m^2} + \Delta}
 \left ( \Delta +\frac{m^2}{\sqrt{|{\vec k}|^2+m^2}} \right ) 
 \nonumber\\&&\nonumber\\
 &\stackrel{m L \gg 1}{\longrightarrow}& 
 \frac{i\, m^2}{24\pi}\sum_{\vec{u}\not=
   \vec{0}}\frac{{\mathrm{e}}^{-u m L}}{ u \,L} 
   {\mathcal{A}} \,,
\end{eqnarray}
where $\vec u=(u_1,u_2,u_3)$ with $u_i\in \mathbb{Z}$, $u \equiv
|\vec{u}|$ and
\begin{eqnarray}
\label{eq:calA_def}
 {\mathcal{A}} &=&\exponential^{(z^{2})} \big [ 
1 - {\mathrm{Erf}}(z)\mbox{ }\big ]
+\left (\frac{1}{u m L} \right ) \bigg [
 \frac{1}{\sqrt{\pi}} \left ( \frac{9z}{4} - 
\frac{z^{3}}{2}\right )
 + \left(\frac{z^{4}}{2}-2\,z^2\right)\exponential^{(z^{2})} 
 \big [ 1 - {\mathrm{Erf}}(z)\mbox{ }\big ]
\bigg ]\\
& &
-\left (\frac{1}{u m L} \right )^{2}\bigg [
\frac{1}{\sqrt{\pi}}\left ( -\frac{39z}{64} + 
\frac{11z^{3}}{32}
  -\frac{9z^{5}}{16} + \frac{z^{7}}{8} \right )
-\left ( -\frac{z^{6}}{2} + \frac{z^{8}}{8}\right )
\exponential^{(z^{2})} \big [ 1 - {\mathrm{Erf}}(z)\mbox{ }\big ]
\bigg ]
+\op\left ( \frac{1}{(u m L)^{3}}\right ) ,
\nonumber
\end{eqnarray}
with
\begin{equation}
 z = \left ( \frac{\Delta}{m} \right )
  \sqrt{\frac{u m L}{2}} .
\end{equation}
Higher order terms in the $1/(u m L)$ expansion in
Eq.~(\ref{eq:calA_def}) are easily calculated.  For convenience, we
also define the functions
\begin{equation}
\label{eq:def_calH}
 \calH(m,\Delta) \equiv \frac{\partial \calF(m,\Delta)}{\partial \Delta}
\qquad {\rm and} \qquad
 \calK (m,\Delta) \equiv \frac{\calF(m,\Delta)-\calF(m,0)}{\Delta} \,,
\end{equation}
and their finite volume counterparts which we denote by the
corresponding roman letter with the superscript $^{\rm FV}$ as in
Eq.~(\ref{eq:decompose_calF}), e.g.  $\calK\to K^{\rm FV}$.

\begin{figure}[!t]
  \centering\hspace*{-1cm} \includegraphics[width=9.5cm]{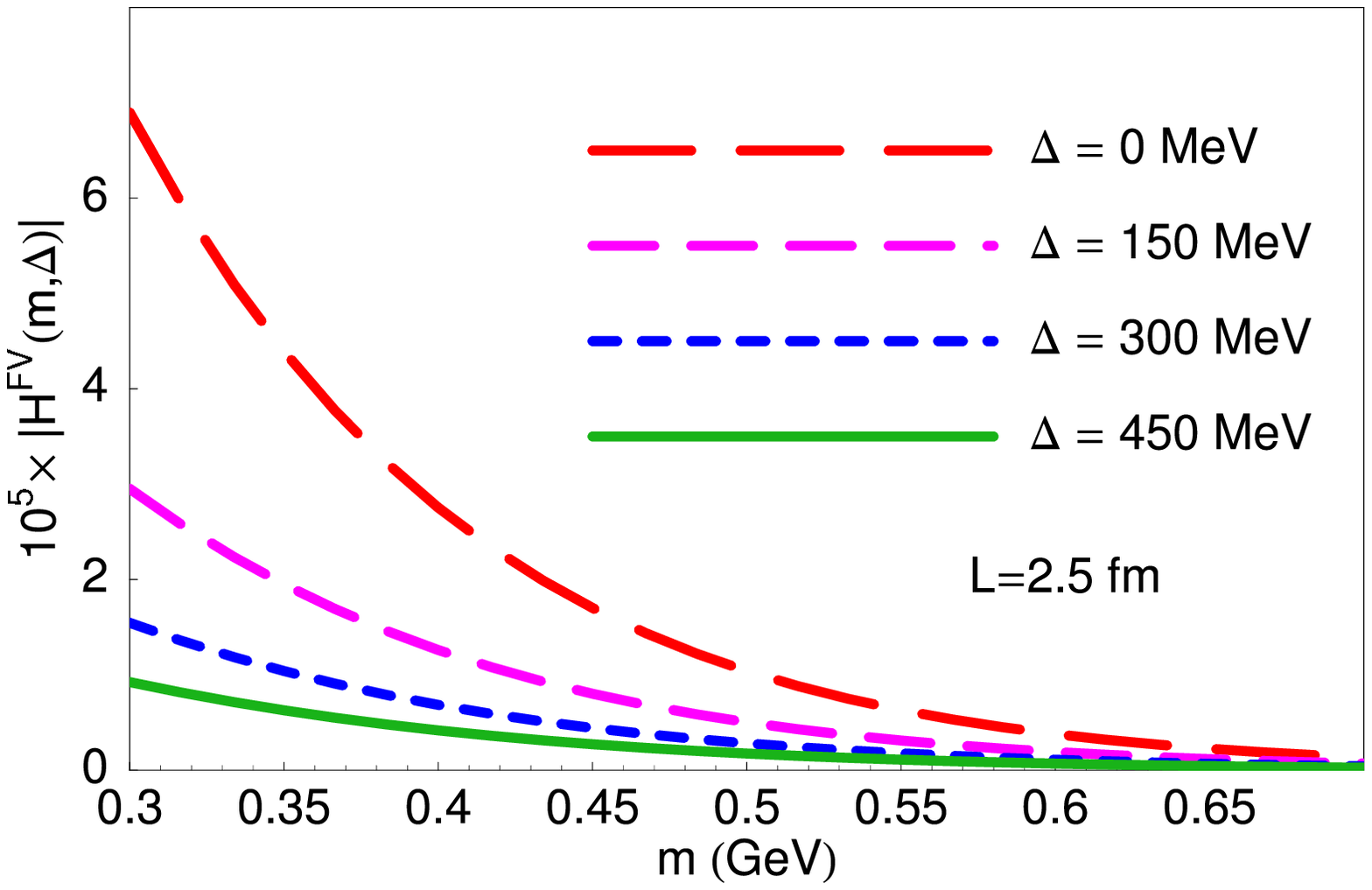} \hspace{-1cm}
  \includegraphics[width=9.5cm]{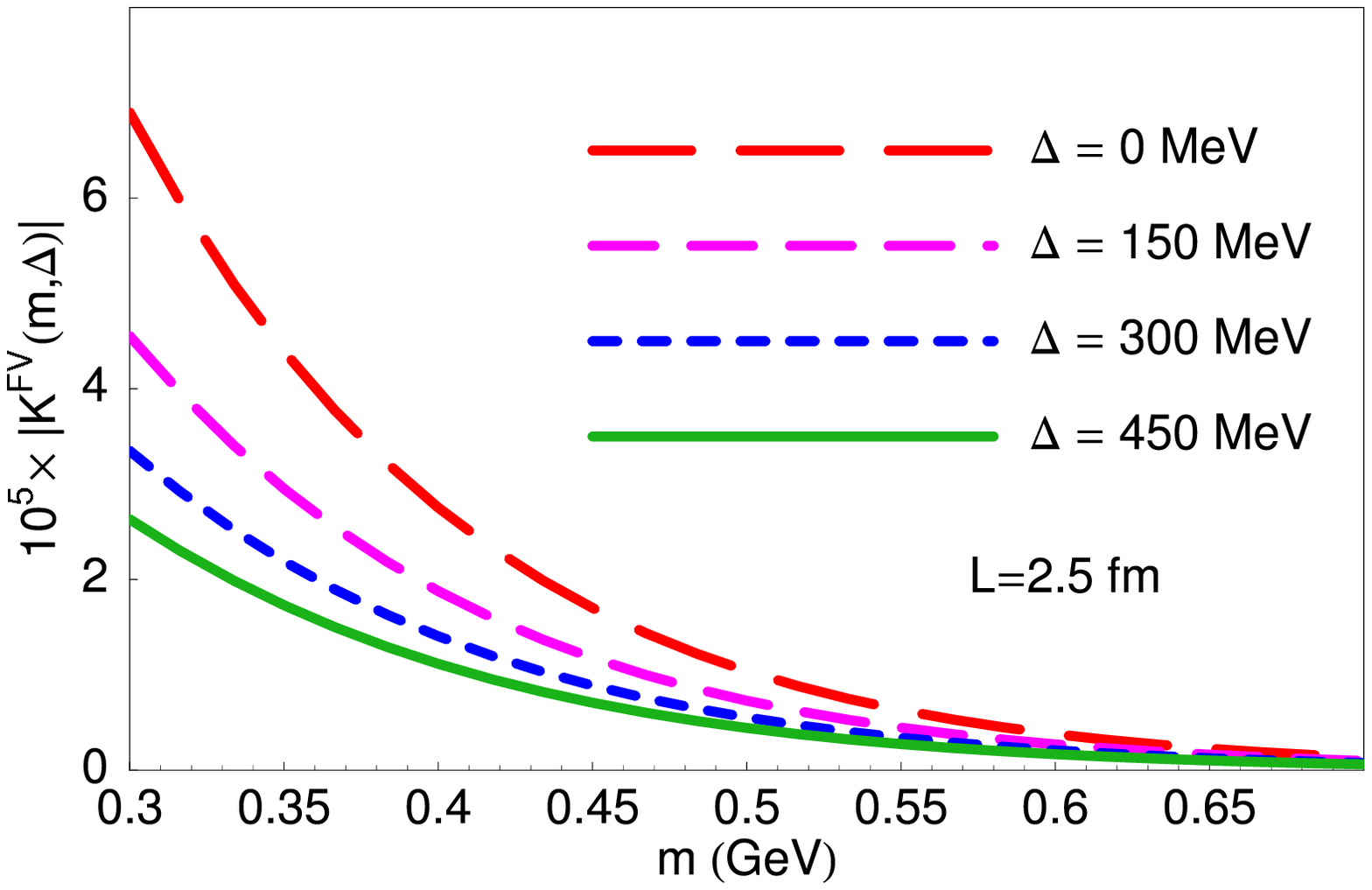}
  \caption{Dependence of finite volume effects on the mass-splitting
    $\Delta$ in individual integrals/sums corresponding to diagrams
    (a)---(f) in Fig.~\protect\ref{fig:nucleondiagrams}. The point
    $m=0.3$~GeV corresponds to $m L=3.8$.}
  \label{fig:FV_inloopsums}
\end{figure}
The function $\cal A$ represents the modification of the FV effects
due to the mass-splitting $\Delta$; in the limit $\Delta\to0$, ${\cal
  A}\to1$.  Figure~\ref{fig:FV_inloopsums} shows the dependence of FV
effects on the scale $\Delta$ for the functions $H^{\rm FV}(m,\Delta)$
and $K^{\rm FV}(m,\Delta)$ that arise from diagrams (a)---(f) in
Fig.~\ref{fig:nucleondiagrams}.  It is clear that finite volume
effects in individual diagrams involving a {\bf 44}-plet are
suppressed relative to those involving only meson and {\bf 70}-plet
baryon propagators, though this can be compensated for by large
coefficients. A very similar result was found in the heavy meson
sector \cite{Arndt:2004bg} where the contributions involving $B^*$
mesons are suppressed compared to those involving the $B$ meson by the
mass difference $\Delta_B=m_{B^*}-m_B$.  However, the origin and
behaviour of the mass difference in the current context is distinct.
In contrast to the heavy meson case where $\Delta_B\sim 1/m_B$ arises
from the breaking of heavy-quark spin symmetry and vanishes in the
heavy quark limit, the mass difference $\Delta$ is generated by
strong-interaction dynamics and remains finite in the chiral limit.
Empirically, $\Delta$ is almost constant over the range of quark
masses considered here.

\begin{figure}[!t]
  \centering \hspace*{-1cm}
  \includegraphics[width=9.5cm]{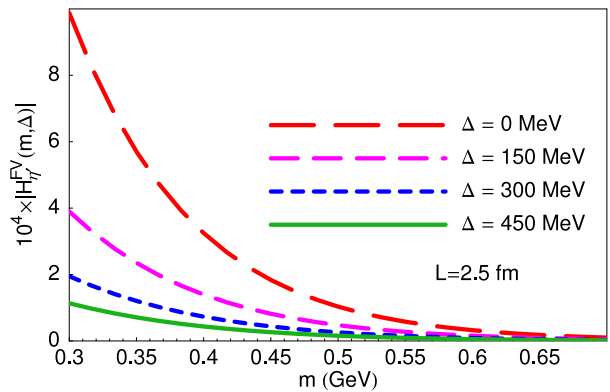}
  \caption{Dependence of finite volume effects in double-pole contributions
    on the mass-splitting $\Delta$. Note that the scale here is ten
    times that in Fig.~\protect{\ref{fig:FV_inloopsums}}. The point
    $m=0.3$~GeV corresponds to $m L=3.8$. }
  \label{fig:FV_inDPloopsums}
\end{figure}

When one considers quenched or partially quenched theories rather than
standard \xpt, one expects somewhat larger finite volume effects
because of the enhanced long-distance behaviour of double-pole
structures in the singlet meson propagators of these theories
\cite{Bernard:1995ez}. These double pole contributions are given by
terms proportional to the functions
\begin{equation}
\label{eq:def_calHdp}
 \calH_{\eta^{\prime}}(m,\Delta) \equiv \frac{\partial
   \calH(m,\Delta)}{\partial m^{2}} \,,
\qquad\qquad\qquad
\calKdp (m,\Delta) \equiv \frac{\partial\calK(m,\Delta)}{\partial
  m^{2}} \,,
\end{equation}
and the double-pole tadpole function $\calIdp(m)$ given in
Appendix~\ref{A1} [and their finite volume analogues constructed as in
Eq.~(\ref{eq:decompose_calF})]. From Figures~\ref{fig:FV_inloopsums}
and \ref{fig:FV_inDPloopsums} it is clear that $H^{\rm
  FV}_{\eta^{\prime}}(m,\Delta)$ is about an order of magnitude larger
than $H^{\rm FV}(m,\Delta)$ in accordance with expectations.

In considering the magnitude of finite volume effects, the standard
chiral power counting can be misleading; the FV effects of a diagram
of a given order in the power counting may be larger than those of
lower orders. For some generic observable, one may consider two
contributions, ${\cal C}_1$ and ${\cal C}_2$, that enter at different
orders, $m_1<m_2$, in infinite volume \xpt. As discussed above, the
dominant finite volume effects in these contributions will typically
be of the form $\delta {\cal C}_i\sim (M_\pi L)^{\ell_i}\exp(-M_\pi
L)$ when $M_\pi L\gg1$ and no multi-particle on-shell intermediate
states can contribute.  In some situations, the presence of additional
meson propagators or other infrared enhancement in the higher order
contribution (${\cal C}_2$) can amplify its finite volume shift
relative to that of the lower order contribution (${\cal C}_1$).  For
some (contemporarily relevant) choices of masses and volumes, the
quantity
\begin{equation} 
(M_\pi
L)^{\ell_2-\ell_1}\left(\frac{M_\pi}{\Lambda_\chi}\right)^{m_2-m_1}>1\,,
\end{equation}
and the formally higher-order contribution will provide the dominant
finite volume effect.  In the current calculation, diagram (g) in
Fig.~\ref{fig:nucleondiagrams}, in which the twist-two operator is
attached to mesonic propagators, may indeed fall into such a category.
The finite volume corrections to these diagrams will be given by
\begin{equation}
  \label{eq:piondiagram_FV}
  \delta I_\pi^{(n)}\sim
  a^{(a)}_{n}\frac{1}{\Lambda_\chi^n}
  \frac{M_\pi^{n+1}}{4f^2}\frac{\partial}{\partial M_\pi^2} F^{\rm FV}(M_\pi,0)\,,
\end{equation}
compared with those of the corresponding baryon operator diagrams
(diagram (c) in Fig.~\ref{fig:nucleondiagrams})
\begin{equation}
  \label{eq:nucleondiagram_FV}
  \delta I_N^{(n)}\sim \frac{\alpha_{n}^{(a)}}{f^2}  H^{\rm FV}(M_\pi,0)\,.
\end{equation}
From this we see that the ratio
\begin{equation}
  \label{eq:pi_nuc_ratio}
  \frac{ \delta I_\pi^{(n)}}{\delta I_N^{(n)}} \sim
 \left(\frac{M_\pi}{\Lambda_\chi}\right)^n
 \frac{1}{4\sqrt{M_\pi L}}\left(1-\frac{M_\pi\, L}{2}\right) \,,
\end{equation}
where there is an undetermined coefficient involving $a_{n}^{(a)}$,
$\alpha_n^{(a)}$ and other numerical factors that we assume to be
${\cal O}(1)$.  Whilst formally the magnitude of this ratio is indeed
greater than unity for any $n$ in the limit $M_\pi\,L\to\infty$, both
$\delta I_\pi^{(n)}$ and $\delta I_N^{(n)}$ are exponentially
suppressed.  For realistic pion masses and volumes used in current
lattice simulations this ratio is consistently smaller than one and
the FV effects of diagrams (g), (h) and (j) in
Fig.~\ref{fig:nucleondiagrams} can be neglected. The only exception to
this is in the $n=0$ unpolarised matrix elements, the isoscalar and
isovector quark numbers. Here, the volume dependence of diagrams in
Fig.~\ref{fig:nucleondiagrams} with mesonic operators exactly cancels
that of those involving baryonic operators and wave-function
renormalisations to give an overall result that is independent of the
volume.

\subsection{Relevance to lattice data}

Lattice calculations of twist-two matrix elements have a long history,
with the first calculations occurring in the 1980s
\cite{Martinelli:1987si}.  Over the last decade, a considerable effort
has been made to study them in detail with major contributions from
the QCDSF, LHP, RBCK and ZeRo collaborations. In Table~\ref{tab:data},
recent simulation parameters are shown.  State-of-the-art lattice
simulations are beginning to enter the region of quark masses and
lattice volumes in which the use of NLO chiral perturbation theory is
justified (naively, this requires $m_\pi/\Lambda_\chi\alt 1/3$ and
$m_\pi L \agt 4$). At the moment however, there is little data in this
region and a realistic fit of the combined mass and volume dependence
in our \pqxpt\ formulae (and thereby a determination of the LECs) is
not possible; only general trends can be extracted.
\begin{table}[t]
  \centering
  \begin{ruledtabular}
    \begin{tabular}{cccl}
      Group & $m_\pi$ [MeV] & Volume [fm$^3$] & Notes \\
      \hline
      \multicolumn{4}{c}{Dynamical simulations ($N_f=2$)} \\
      \hline
      QCDSF \cite{Gockeler:2004vx}     &  560--940  &     1.1$^3$,
      1.5$^3$, 2.2$^3$             & Clover, a$\sim$0.08--0.12~fm\\ 
      LHP \cite{Renner:2004ck}       &  340       & 3.5$^3$          & Staggered sea, DW valence,
      a$\sim$0.13 fm\\
      LHP \cite{Renner:2004ck} &  340--730  & 2.6$^3$          & Staggered sea, DW
                 valence, a$\sim$0.13 fm\\
      LHP-SESAM \cite{Dolgov:2002zm} &  730--900  & 1.6$^3$          & Wilson, a$\sim$0.1~fm\\
      LHP-SCRI \cite{Dolgov:2002zm}  &  480--670  & 1.5$^3$          & Wilson, a$\sim$0.1~fm\\
      RBCK \cite{Ohta:2004mg}      &  560--700  &    1.9$^3$         & DW, a$\sim$0.13~fm \\
      \hline
      \multicolumn{4}{c}{Quenched simulations} \\
      \hline
      QCDSF \cite{Gockeler:2004wp}     & 580--1200  & 1.6$^3$ & Clover, a$\sim$0.05, 0.07, 0.09 fm \\
      QCDSF \cite{Gockeler:1995wg}           & 310--1000     & 1.5$^3$          & Wilson, a$\sim$0.09 fm \\
      QCDSF \cite{Gurtler:2004ac}       & 440--950   & 1.5$^3$, 2.3$^3$ & Overlap, a$\sim$0.095 fm  \\
      LHP \cite{Dolgov:2002zm}  & 580--820   & 1.6$^3$          & Wilson,a$\sim$0.1~fm\\
      RBCK \cite{Sasaki:2003jh}       & 390--850   &
      1.2$^3$,1.6$^3$,2.4$^3$               & DW, a$\sim$0.15~fm \\ 
      ZeRo \cite{Wetzorke:2004xv}      & 750--910   &
      1.1$^3$,1.5$^3$,2.2$^3$,3.0$^3$ & Clover, a$\sim$0.093 fm\\ 
    \end{tabular}
  \end{ruledtabular}
  \caption{Summary of recent lattice calculations of nucleon twist two
    matrix elements. Not all calculations involve the full set of
    twist-two operators.  }
  \label{tab:data}
\end{table}

In order to address the expected size of finite volume corrections
arising from our calculation, we first define the finite volume
correction
\begin{equation}
  \label{eq:ME_FV_def}
  \langle N | {\cal O} | N \rangle^{\rm FV} = \frac{\langle N | {\cal O} | N
  \rangle_{L}^{\rm one-loop} - \langle N | {\cal O} | N \rangle_{L=\infty}^{\rm one-loop}}{ \langle N
  | {\cal O} | N \rangle^{\rm tree-level}}\,,
\end{equation}
for each of the operator matrix elements we calculate. These
corrections depend on a number of low-energy constants and couplings,
some of which involve the $\Delta$ resonance. In principle, all of
these parameters can be extracted from fits of the \xpt\ forms to
lattice data on nucleon matrix elements (thereby bypassing issues of
the structure of unstable particles and transition matrix elements),
however this is not practical at the present stage. Therefore, to fix
the twist-two low energy constants ($\alpha_n^{(a)}$, $\delta\gamma_n$
etc) we assume that large-$N_c$ relations \cite{Chen:2001et} amongst
the parton distributions in the nucleon, $\Delta$-isobar and
$N$--$\Delta$ transition are valid, leading to
\begin{eqnarray}
  \label{eq:largeNC_constraints}
  \gamma_n^{(a)} &= 2\alpha_n^{(a)} - \beta_n^{(a)}\,,
\qquad\qquad
\Delta\gamma_n^{(a)} &= \frac{1}{5}\left(2\Delta\alpha_n^{(a)} 
- \Delta\beta_n^{(a)}\right)\,,
\qquad
\Delta c_n^{(a)} = \frac{1}{2}\left(2\Delta\alpha_n^{(a)} -
  \Delta\beta_n^{(a)}\right)\,,
\\
\nonumber   
  \gamma_n^{(s)} &= \alpha_n^{(s)} + \beta_n^{(s)}\,,
\qquad\qquad
  \Delta\gamma_n^{(s)} &= \Delta\alpha_n^{(s)} + \Delta\beta_n^{(s)}\,,
\qquad\qquad\quad
\delta\alpha_n = - 4\delta\beta_n=\frac{4}{5}\delta\gamma_n =
\frac{8}{9}\delta c_n\,.
\end{eqnarray}
The remaining LECs are not constrained by large-$N_c$ relations in
QCD, and for want of accurate lattice data with which to fit them, we
set $\beta_n^{(a,s)}=\alpha_n^{(a,s)}$,
$\Delta\beta_n^{(a,s)}=\Delta\alpha_n^{(a,s)}$ and $\sigma_n^{(a,s)} =
\Delta\sigma_n^{(a,s)} = \delta\sigma_n=0$.  Throughout, we use
$f=0.132$~GeV, and keep $\Delta=0.3$~GeV fixed independent of the
quark mass. For the parameters appearing in the the flavour matrices
$\bar\tau_3$ and $\bar\tau_T$, we set $q_j=q_l=q_k=0$ and
$y_j=y_k=y_l=y_m=0$, and set $y_i$ to be either $\pm 1$.  As discussed
above, if one is using lattice data to determine the LECs, the $q$'s
and $y$'s are fixed by the details of the lattice calculation.  After
making all of the above substitutions, the isospin limit results
become proportional to the corresponding bare matrix elements and the
finite volume effects given by Eq.~(\ref{eq:ME_FV_def}) are easily
studied.
\begin{figure}[!t]
  \centering
  \hspace*{-15mm}
  \includegraphics[width=10cm]{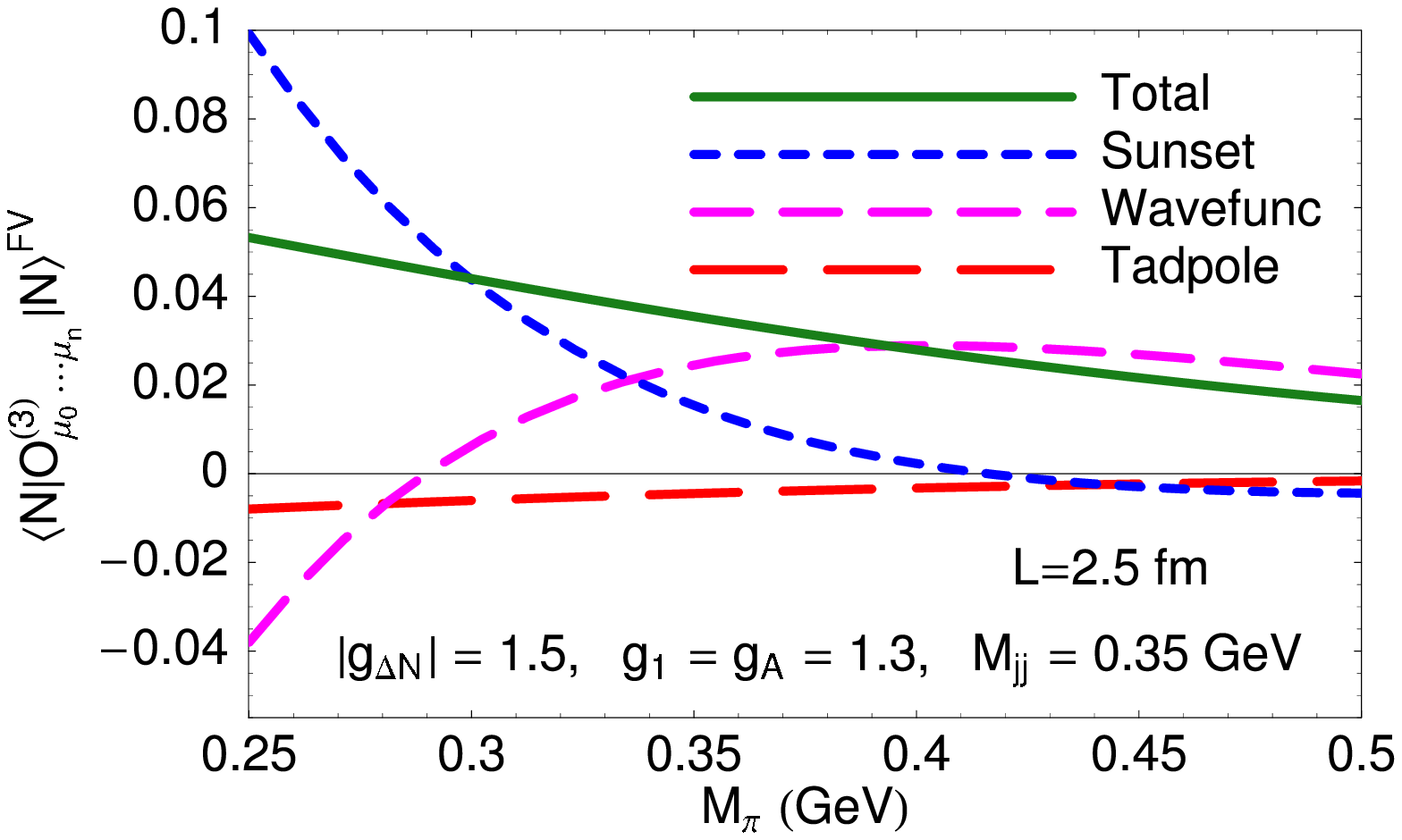}
  \hspace{-15mm}
  \includegraphics[width=10cm]{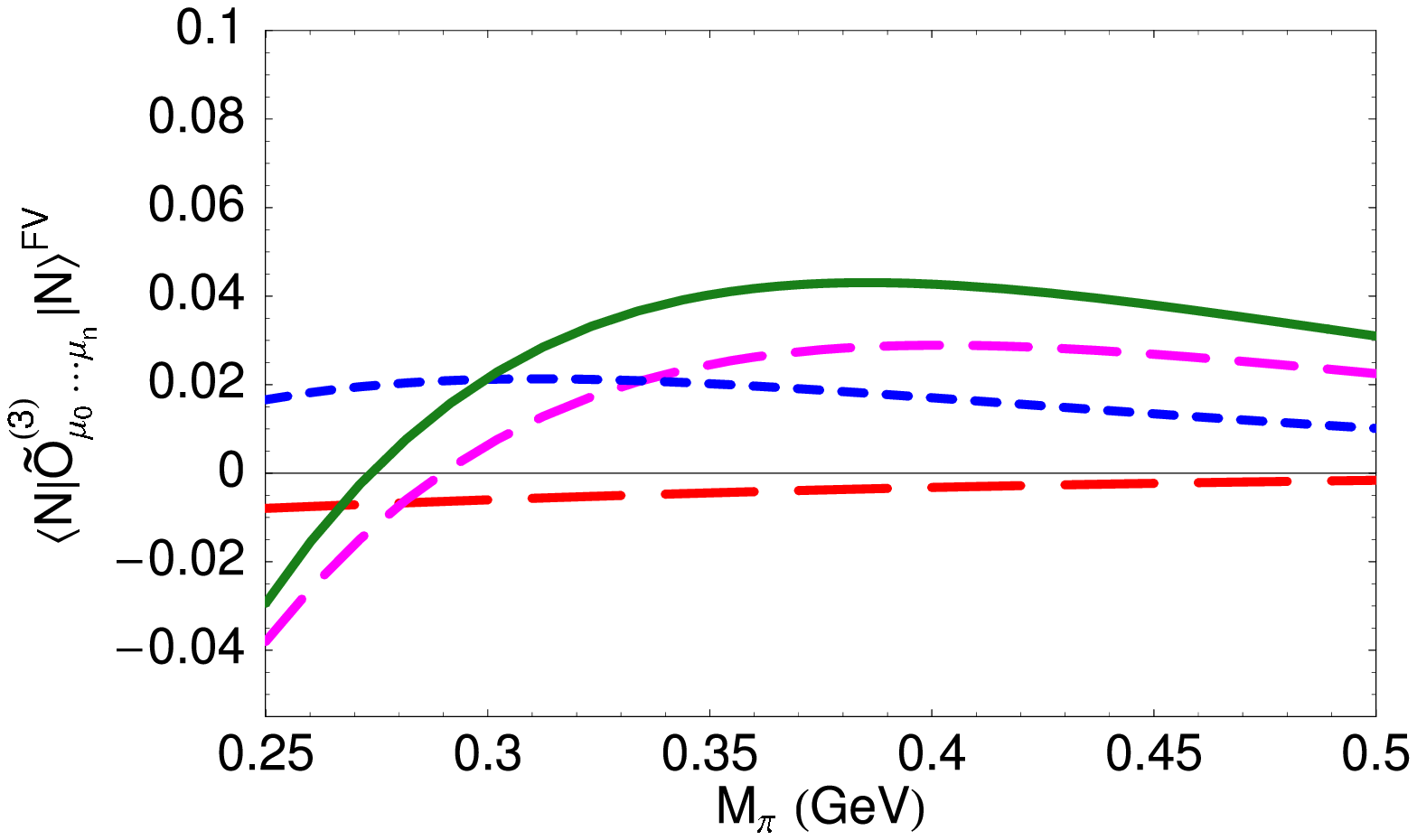} \\
  \hspace*{-15mm}
  \includegraphics[width=10cm]{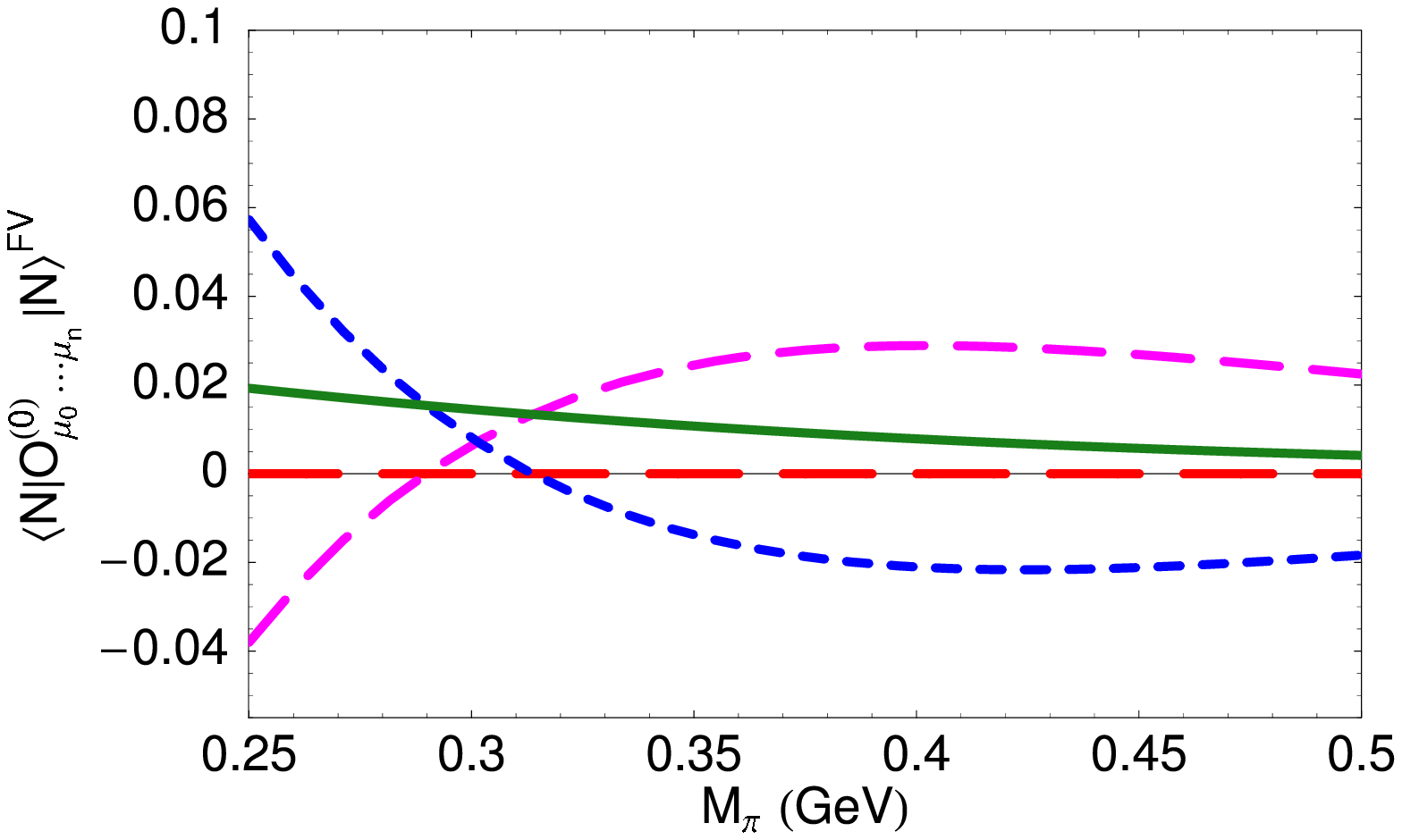}
  \hspace{-15mm}
  \includegraphics[width=10cm]{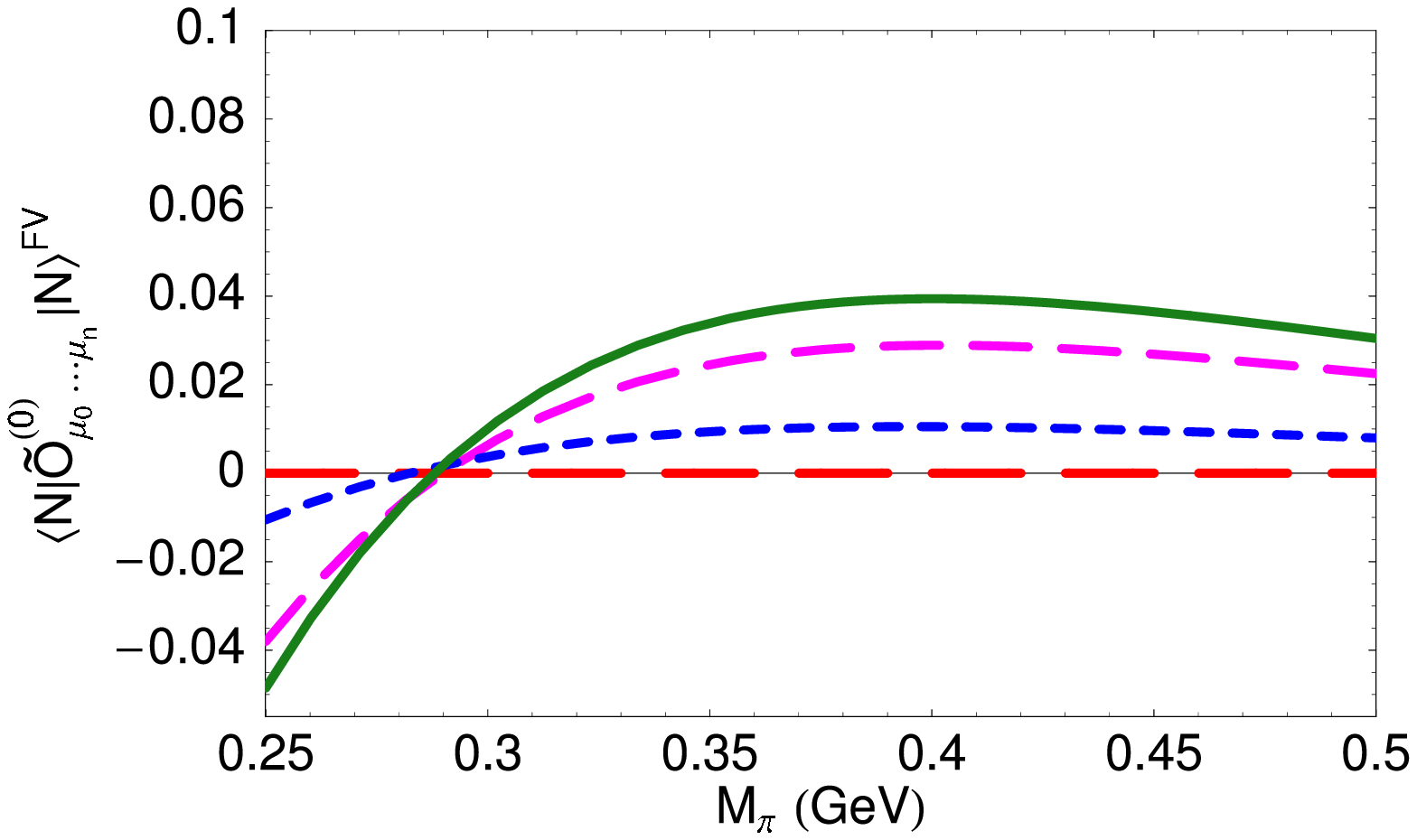} \\
  \hspace*{-15mm}
  \includegraphics[width=10cm]{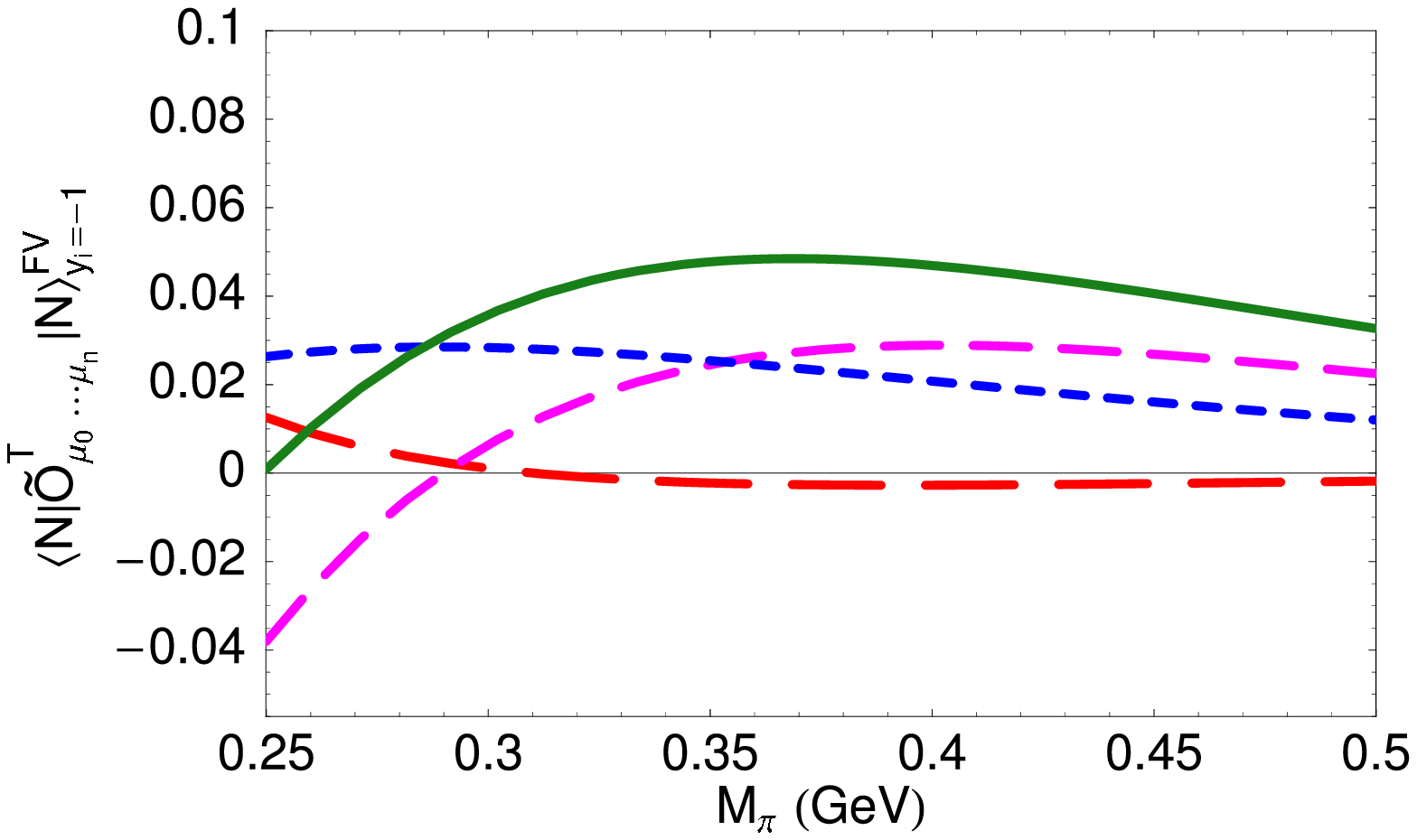}
  \hspace{-15mm}
  \includegraphics[width=10cm]{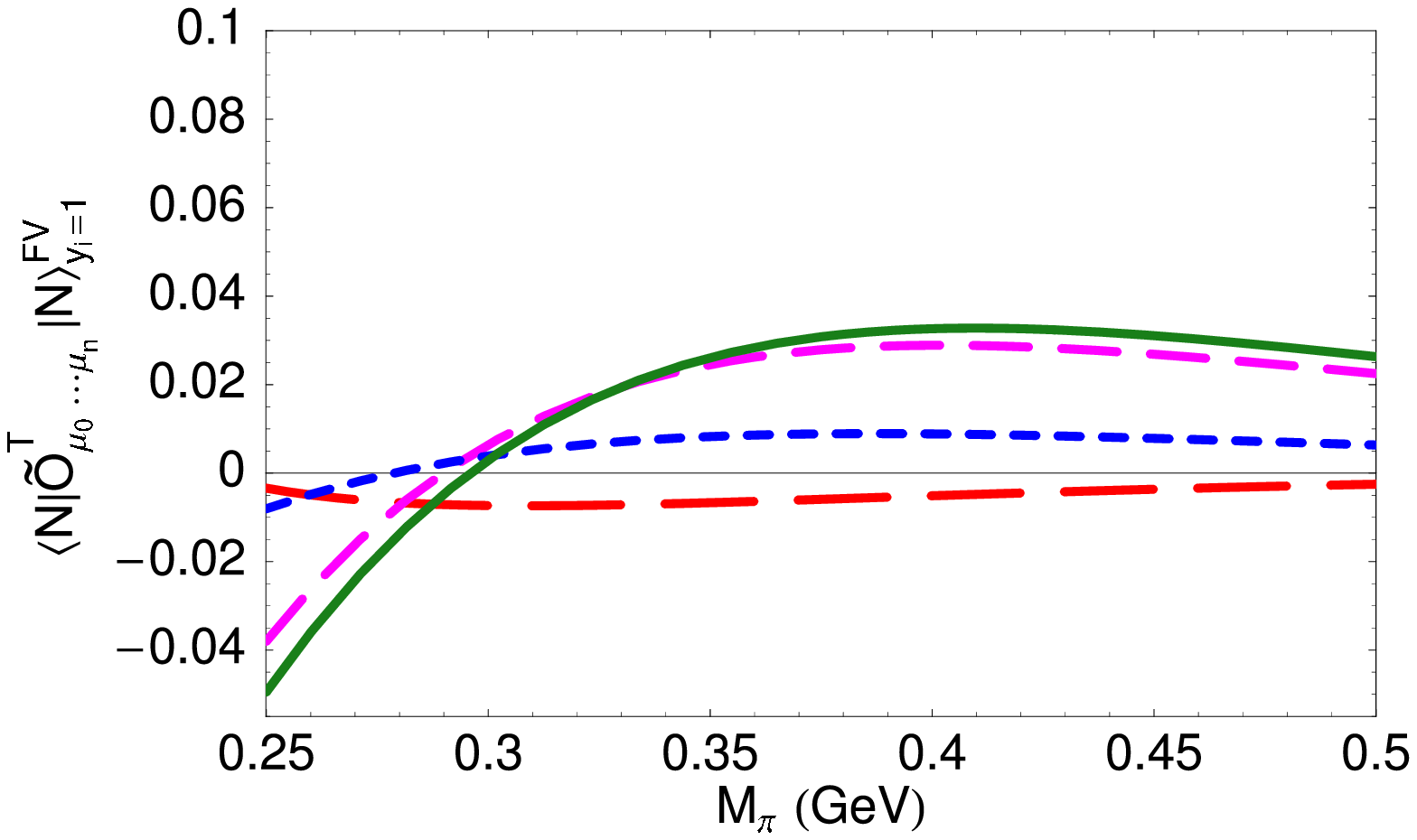}
  \caption{Indicative finite volume effects in SU(4$|$2)
    matrix elements. The results in the first row are for the
    isovector unpolarised (left) and helicity (right) operators and
    those in the second row are similarly for the isoscalar
    unpolarised (left) and helicity (right) operators. The third row
    corresponds to the transversity ``isovector'' $y_i=-1$ (left) and
    ``isoscalar'' $y_i=1$ (right) matrix elements. In each plot, the
    solid curve shows the total result, whilst the short-, medium- and
    long-dashed curves correspond to the individual FV effects arising
    from diagrams (c)--(f), diagrams (a) and (b), and diagram (i) in
    Fig.~\protect~\ref{fig:nucleondiagrams}. In all of these results,
    we have considered a (2.5 fm)$^3$ box and set $g_A=g_1=1.3$ and
    $|g_{\Delta N}|=1.5$. $M_\pi=0.25$~GeV corresponds to $M_\pi
    L=3.2$.}
  \label{fig:FV_generic_g1gA}
\end{figure}

The axial couplings $g_A$, $g_1$, and $g_{\Delta N}$ occurring in our
results are the chiral limit couplings and there is some uncertainty
in their values. We will fix $g_A=1.3$ (though even the chiral limit
value of this is not well known \cite{Beane:2004ks}), $|g_{\Delta
  N}|=1.5$ and vary $g_1=\pm g_A$.  In the QCD limit, results are
independent of $g_1$ since in this case it only involves couplings to
the $\eta^\prime$ meson which remains massive in the chiral limit and
can be integrated out.

\begin{figure}[!t]
  \centering
  \hspace*{-15mm}
  \includegraphics[width=10cm]{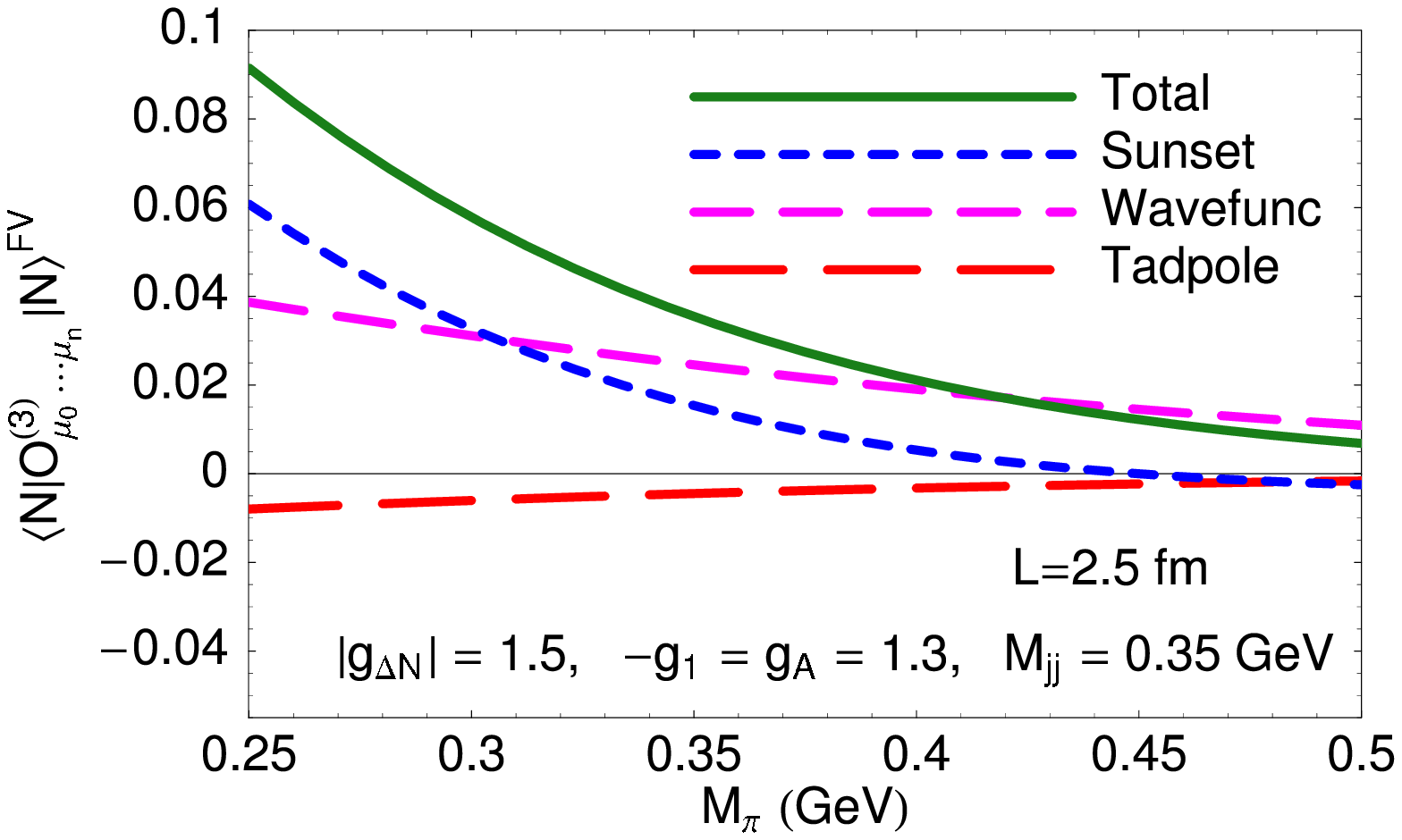}
  \hspace{-15mm}
  \includegraphics[width=10cm]{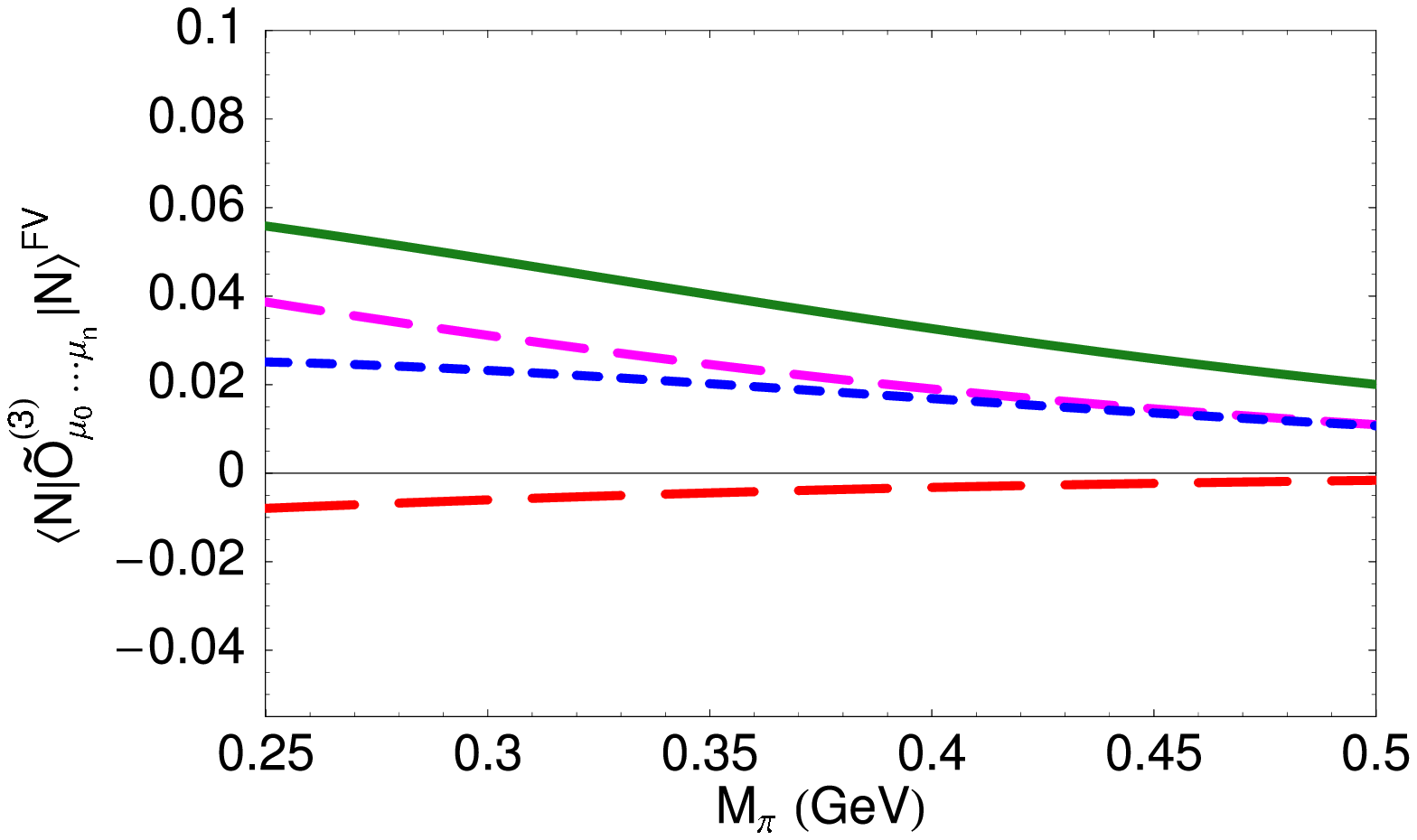} \\
  \hspace*{-15mm}
  \includegraphics[width=10cm]{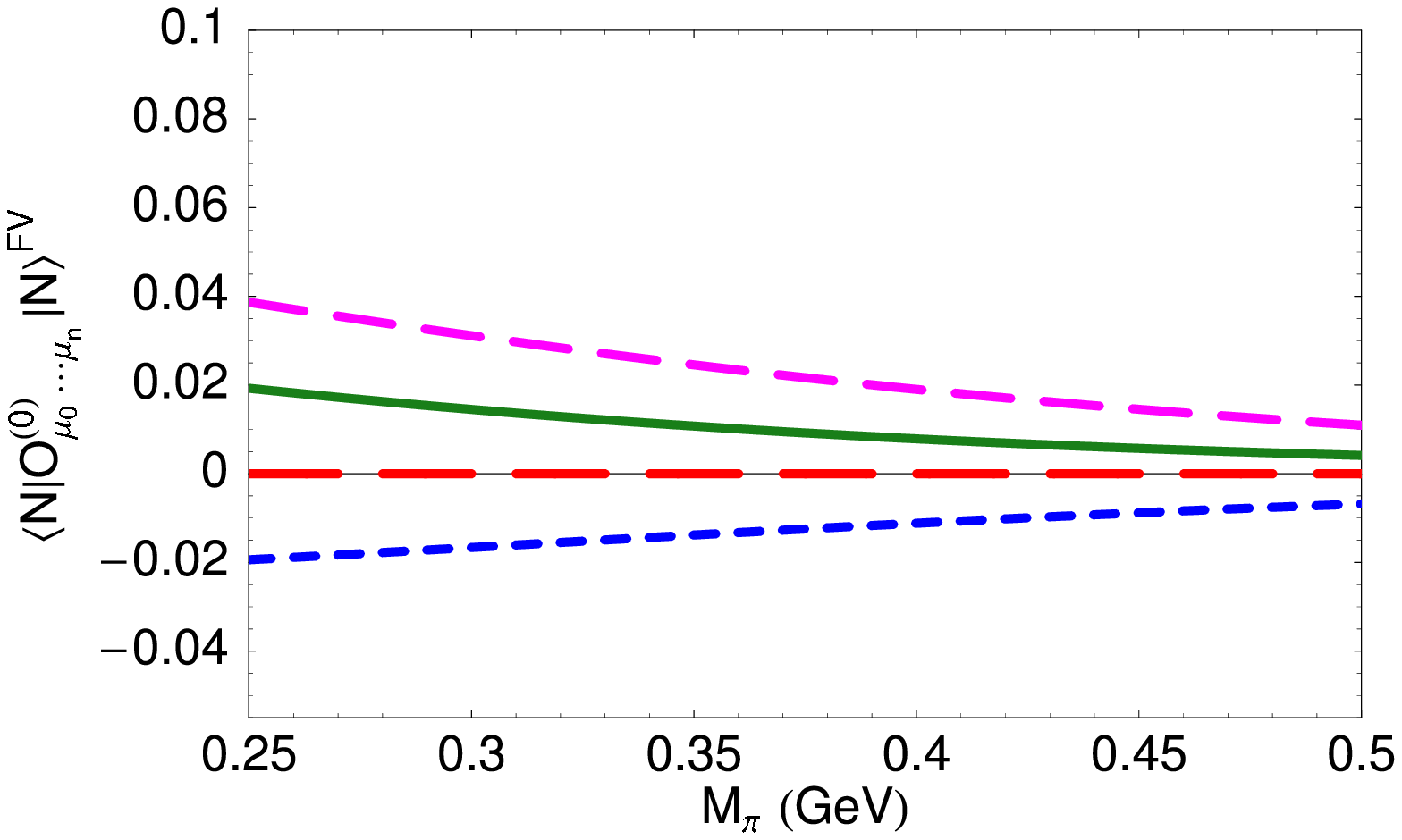}
  \hspace{-15mm}
  \includegraphics[width=10cm]{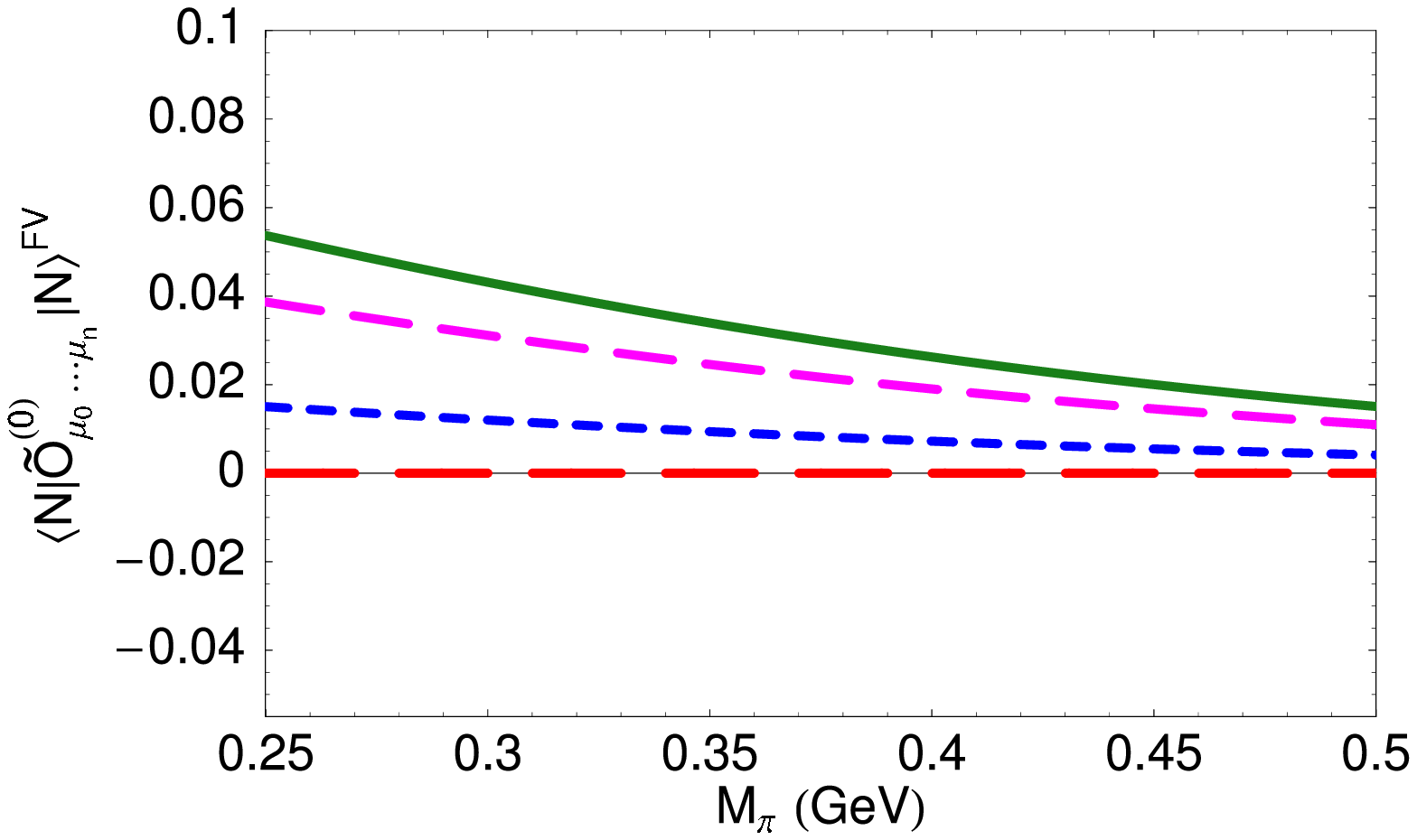} \\
  \hspace*{-15mm}
  \includegraphics[width=10cm]{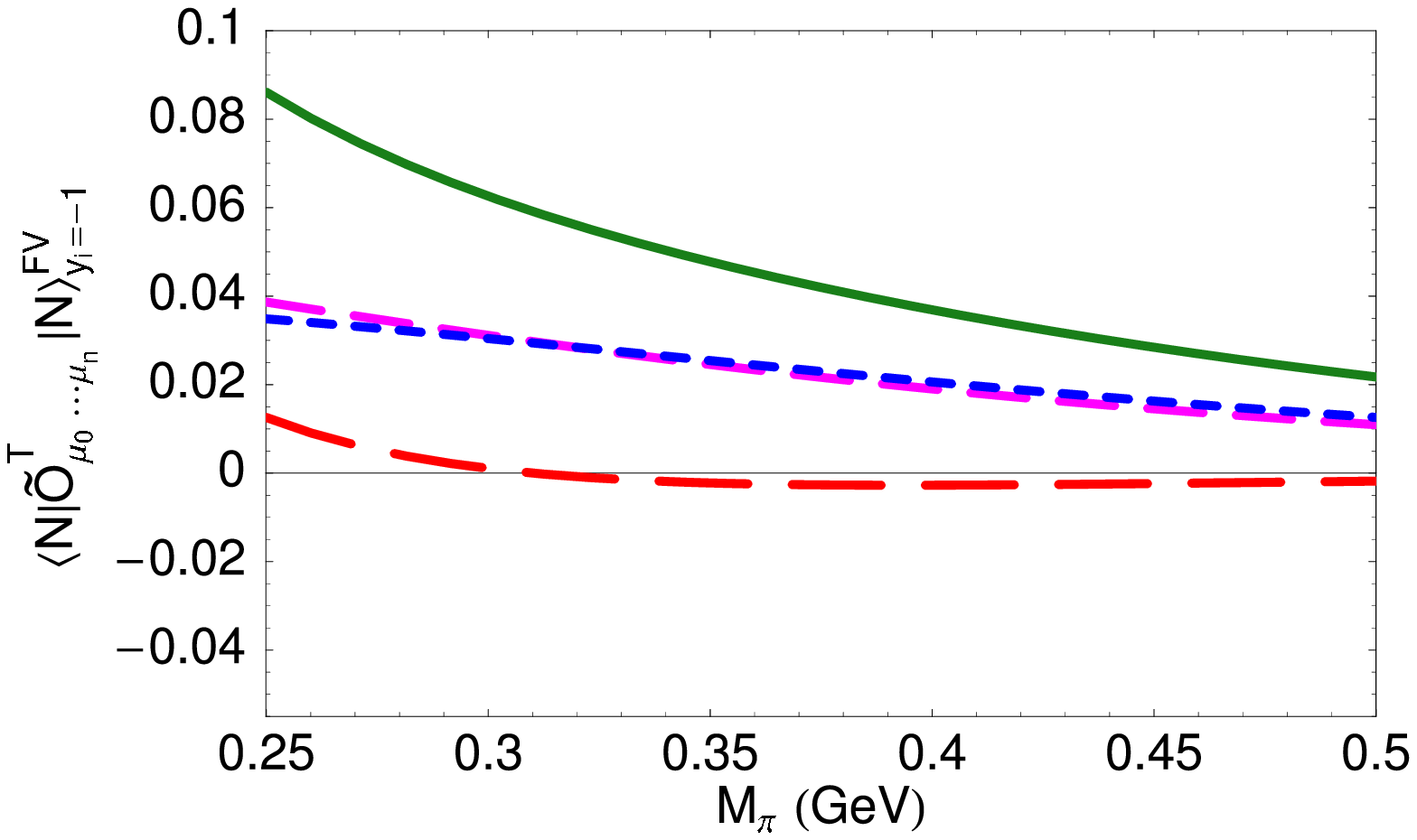}
  \hspace{-15mm}
  \includegraphics[width=10cm]{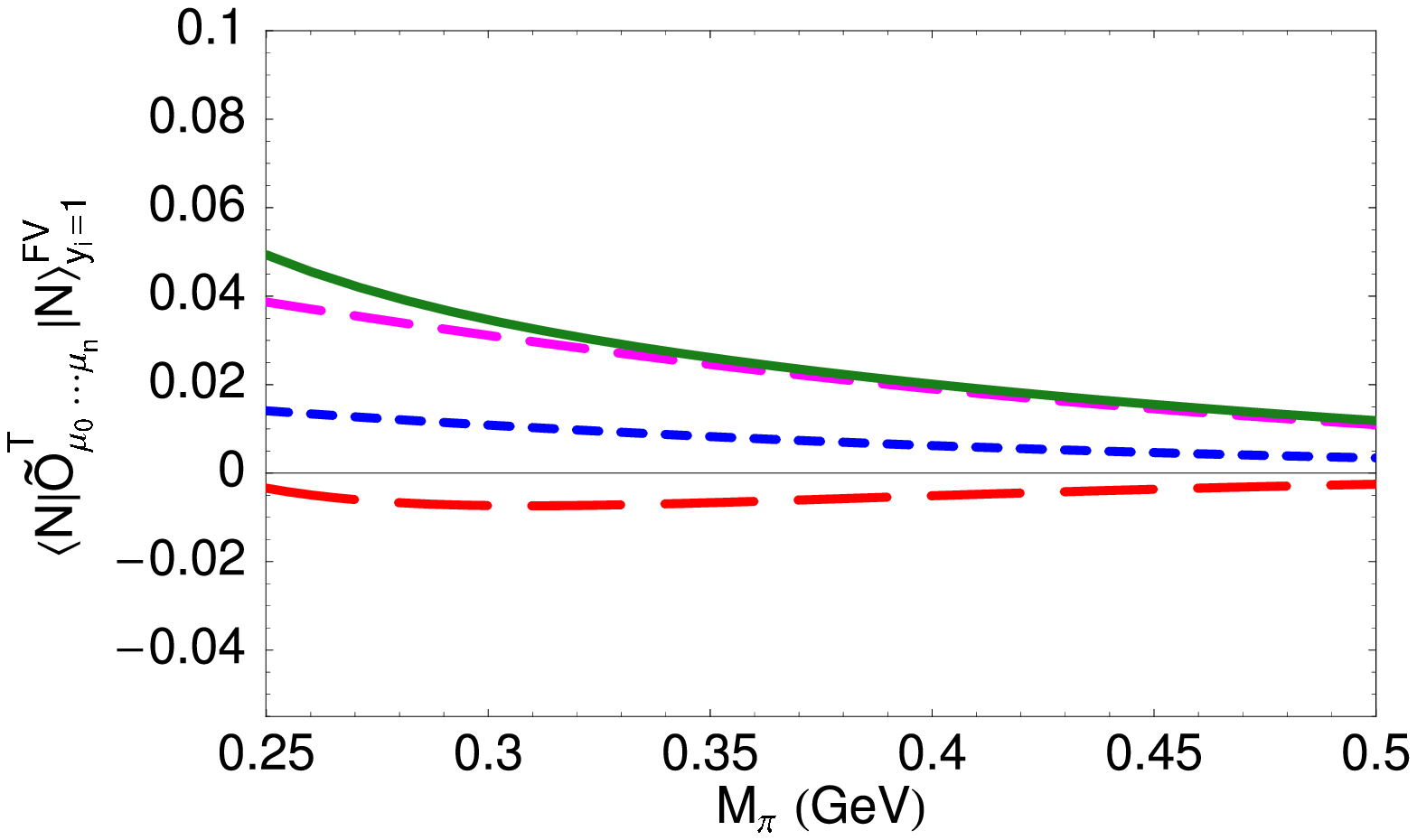}
  \caption{As in Fig.~\protect\ref{fig:FV_generic_g1gA} except with
    $g_1=-g_A$. $M_\pi=0.25$~GeV corresponds to $M_\pi L=3.2$.}
  \label{fig:FV_generic_g1minusgA}
\end{figure}
Using these parameters, Figures~\ref{fig:FV_generic_g1gA} and
\ref{fig:FV_generic_g1minusgA} illustrate the typical size of finite
volume effects in the various matrix elements. In
Fig.~\ref{fig:FV_generic_g1gA}, we consider a (2.5 fm)$^3$ box with a
sea quark mass set such that the corresponding sea--sea Goldstone
boson has a mass $M_{jj}=0.35$~GeV and take $g_1=g_A$.
Fig.~\ref{fig:FV_generic_g1minusgA} is similar except here we take
$g_1=-g_A$ to show the effect of this undetermined parameter.
Variation of $g_A$ and $g_{\Delta N}$ within reasonable bounds leads
to similar modifications as those for varying $g_1$. At the smallest
pion mass in these plots, $M_\pi L \sim 3$ and one must start to worry
that infinite volume $p$-counting is no longer appropriate; at the
largest pion mass in the plots, $M_\pi / \Lambda_\chi\sim 1/2$ and one
must worry that the convergence of the chiral expansion becomes
questionable. From these figures, it is nevertheless apparent that NLO
\pqxpt\ predicts finite volume effects in twist-two matrix elements
that are generically $\alt 5$--10\% for the range of masses and
volumes studied here. However, there is some evidence that finite
volume effects from higher orders of the standard chiral
power-counting can be significant
\cite{Colangelo:2003hf,Colangelo:2004xr}.

\begin{figure}[!t]
  \centering\hspace*{-2.5cm}
  \includegraphics[width=11.5cm]{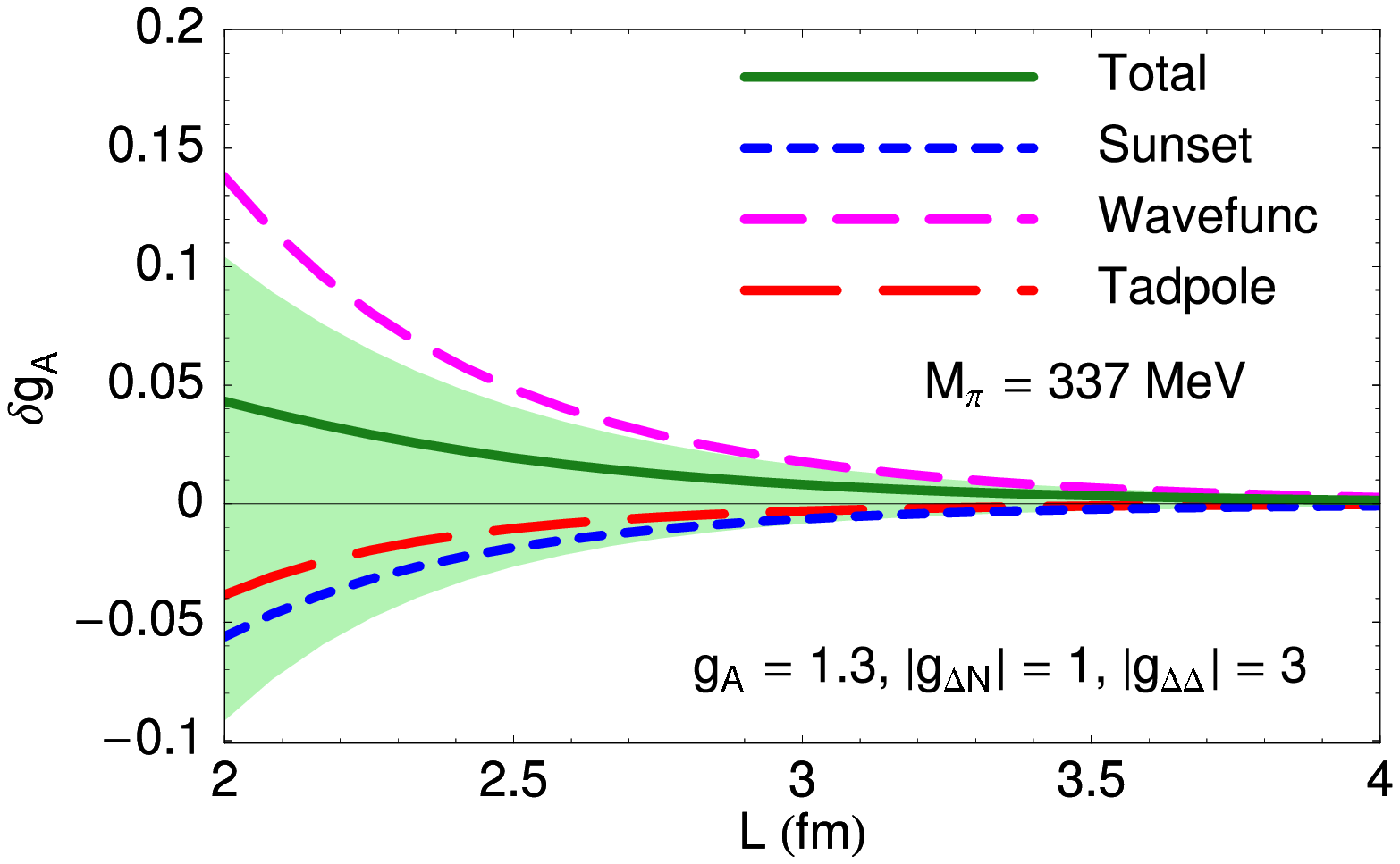}
  \caption{Finite volume effects in QCD calculations of $g_A$ at
    $M_\pi=337$~MeV (as appropriate for the recent LHP simulations
    \protect\cite{Renner:2004ck}).  $L=2$~fm corresponds to $M_\pi
    L=3.4$. The shaded region corresponds to varying $1.0\leq g_A \leq
    1.5$, $0 \leq |g_{\Delta N}|\leq 2$.}
  \label{fig:FV_LHP}
\end{figure}

Recently, staggered-sea, domain-wall-valence results have become
available from the LHP collaboration \cite{Renner:2004ck} for very
large volume ($L=3.5$~fm) calculations of $g_A$ at a pion mass of 337
MeV. Also available are results on a somewhat smaller lattice ($L=2.6$
fm) at the same pion mass. Although these two data points are
consistent within their statistical errors (which will be reduced by
ongoing calculations), their central values differ by $\sim15$\%. If
we ignore the issues of non-locality due to the ``fourth-root'' trick
used in calculating the staggered quark configurations and possible
unitarity violations arising from the different valence and sea quark
actions (which must vanish in the $a\to0$ limit that we have assumed),
one can ask whether the NLO \pqxpt\ formulae presented here can
describe this dependence.  To address this question, we consider the
isospin symmetric QCD limit ($m_u=m_d=m_j=m_l$) and define
\begin{equation}
  \label{eq:deltagA_def}
  \delta g_A = \frac{\langle N|\widetilde{\cal
      O}^{(3)}_{\mu}|N\rangle_{L}^{\rm one-loop} - \langle N|\widetilde{\cal
      O}^{(3)}_{\mu}|N\rangle_{L=\infty}^{\rm one-loop}}{\langle N|\widetilde{\cal
      O}^{(3)}_{\mu}|N\rangle^{\rm tree-level}}\,.
\end{equation}
For this case, the LECs can be expressed in terms of the axial
couplings through Eq.~(\ref{eq:n0spin}) and the FV shift, $\delta
g_A$, depends only on the pion mass, the volume and the chiral limit
couplings $g_A$, $g_{\Delta N}$ and $g_{\Delta\Delta}$.  In
Fig.~\ref{fig:FV_LHP}, we show the FV shift in $g_A$ that NLO \xpt\ 
predicts at the LHP pion mass, $M_\pi=337$~MeV.  To illustrate
uncertainties in the results, we vary the different axial couplings.
In the central fits (indicated by the curves in the plot), we set
$g_A=1.3$, $g_{\Delta N}=-1$ and $g_{\Delta\Delta}=-3$ whilst the
shaded band corresponds to $\delta g_A$ for $1.0\leq g_A \leq 1.5$, $0
\leq |g_{\Delta N}|\leq 2$ and $g_{\Delta\Delta}=-3$.  Whilst, a shift
of 15\% between $L=2.6$ and 3.5~fm is not predicted, the FV effects
are substantial. However, without accurate knowledge of the chiral
limit couplings, even the sign of the finite volume correction to
$g_A$ is not well determined.

\begin{figure}[!t]
  \centering\hspace*{-2.5cm}
  \includegraphics[width=11.5cm]{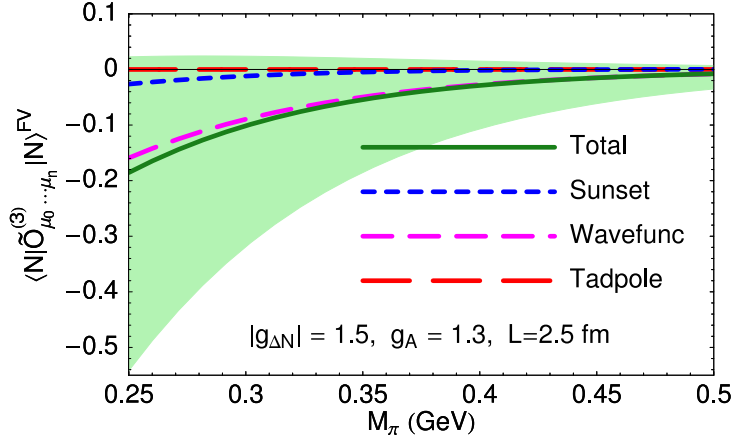}
  \caption{Finite volume effects in the isovector, helicity matrix
    elements in the proton in SU(2$|$2) quenched \xpt. The shaded
    region corresponds to variation of the axial couplings $g_1$ and
    $\gamma$ between $\pm g_A$ with $g_A=1.3$, $g_{\Delta N}=1.5$ and
    assuming large $N_c$ relations for the operator LECs. The point
    $M_\pi=0.25$~GeV corresponds to $M_\pi L=3.2$.}
  \label{fig:FV_quenched}
\end{figure}

As discussed in the previous subsection, in quenched lattice
calculations, FV effects will be enhanced because of the double-pole
contributions to singlet meson propagators. In
Fig.~\ref{fig:FV_quenched}, we plot the volume dependence of the
polarised, isovector twist-two matrix elements in SU(2$|$2) quenched
\xpt\ (the analytic forms of these results are presented in
Appendix~\ref{A4}). In contrast to partially-quenched \xpt, in the
quenched theory the LECs occurring in the Lagrangian and the twist-two
operators are unrelated to those in standard \xpt (though we denote
them by the same symbols for convenience). To be definite, we choose
$m_0=0.7$~GeV and the quenched operator LECs to satisfy
$\Delta\alpha^{(a)}_n = \Delta\beta^{(a)}_n = 5\Delta\gamma^{(a)}_n =
2\Delta c_n^{(a)}$ (as in the partially-quenched case), set the
quenched axial couplings to $g_A=1.3$, $|g_{\Delta N}|=1.5$ and let
$g_1$ and $\gamma$ vary between $\pm g_A$ as indicated by the shaded
region.  The curves in the figure correspond to $g_1=g_A/2$ and
$\gamma=0$. As expected, the volume dependence here is enhanced over
that in the \pqxpt\ and \xpt\ cases.

\subsection{Off-forward matrix-elements}
\label{S5}

The off-forward matrix elements (in which the incoming and outgoing
hadrons carry different momenta) of the twist-two operators correspond
to moments of generalised parton distributions. Very little is known
from experiment about GPDs though major programs are underway at
HERMES, Jefferson Lab and COMPASS to investigate them. As such,
moments of GPDs are important quantities to extract from lattice
calculations and much progress is being made in this direction
\cite{latticeGPDs}. It is therefore important to investigate the quark
mass dependence\footnote{Refs.~\cite{Chen:2001pv,Belitsky:2002jp}
  address this issue for the infinite volume $n=1$ matrix element
  relevant for the spin content of the proton in SU(2) \hbxpt.} and
size of finite volume effects in these calculations.  Here we shall
only discuss the novel features that appear in regard to finite volume
effects when off-forward matrix elements are considered.  A full
analysis of the low-energy behaviour of these matrix elements will be
given elsewhere \cite{offforward}.

To be specific, we shall consider the proton matrix elements $\langle
p^\prime|^{\rm QCD}{\cal O}_{\mu_0\ldots\mu_n}^{(A)}|p \rangle$ in
which four momentum $q^\mu=(p^\prime-p)^\mu$ (with
$|q^2|\ll\Lambda_\chi^2$) is injected through the twist-two operator.
The analysis of these matrix elements is significantly more
complicated than in the forward limit.  This is primarily because the
number of possible independent tensor structures in the matrix element
grows with $n$; for example,
\begin{eqnarray}
\label{eq:off_forward_ME}
\langle p^\prime|^{\rm QCD}{\cal O}_{\mu_0\ldots\mu_n}^{(A)}|p \rangle &=&
\overline{U}_p(p^\prime)
\Bigg[ \sum_{\substack{i=0 \\ {\rm even}}}^{n}\Bigg\{ \gamma
^{\{\mu_0 }q ^{\mu _{1}}\ldots q ^{\mu _{i}}\pbar^{\mu
_{i+1}}\ldots \pbar^{\mu _{n}\}}A^{(A)}_{n,i}(q ^{2})
\\
&& -i\frac{q _{\nu }\sigma ^{\nu \{\mu_0 }}{2m}q ^{\mu_{1}}
\ldots q ^{\mu _{i-1}}\pbar^{\mu _{i}}\ldots \pbar
^{\mu _{n}\}}B^{(A)}_{n,i}(q ^{2})\Bigg\}
 +\left.\frac{q ^{\mu_0 }\ldots q ^{\mu _{n}}}{m}
C^{(A)}_{n,0}(q ^{2})\right| _{n\text{ even}}\Bigg]
U_p(p)\,,
\nonumber
\end{eqnarray}
where $\pbar=\frac{1}{2}(p^\prime+p)$.  For each of the coefficient
functions $A^{(A)}_{n,i}(q ^{2})$, $B^{(A)}_{n,i}(q ^{2})$ and
$C^{(A)}_{n,0}(q ^{2})$, there is an independent finite volume \xpt\ 
expansion.

Another complication enters when one considers the operators that
match onto the twist-two operators in the low energy effective theory.
The presence of the new scale $q$ means that considerably more
operators must be included in
Eqs.~(\ref{eq:hadron_op})--(\ref{eq:hadron_sigmaop}); for example, the
term proportional to $\alpha^{(r_A)}_n$ in Eq.~(\ref{eq:hadron_op}) is
replaced by
\begin{eqnarray}
\label{eq:some_terms}
 \Bigg[\sum_{j=0}^n \alpha^{(r_A)}_{n,j}
  \left[v^{\{\mu_0}\ldots v^{\mu_j}(-i\partial)^{\mu_{j+1}}\ldots
    (-i\partial)^{\mu_n\}}  
    - {\rm tr}\right]
+ \alpha^{(r_A)}_{n,-1}
  \left[(-i\partial)^{\mu_{0}}\ldots
    (-i\partial)^{\mu_n}  
    - {\rm tr}\right]
\Bigg]
  (\overline{\cal B}\tau^A_{\xi^+} {\cal B} ) \,.
\end{eqnarray}
Note that only terms with an even number of derivatives contribute
here.  Each LEC in the forward case is replaced by ${\cal O}(n)$ LECs.
Additionally, the vector $q$ allows more tensor structures to enter
and we must also include
\begin{eqnarray}
\label{eq:some_more_terms}
  \sum_{j=0}^n \hat\alpha^{(r_A)}_{n,j}
  \left[v^{\{\mu_0}\ldots v^{\mu_j} (-i\partial)^{\mu_{j+1}}\ldots
    (-i\partial)^{\mu_{n-1}} (\overline{\cal B}
    \frac{\left[S_{\mu_n\}},S\cdot (-i\partial)\right]}{M}
    \tau^A_{\xi^+} {\cal B})  - {\rm  tr}\right] \,,
\end{eqnarray}
since $S\cdot q\ne0$. When we also take into account the {\bf 44}-plet
fields ${\cal T}^\mu$, many more structures are possible since $q\cdot
{\cal T}\ne0$. Even for the $n<3$ matrix elements that have been
calculated on the lattice, a large number of LECs need to be
determined. This makes a reliable extraction of the physical matrix
elements from finite volume, unphysical mass, lattice calculations a
challenging proposition.

In terms of FV effects, the modifications for the off-forward case are
relatively simple and it is worthwhile to investigate them in some
detail. There are essentially two classes of diagrams: ones where the
twist-two operator injects momentum into a (heavy) baryon field (e.g.,
(c) in Fig.~\ref{fig:nucleondiagrams}); and ones where a meson
receives the additional momentum (e.g., (h) in
Fig.~\ref{fig:nucleondiagrams}). The former class is relatively
uninteresting since for heavy fields, derivatives in the twist-two
operators pick out only the momentum transfer between the external
states, $q$, and can therefore be factored out of the integral. For
this type of diagram, finite volume effects will be the same as those
in the forward limit as we are free to work in the Breit frame where
$q\cdot v=0$.

For diagrams in which the twist-two operator is on a meson line, the
situation is different since the derivatives in the operator can
result in powers of the integration momentum. The relevant integrals
are of the form
\begin{eqnarray}
  \label{eq:a_big_mess}
\frac{1}{L^3}\sum_{\vec
  k}\int\frac{dk_0}{2\pi} 
\frac{S\cdot(-k)S\cdot(k+q) \left[(-k)^{\{\mu_0}\ldots
      (-k)^{\mu_{j}}(k+q)^{\mu_{j+1}}\ldots(k+q)^{\mu_n\}} - {\rm
        tr}\right]}{(k\cdot
    v-\Delta + i \epsilon) \,(k^2-m^2+i\epsilon)((k+q)^2-m^2+i\epsilon)}
\hspace*{4.7cm}
\\
\hspace*{0.55cm}
=4\int_0^1 dx \int_0^\infty d\lambda \frac{1}{L^3}\sum_{\vec\ell}\int\frac{d\,\ell_0}{2\pi}
\frac{S\cdot(\ell+a)S\cdot(\ell+b)
}{(\ell^2 -{\cal M}^2)^3} 
(-1)^{j}\left[(\ell+a)^{\{\mu_0}\ldots
      (\ell+a)^{\mu_{j}}(\ell+b)^{\mu_{j+1}} \ldots (\ell+b) ^{\mu_n\}} - {\rm
        tr}\right]
\nonumber 
\end{eqnarray}
after introducing Feynman and Schwinger parameters and shifting the
momentum integration $k \to \ell = k + x\,q - \lambda\, v$. Here
$a=-x\, q+\lambda\, v$, $b=(1-x)\,q+\lambda\, v$ and
\begin{eqnarray}
\label{eq:M_of_q}
{\cal M}^2 \equiv {\cal M}^2(x,\lambda,m^2,q^2,\Delta)
 = m^2 - x(1-x) q^2 +\lambda^2 + 2\lambda \Delta\,.
\end{eqnarray}
The trace subtractions prevent any of the $\ell^{\mu_i}$'s arising
from the operator from contracting with one another, consequently the
non-vanishing scalar integrals/sums whose finite volume effects we are
interested in will be of the form
\begin{eqnarray}
  \label{eq:less_of_a_mess}
\int_0^1 dx \int_0^\infty d\lambda
\frac{1}{L^3}\sum_{\vec\ell}\int\frac{d\,\ell_0}{2\pi} 
\frac{\left(\ell^2\right)^r}{(\ell^2 -{\cal M}^2)^3} \,,
\end{eqnarray}
where $r=0$, 1 or 2.

Without going into further details of the tensor structure
\cite{offforward}, it is already clear from Eq.~(\ref{eq:M_of_q}) that
the effect of the momentum injection on overall finite volume shifts
is very similar to the effect of the N--$\Delta$ mass splitting.
Since $x(1-x)>0$, when space-like momentum ($q^2=-|{\vec q}|^2<0$ in
Minkowski space) is injected, FV effects are suppressed as the meson
receiving the momentum injection moves further away from its mass
shell.  However, if time-like momentum is injected the situation is
more complicated.  Provided the virtual particle cannot reach its mass
shell, finite volume effects are enhanced over the forward case but
will still remain formally exponentially suppressed. However, if the
injected energy-momentum is enough to put the intermediate particles
on shell, it leads to a cut in Minkowski space in infinite volume. In
this case, finite volume effects are only suppressed by powers of
$1/L$ in QCD.  In (partially) quenched QCD, volume corrections for
isoscalar twist-two matrix elements may be proportional to positive
powers of $L$ whereby the infinite volume limit will be undefined.
This suppression of finite volume effects with space-like momentum
injection and enhancement in the time-like case (which is relevant for
twist-two matrix elements between states of different masses, {\it
  e.g.} $N\to\Delta$ transitions) will occur in hadronic form-factors
that are specific cases of twist-two matrix elements.


\section{Conclusions}
\label{S6}

We have studied the matrix elements of twist-two operators that
determine the moments of the unpolarised, helicity and transversity
quark distributions to NLO in (partially) quenched chiral perturbation
theory in both infinite and finite volumes. We have performed our
calculations in $N_f=2$ and $N_f=2+1$ partially quenched heavy baryon
chiral perturbation theory and also studied the SU(2$|$2) quenched
theory. These results will be relevant for extrapolations of lattice
calculations of these matrix elements in the proton and other octet
baryons (e.g., the $\Lambda$ hyperon \cite{Gockeler:2002uh}).

We have focused primarily on the effects of the finite volumes used in
lattice calculations. Without accurate data in the chiral regime with
which to fit the various low energy constants on which the results
depend, it is difficult to be specific, however it is clear that for
most current simulations FV effects are not negligible. For typical
full- or partially-quenched- QCD calculations, they are $\alt 5$--10\%
but may be significantly larger in quenched simulations.

In the case of the off-forward matrix elements relevant to generalised
parton distributions, we have not presented full results for arbitrary
moments \cite{offforward}. However, we have analysed the finite volume
effects in these matrix elements. We find that they should decrease
with respect to the forward matrix elements if space-like momentum is
injected. On the other hand if time-like momentum is transferred,
finite volume effects will be enhanced; in QCD they may become only
$1/L$ suppressed, and in (partially) quenched QCD finite volume
isoscalar matrix elements may even be proportional to powers of $L$.


\acknowledgments We are grateful to J.-W.~Chen, R.~Edwards,
W.~Melnitchouk, M.~Savage, S.~R.~Sharpe, A.~W.~Thomas and
A.~Walker-Loud for discussions and particularly to S.~Beane for
discussions and helpful correspondence.  We also thank M.~Golterman,
S.~Peris and the Benasque Centre for Science, Spain for organising the
workshop {\it Matching Light Quarks to Hadrons} at which this work
began and CJDL acknowledges the hospitality of the National Center for
Theoretical Sciences at Taipei.  WD and CJDL are supported by DOE
grants DE-FG03-97ER41014, DE-FG03-00ER41132 and DE-FG03-96ER40956.


\appendix
\section{Tadpole integrals and finite volume sums}
\label{A1}
The sums that appear in tadpole diagrams, after subtracting
$\bar{\lambda}$, are
\begin{eqnarray}
\label{eq:def_tadint}
\calI (m) &=& \frac{1}{L^3} \sum_{\vec{k}} \int \frac{d k_{0}}{2 \pi} 
 \frac{i}{k^{2} - m^{2} + i \epsilon} + \frac{m^{2}}{16\pi^{2}}
 \bar{\lambda}\,,
\end{eqnarray}
and 
\begin{eqnarray}
\calIdp (m) &=& \frac{1}{L^3} \sum_{\vec{k}} \int \frac{d k_{0}}{2 \pi} 
 \frac{i}{(k^{2} - m^{2} + i \epsilon)^{2}} + \frac{1}{16\pi^{2}}\bar{\lambda}
  =  \frac{\partial \calI (m)}{\partial m^{2}} .
\end{eqnarray}
Using Poisson's summation formula, it is straightforward to show that
\begin{equation}
 \calI (m) = I(m) + I^{\mathrm{FV}}(m) ,
\end{equation}
where
\begin{equation}
\label{eq:InfV_tadint}
 I (m) = \mu^{4-d} \int \frac{d^{d}k}{(2\pi)^{d}} 
   \frac{i}{k^{2} - m^{2} + i \epsilon} 
   + \frac{m^{2}}{16 \pi^{2}} \bar{\lambda}
 = \frac{m^{2}}{16 \pi^{2}} 
   {\mathrm{log}}\left ( \frac{m^{2}}{\mu^{2}} \right )  ,
\end{equation}
is the infinite-volume limit of $\calI (m)$, and
\begin{eqnarray}
 I^{\mathrm{FV}} (m) &=& \frac{m}{4\pi^{2}} \sum_{\vec{u}\not= \vec{0}}
   \frac{1}{u L} K_{1} (umL)\nonumber\\ 
\label{eq:FV_tadint}
 &\stackrel{m L \gg 1}{\longrightarrow}& \frac{1}{4\pi^{2}} 
  \sum_{\vec{u}\not= \vec{0}} \sqrt{\frac{m\pi}{2 u L}}
  \left ( \frac{1}{u L}\right ) {\mathrm{e}}^{-u m L}
  \left \{ 1 + \frac{3}{8 u m L} - \frac{15}{128 (u m L )^{2}}
   + \op \left ( \frac{1}{(u m L)^{3}}\right ) \right \} \,.
\end{eqnarray}
%


\section{Results for SU(4$|$2) \pqxpt}
\label{A2}

In this section, we present the results for twist-two matrix elements
in the isospin limit in $SU(4|2)$ \pqxpt.  The various masses and the
mass-splitting $\delta$ are defined in Sections~\ref{S2} and
~\ref{S3}.

The nucleon wave-function renormalisation is
\begin{eqnarray}
{\cal W}_{SU(4|2)}&=& \frac{i}{2f^2}\Big\{  {\cal H}({M_{\pi }},0)\,\left(
  -5\,g_1^2 - 4\,{g_1}\,{g_A} + g_A^2 \right)  +  
  {\cal H}({M_{uj}},0)\,\left( 5\,g_1^2 + 4\,{g_1}\,{g_A} + 8\,g_A^2
  \right) 
\nonumber \\ &&+ 
  4\,\left( {\cal H}({M_{\pi }},\Delta ) + {\cal H}({M_{uj}},\Delta ) \right)
  \,g_{\Delta N}^2 +  
  \left( -6\,\,g_1^2 - 12\,\,{g_1}\,{g_A} - 6\,\,g_A^2
  \right){\delta }^2 \, 
   {{\cal H}_{\eta^\prime}}({M_{\pi }},0) \big\} \,.
\label{eq:waverenorm_su42}
\end{eqnarray}

The isovector, unpolarised nucleon matrix element is
\begin{eqnarray}
 \langle N|{\cal O}^{(3)}_{\mu_0\ldots\mu_n}|N\rangle &=&  
\frac{1}{3}{\overline{U}_N v_{\mu_0}\ldots
  v_{\mu_n}}U_N(2\alpha_n^{(a)}-\beta_n^{(a)})\times(1+(1-\delta_{n0}){\cal W}_{SU(4|2)})
\nonumber \\&&+\frac{i(1-\delta_{n0}) }{12f^2}\,{\overline{U}_N v_{\mu_0}\ldots v_{\mu_n}U_N}\,
    \Bigg\{ \frac{4}{3}\,g_{\Delta N}^2\,
         \left( {{\gamma }_n^{(a)}} - \frac{{{\sigma }_n^{(a)}}}{3} \right)
         \times\
\nonumber \\&& \hspace*{1.5cm}
         \Bigg[ -3\,{\cal H}({M_{\pi }},\Delta )\,\left( -4 + {q_j} + {q_l} \right)  + 
           {\cal H}({M_{uj}},\Delta )\,\left( 8 + 3\,{q_j} + 3\,{q_l} \right)  \Bigg] \,
\nonumber \\&&
       + {{\alpha }_n^{(a)}}\,\Bigg[ -4\,i \,{\cal I}({M_{uj}})\,
          \left( -4 + 3\,{q_j} + 3\,{q_l} \right)  +  4\,i \,{\cal
            I}({M_{\pi }})\,\left( {q_j} + {q_l} \right) 
\nonumber \\&& \hspace*{1cm} + 
         3\,\Big( -\,{\cal H}({M_{uj}},0)\,
               \Big[ 4\,g_A^2\,\left( {q_j} + {q_l} \right)  +
                 2\,{g_1}\,{g_A}\,\left( 2 + {q_j} + {q_l} \right)  +                 
                 g_1^2\,\left( 6 + {q_j} + {q_l} \right)  \Big]
\nonumber \\&& \hspace*{1.75cm}
           + {\cal H}({M_{\pi }},0)\,\Big[ 4\,g_A^2\,\left( 1 + {q_j}
                + {q_l} \right)  +  
               2\,{g_1}\,{g_A}\,\left( 2 + {q_j} + {q_l} \right)  +
               g_1^2\,\left( 6 + {q_j} + {q_l} \right)  \Big]  
\nonumber \\&&   \hspace*{1.75cm}
              + 8\,{\delta }^2\,{\left( {g_1} + {g_A} \right)
               }^2\,{{\cal H}_{\eta^\prime}}({M_{\pi }},0)
           \Big)  
         \Bigg]  
\nonumber \\&&
 -  {{\beta }_n^{(a)}}\,\Bigg[ 4\,i \,{\cal I}({M_{uj}})\,
          \left( 2 + 3\,{q_j} + 3\,{q_l} \right)  
 -4\,i \,{\cal I}({M_{\pi }})\,\left( {q_j} + {q_l} \right) 
\nonumber \\&& \hspace*{1cm}
        + 3\,\Big( {\cal H}({M_{uj}},0)\,{g_1}\,
             \left[ -8\,{g_A} + 3\,{g_1}\,\left( {q_j} + {q_l} \right)  \right]  + 
            {\cal H}({M_{\pi }},0)\,\left[ 8\,{g_1}\,{g_A} + 2\,g_A^2 -
              3\,g_1^2\,\left( {q_j} + {q_l} \right)  \right] 
\nonumber \\&&    \hspace*{1.75cm}  
             +  4 \,{\delta }^2\,{\left( {g_1} + {g_A} \right)
               }^2\,{{\cal H}_{\eta^\prime}}({M_{\pi }},0)
           \Big)  \Bigg]  
\Bigg\} \,.
\label{eq:unpol_isovec_su42}
\end{eqnarray}
The isovector, helicity matrix element in the nucleon is
\begin{eqnarray}
 \langle N|\widetilde{\cal O}^{(3)}_{\mu_0\ldots\mu_n}|N\rangle &=&  
\frac{1}{3}{\overline{U}_N v_{\{\mu_0}\ldots
  v_{\mu_{n-1}}S_{\mu_n\}}U_N}(2\Delta\alpha_n^{(a)}-\Delta\beta_n^{(a)})
\times(1+{\cal W}_{SU(4|2)}) 
\nonumber \\&&
+\frac{i }{12f^2}\,\,{\overline{U}_N v_{\{\mu_0}\ldots
  v_{\mu_{n-1}}S_{\mu_n\}}U_N}\,
    \Bigg\{ \frac{16}{3}\,{\sqrt{\frac{2}{3}}}\,{g_{{\Delta N}}}\, 
         {\Delta c_n^{(a)}} \times
\nonumber \\&&
         \Bigg[ {\cal K}({M_{\pi }},\Delta )\,
            \Big( 8\,{g_A} + g_1\,\left( 2 - 3\,{q_j} - 3\,{q_l} \right)  \Big)  + 
           {\cal K}({M_{uj}},\Delta )\,
            \Big( 8\,{g_A} + g_1\,\left( -2 +
                3\,{q_j} + 3\,{q_l} \right)  \Big)  \Bigg]
\nonumber \\&&
- \frac{20 }{27}\,g_{\Delta N}^2\,
         \Big[ 3\,{\cal H}({M_{\pi }},\Delta )\,\left( -4 + {q_j} + {q_l} \right)  - 
           {\cal H}({M_{uj}},\Delta )\,\left( 8 + 3\,{q_j} + 3\,{q_l} \right)  \Big] \,
         \left( {\Delta \gamma_n^{(a)}} - \frac{\Delta\sigma_n^{(a)}}{5} \right)
\nonumber \\&&
 +       {\Delta \alpha_n^{(a)}}\,\Bigg[12\,i\,{\cal I}({M_{\pi }})\, 
          \left( {q_j} + {q_l} \right)  -  4\,i \,{\cal I}({M_{uj}})\, 
          \left( -4 + 3\,{q_j} + 3\,{q_l} \right) 
\nonumber \\&&  \hspace*{1.5cm} 
        + {\cal H}({M_{uj}},0)\,\Big( 4\,g_A^2\,\left( {q_j} + {q_l} \right)  + 
            2\,{g_1}\,{g_A}\,\left( 2 + {q_j} + {q_l} \right)  +
            g_1^2\,\left( 6 + {q_j} + {q_l} \right)  \Big) 
\nonumber \\&&  \hspace*{1.5cm} 
       -  {\cal H}({M_{\pi }},0)\,\Big( 4\,g_A^2\,\left( 1 + {q_j} + {q_l} \right)  + 
            2\,{g_1}\,{g_A}\,\left( 2 + {q_j} + {q_l} \right)  +
            g_1^2\,\left( 6 + {q_j} + {q_l} \right)  \Big)
\nonumber \\&&   \hspace*{1.5cm} 
       -  8\,{\delta }^2\,{\left( {g_1} + {g_A} \right)
         }^2\,{{\cal H}_{\eta^\prime}}({M_{\pi }},0) \Bigg]
\nonumber \\&&
  +        {\Delta \beta_n^{(a)}}\,\Bigg[12\,i\,{\cal I}({M_{\pi }})\, 
          \left( {q_j} + {q_l} \right)  - 4\,i  \,{\cal I}({M_{uj}})\, 
          \left( 2 + 3\,{q_j} + 3\,{q_l} \right)
\nonumber \\&&  \hspace*{1.5cm}+ 
         {\cal H}({M_{uj}},0)\,{g_1}\,\Big( -8\,{g_A} +
           3\,{g_1}\,\left( {q_j} + {q_l} \right)  \Big) 
\nonumber \\&& \hspace*{1.5cm}+  
         {\cal H}({M_{\pi }},0)\,\Big( 8\,{g_1}\,{g_A} + 2\,g_A^2 -
           3\,g_1^2\,\left( {q_j} + {q_l} \right)  \Big)  +  
         4\,{\delta }^2\,{\left( {g_1} + {g_A} \right)
         }^2\,{{\cal H}_{\eta^\prime}}({M_{\pi }},0) \Bigg]
  \Bigg\}    \,,
\label{eq:helicity_isovec_su42}
\end{eqnarray}
and the isoscalar, unpolarised matrix element is
\begin{eqnarray}
 \langle N|{\cal O}^{(0)}_{\mu_0\ldots\mu_n}|N\rangle &=&  
{\overline{U}_N v_{\mu_0}\ldots
  v_{\mu_n}U_N}(\alpha_n^{(s)}+\beta0_n^{(s)})\times(1+(1-\delta_{n0})\,{\cal W}_{SU(4|2)})
\nonumber \\&&  + \frac{i }{2f^2}\,{\overline{U}_N v_{\mu_0}\ldots
  v_{\mu_n}U_N}(1-\delta_{n0})\,
    \Bigg\{4\,\Big( {\cal H}({M_{\pi }},\Delta ) + {\cal H}({M_{uj}},\Delta ) \Big) \,
         g_{\Delta N}^2\,\left( \gamma_n^{(s)} -
           \frac{\sigma_n^{(s)}}{3} \right) 
\nonumber \\&&+  
      \left( \alpha_n^{(s)} + \beta_n^{(s)} \right) \,
       \Bigg[ {\cal H}({M_{\pi }},0)\,\left( 5\,g_1^2 + 4\,{g_1}\,{g_A} - g_A^2 \right)   
\nonumber \\&& \hspace*{1cm}
-   {\cal H}({M_{uj}},0)\,\left( 5\,g_1^2 + 4\,{g_1}\,{g_A} +
           8\,g_A^2 \right)+ 
         6\,{\delta }^2\,{\left( {g_1} + {g_A} \right)
         }^2\,{{\cal H}_{\eta^\prime}}({M_{\pi }},0) \Bigg]  \Bigg\} \,.
\label{eq:unpol_isoscal_su42}
\end{eqnarray}
The isoscalar, helicity matrix element is
\begin{eqnarray}
 \langle N|\widetilde{\cal O}^{(0)}_{\mu_0\ldots\mu_n}|N\rangle &=&  
{\overline{U}_N v_{\{\mu_0}\ldots
  v_{\mu_{n-1}}S_{\mu_n\}}U_N}(\Delta\alpha_n^{(s)}+\Delta\beta_n^{(s)})\times(1+{\cal W}_{SU(4|2)})
\nonumber \\&& 
+ \frac{i }{12f^2}\,{\overline{U}_N v_{\{\mu_0}\ldots
  v_{\mu_{n-1}}S_{\mu_n\}}U_N}\,
    \Bigg\{ \frac{40}{3}\,\Big( {\cal H}({M_{\pi }},\Delta ) + {\cal
      H}({M_{uj}},\Delta ) \Big)  \,
         g_{\Delta N}^2\,\left( \Delta \gamma_n^{(s)} - \frac{\Delta \sigma_n^{(s)}}{5} 
           \right) 
\nonumber \\&&+ \left(\Delta\alpha_n^{(s)}+\Delta\beta_n^{(s)}\right) \, 
       \Bigg[ -2\,{\cal H}({M_{\pi }},0)\,\left( 5\,g_1^2 +
         4\,{g_1}\,{g_A} - g_A^2 \right)  
\nonumber \\&& \hspace*{1cm} 
+  2\,{\cal H}({M_{uj}},0)\,\left( 5\,g_1^2 + 4\,{g_1}\,{g_A} +
           8\,g_A^2 \right)- 
         12\,{\delta }^2\,{\left( {g_1} + {g_A} \right)
         }^2\,{{\cal H}_{\eta^\prime}}({M_{\pi }},0) \Bigg]  \Bigg\} \,.
\label{eq:helicity_isoscal_su42}
\end{eqnarray}

Finally, the transversity matrix elements are
\begin{eqnarray}
\langle N|\widetilde{\cal O}^{T}_{\mu_0\ldots\mu_n\alpha}|N\rangle &=&
\nonumber
\\  
&&\hspace*{-2cm} 
\frac{1}{6}{\overline{U}_N v_{\{\mu_0}\ldots
  v_{[\mu_{n}\}}S_{\mu_\alpha]}U_N}((5+{y_i})\delta\alpha_n +
2(1+2{y_i})\delta\beta_n) \times (1+{\cal W}_{SU(4|2)}) 
\nonumber \\&& \hspace*{-2cm} 
+ \frac{i }{12f^2}\,{\overline{U}_N v_{\{\mu_0}\ldots
  v_{[\mu_{n}\}}S_{\mu_\alpha]}U_N}\,
    \Bigg\{ \frac{16}{3}\,{\sqrt{\frac{2}{3}}}\,{g_{\Delta N}}\,
         \Bigg[ {\cal K}({M_{uj}},\Delta )\,
            \Big( 4g_A\,\left( 1 - {y_i} \right)  + 
              g_1\,\left( -4 - 2\,{y_i} + 3\,{y_j} + 3\,{y_k} \right)
            \Big) 
\nonumber \\&& \hspace*{-2cm}  \hspace*{1.5cm}
  +  {\cal K}({M_{\pi }},\Delta )\,\Big( 4g_A\,\left( 1 - {y_i} \right)  + 
              g_1\,\left( 4 + 2\,{y_i} - 3\,{y_l} - 3\,{y_m} \right)  \Big)  \Bigg] \,
         {\delta c_n}
\nonumber \\&& \hspace*{-2cm} 
+ \frac{20 }{27}\,g_{\Delta N}^2\,
         \Big[ {\cal H}({M_{uj}},\Delta )\,\left( 10 + 2\,{y_i} + 3\,{y_j} + 3\,{y_k} \right)  + 
           3\, {\cal H}({M_{\pi }},\Delta )\,\left( 6 + 2\,{y_i} - {y_l} -
             {y_m} \right)  \Big] \, 
         \left( {\delta \gamma_n} - \frac{3\,{\delta \sigma_n}}{5}
         \right)
\nonumber \\&& \hspace*{-2cm} 
     + {\delta \alpha_n}\,\Bigg[ 2\, i \left[\,{\cal I}({M_{uj}})\,
          \left( 10 + 2\,{y_i} + 6\,{y_j} + 6\,{y_k} \right) 
 +  \,{\cal I}({M_{\pi }})\,\left( 1 + 5\,{y_i}
              - 6\,{y_l} - 6\,{y_m} \right)  \right]
 -4 \left( 5 + {y_i} \right) {\delta }^2\,{\cal I}_{\eta^\prime}(M_\pi)
\nonumber \\&& \hspace*{-2cm} 
  \hspace*{1.5cm} 
        + {\cal H}({M_{uj}},0)\,\left( g_1^2\,\left( 7 + {y_i} + {y_j} + {y_k} \right)  + 
            2\,{g_1}\,{g_A}\,\left( 2 + {y_j} + {y_k} \right)  + 
            4\,g_A^2\,\left( 1 + {y_i} + {y_j} + {y_k} \right)  \right)  
\nonumber \\&& \hspace*{-2cm} 
  \hspace*{1.5cm} 
        + {\cal H}({M_{\pi }},0)\,\left( g_A^2\,\left( 3 + 7\,{y_i} -
             4\,{y_l} - 4\,{y_m} \right)  -2\,  
            {g_1}\,{g_A}\,\left( 2 + {y_l} + {y_m} \right)  -
            g_1^2\,\left( 7 + {y_i} + {y_l} + {y_m} \right)  
            \right)  
\nonumber \\&&\hspace*{-2cm} 
 \hspace*{1.5cm} 
     -2\,\left(g_1+g_A\right)^2\,\left(5+{y_i}\right)\,       
             {\delta }^2\,{{\cal H}_{\eta^\prime}}({M_{\pi }},0)
             \Bigg] 
\nonumber \\&& \hspace*{-2cm} 
+ 
      {\delta \beta_n}\,\Bigg[ 4\,i \left[\,{\cal I}({M_{uj}})\,
          \left( 2 + 4\,{y_i} + 3\,{y_j} + 3\,{y_k} \right) + 
         {\cal I}({M_{\pi }})\,\left( 2 + {y_i} - 3\,{y_l} - 3\,{y_m} \right)  \right] 
         -8\left( 1 +2\,{y_i} \right) \,{{\delta }^2\,\cal I}_{\eta^\prime}(M_\pi)
\nonumber \\&&  \hspace*{-2cm} 
 \hspace*{1.5cm}
        + {\cal H}({M_{uj}},0)\,\left( 8\,{g_1}\,{g_A}\,{y_i} +
           8\,g_A^2\,\left( 1 + {y_i} \right)  +  
            g_1^2\,\left( 2 + 2\,{y_i} + 3\,{y_j} + 3\,{y_k} \right)  \right)  
\nonumber \\&& \hspace*{-2cm} 
 \hspace*{1.5cm}
        + {\cal H}({M_{\pi }},0)\,\left( 2\,g_A^2 - 8\,{g_1}\,{g_A}\,{y_i} - 
            g_1^2\,\left( 2 + 2\,{y_i} + 3\,{y_l} + 3\,{y_m} \right)  \right)  
\nonumber \\&& \hspace*{-2cm} 
 \hspace*{1.5cm}
     -4\,\left(g_1+g_A\right)^2\,\left(1+2{y_i}\right)       
             \,{\delta }^2\,{{\cal H}_{\eta^\prime}}({M_{\pi }},0) \Bigg] 
      \Bigg\} \,.
\label{eq:transversity_su42}
\end{eqnarray}
%


\section{Results for SU(6$|$3) \pqxpt\ with $m_u=m_d\ne m_s$}
\label{A3}

In three flavour QCD, one seeks to determine the up, down and strange
quark, unpolarised, helicity, and transversity distributions in the
octet baryons. Consequently, the goal of lattice calculations is to
determine the corresponding twist-two operator matrix elements for
each of these flavours in the octet baryons. The SU(6$|$3) results
presented in this appendix will be relevant for chiral and infinite
volume extrapolations of lattice calculations of moments of the the
strange-quark distributions in the nucleon\footnote{One can also study
  strangeness in the nucleon using two-flavour \xpt\ 
  \cite{Chen:2002bz} thereby sidestepping issues of the slow(er)
  convergence of the chiral expansion around the physical
  strange-quark mass.} and the various parton distributions in, for
example, the $\Lambda$ hyperon \cite{Gockeler:2002uh}.

If we consider the extensions of three-flavour QCD to partially
quenched theories, we are naturally led to SU(6$|$3) \hbxpt.  The
Lagrangian of this theory is very similar to that described in
Sec.~\ref{S2} with some simple modifications. Obviously the meson
field $\Phi$ is enlarged, becoming a 9$\times$9 matrix encoding the
${\bf 80}$-plet of pseudo-Goldstone mesons. The octet baryons are now
embedded in a ${\bf 240}$ representation and the decuplet baryons in a
${\bf 138}$ representation.  Additionally, the couplings $\alpha$,
$\beta$ and ${\cal C}$ in Eq.~(\ref{eq:free_lagrangian}) are replaced
by $\alpha\to \frac{2}{3}D+2F$, $\beta\to -\frac{5}{3}D+F$ and ${\cal
  C}\to C$ so that the nomenclature is the same as in SU(3) \xpt. For
definiteness, the quark masses we consider are $m_Q = {\rm
  diag}(m_u,m_d,m_s,m_j,m_l,m_r,m_u,m_d,m_s)$ and ${\cal Q}$ after
Eq.~(\ref{eq:mesonmass_def}) is replaced by ${\cal
  Q}=(u,\,d,\,s,\,j,\,l,\,r,\,\tilde{u},\,\tilde{d},\,\tilde{s})$.
Further details are given in Ref.~\cite{Chen:2001yi}.

To calculate the independent moments of the PDFs in three flavour QCD,
one constructs three independent flavour combinations of operators.
The standard choice is
\begin{equation}
\label{eq:334}
i^n\overline\psi\Gamma_{\{\mu_0} \tensor{D}_{\mu_1}\ldots
\tensor{D}_{\mu_n\}}\left\{{\bf 1},\lambda_3,\lambda_8\right\}\psi\,,
\end{equation}
where $\Gamma=\gamma,\,\gamma\gamma_5,\,\sigma$ represents the
appropriate Dirac structure and the $\lambda_i$ are the usual
Gell-Mann basis for SU(3). The unpolarised and helicity operators are
in the singlet ({\bf 1},{\bf 1}) or adjoint ({\bf 8},{\bf
  1})$\oplus$({\bf 1},{\bf 8}) representations of
SU(3)$_L\times$SU(3)$_R$. From a lattice practitioner's point of view,
$\lambda_3$ is somewhat special since in the limit $m_u=m_d$ there are
no disconnected contributions to matrix elements of such operators. On
the other hand, both the singlet and $\lambda_8$ operator require such
contributions and no choice of flavour basis can ameliorate the
situation. In partially quenched QCD, there is again freedom in the
extension of the above QCD operators; a natural choice with a smooth
QCD limit is
\begin{equation}
\bar\lambda_3={\rm diag}(1,-1,0,1,-1,0,1,-1,0)
  \label{eq:4}
\qquad {\rm and} \qquad
\bar\lambda_8={\rm diag}(1,1,-2,1,1,-2,1,1,-2)\,.
\end{equation}
In our results, only a single adjoint representation operator is
presented, corresponding to
\begin{equation}
\bar\lambda_{Adj}={\rm
  diag}(1,q_1,-1-q_1,1,q_2,-1-q_2,1,q_1,-1-q_1)\,, 
\end{equation}
and to determine $\bar\lambda_3$ and $\bar\lambda_8$, we set
$q_2=q_1=\mp 1$. Keeping $q_2\ne q_1$, will allow SU(3) breaking
effects to be analysed in detail. However, the singlet operator is
uniquely defined
\begin{equation}
\bar\lambda_0={\rm diag}(1,1,1,1,1,1,1,1,1)
  \label{eq:45}
\end{equation}
and disconnected loops involving sea quarks are unavoidable.

The transversity operators in QCD belong to the $({\bf
  3},\overline{{\bf 3}})\oplus(\overline{{\bf 3}},{\bf 3})$
representation of SU(3)$_L\times$SU(3)$_R$ irrespective of the choice
of flavour structure in Eq.~(\ref{eq:334}). For the partially quenched
QCD extension of these operators, we choose the operators built from
\begin{equation}
\bar\lambda_T={\rm diag}(1,y_1,y_2,0,0,0,1,y_1,y_2)
  \label{eq:46}
\end{equation}
and then setting $\{y_1,y_2\}=\{1,1\},\,\{-1,0\},\,\{1,-2\}$ gives the
required flavour combinations.

In this section, we give results for the matrix elements in the
proton, $\Lambda^0$, $\Sigma^+$, $\Xi^-$ as well as the
$\Sigma^0$--$\Lambda^0$ transition.  Other octet matrix elements are
simply related to these by isospin symmetry. In the results of this
section, the low-energy coefficients ($\alpha_n^{(a)}$,
$\Delta\gamma_n^{(s)}$, etc) occurring in the SU(6$|$3) versions of
Eqs.~(\ref{eq:hadron_op})--(\ref{eq:hadron_sigmaop}) are different
from those in SU(4$|$2). With this caution, we use the same notation.

The SU(6$|$3) tree level meson masses, $M_{ss}$, $M_{sr}$, $M_{ur}$
and $M_{sj}$, are defined through Eq.~(\ref{eq:mesonmass_def}),
$\delta$ is defined in Eq.~(\ref{eq:delta_def}),
\begin{equation}
\label{eq:tilde_delta_def}
\tilde{\delta}^2 = M_{ss}^{2} - M_{sr}^{2} = \lambda (m_{s} - m_{r}) \, ,
\end{equation}
and
\begin{equation}
\label{eq:M_X_def}
 M_{X}^{2} = \frac{1}{3} 
\left [ M_{\pi}^{2} + 2 M_{ss}^{2} 
 - 2 \left (\delta^{2} + 2 \tilde{\delta}^{2}\right ) \right ] \,.
\end{equation}
The QCD limit is easily recovered, taking $\delta\to0$,
$\tilde\delta\to0$, $j\to u$, $l \to d$ and $r\to s$.  To make the
presentation succinct, we define the following ratios,
\begin{equation}
\label{eq:frakA_def}
 \frakA = -i \frac{\delta^{2} \left (M_{\pi}^{2} - M_{ss}^{2} 
  + 2 \tilde{\delta}^{2}\right )}{M_{\pi}^{2} - M_{ss}^{2} 
  + \delta^{2} + 2 \tilde{\delta}^{2}} ,
\end{equation}
\begin{equation}
\label{eq:frakB_def}
 \frakB = -i \frac{\left (M_{\pi}^{2} - M_{ss}^{2} 
  + 2 \tilde{\delta}^{2}\right )^{2} + 2 \delta^{4}}{2
   (M_{\pi}^{2} - M_{ss}^{2} 
  + \delta^{2} + 2 \tilde{\delta}^{2})^{2}} ,
\end{equation}
\begin{equation}
\label{eq:frakC_def}
 \frakC = -i \frac{\tilde{\delta}^{2} \left (M_{\pi}^{2} - M_{ss}^{2} 
  - 2 \delta^{2}\right )}{M_{\pi}^{2} - M_{ss}^{2} 
  -2  \delta^{2} - 4 \tilde{\delta}^{2}} ,
\end{equation}
\begin{equation}
\label{eq:frakD_def}
 \frakD = -i \frac{\left (M_{\pi}^{2} - M_{ss}^{2} 
  - 2 \delta^{2}\right )^{2} + 8 \tilde{\delta}^{4}}{M_{\pi}^{2} - M_{ss}^{2} 
  -2  \delta^{2} - 4 \tilde{\delta}^{2}} ,
\end{equation}
\begin{equation}
\label{eq:frakE_def}
 \frakE = -i \frac{\delta^{2} \left ( M_{\pi}^{2} - M_{ss}^{2}
  + 2 \tilde{\delta}^{2}\right )}{\left ( M_{\pi}^{2} - M_{ss}^{2}
  \right ) \left ( M_{\pi}^{2} - M_{ss}^{2} + \delta^{2} 
  + 2 \tilde{\delta}^{2} \right )} ,
\end{equation}
\begin{equation}
\label{eq:frakF_def}
 \frakF = 2 i \frac{\tilde{\delta}^{2} \left ( M_{\pi}^{2} - M_{ss}^{2}
  - 2 \delta^{2}\right )}{\left ( M_{\pi}^{2} - M_{ss}^{2}
  \right ) \left ( M_{\pi}^{2} - M_{ss}^{2} - 2 \delta^{2} 
  - 4 \tilde{\delta}^{2} \right )} ,
\end{equation}
and the functions,
\begin{eqnarray}
 \quu &=& \frakA \calI_{\eta^\prime,uu} + \frakB \calI_{uu}
   + (1-\frakB) \calI_X ,
\nonumber\\ 
 \qss &=& \frakC \calI_{\eta^\prime,ss} + \frakD \calI_{ss}
   + (1-\frakD) \calI_X ,
\label{eq:quu_def}
\end{eqnarray}
\begin{eqnarray}
 \ruu &=& \frakA \calH_{\eta^\prime,uu} + \frakB \calH_{uu}
   + (1-\frakB) \calH_X ,
\nonumber \\ 
 \rss &=& \frakC \calH_{\eta^\prime,ss} + \frakD \calH_{ss}
   + (1-\frakD) \calH_X ,
 \nonumber \\
 \rus &=& \frakE \calH_{uu} + \frakF \calH_{ss}
   + (1 - \frakE - \frakF) \calH_X ,
\label{eq:ruu_def}
\end{eqnarray}
\begin{eqnarray}
 \suu &=& \frakA \calK_{\eta^\prime,uu} + \frakB \calK_{uu}
   + (1-\frakB) \calK_{X} ,
\nonumber \\ 
 \sss &=& \frakC \calK_{\eta^\prime,ss} + \frakD \calK_{ss}
   + (1-\frakD) \calK_{X} ,
\nonumber\\ 
 \sus &=& \frakE \calK_{uu} + \frakF \calK_{ss}
   + (1 - \frakE - \frakF) \calK_{X} ,
\label{eq:suu_def}
\end{eqnarray}
and
\begin{eqnarray}
 \tuu &=& \frakA \calH_{\eta^\prime,uu}^\Delta + \frakB \calH_{uu}^\Delta
   + (1-\frakB) \calH_{X}^\Delta ,
\nonumber \\
 \tss &=& \frakC \calH_{\eta^\prime,ss}^\Delta + \frakD \calH_{ss}^\Delta
   + (1-\frakD) \calH_{X}^\Delta ,
\nonumber \\  
 \tus &=& \frakE \calH_{uu}^\Delta + \frakF \calH_{ss}^\Delta
   + (1 - \frakE - \frakF) \calH_{X}^\Delta .
\label{eq:tuu_def}
\end{eqnarray}
where
\begin{eqnarray}
\calI_{ij}=\calI(M_{ij})\,,
&\quad&
\calI_{\eta^\prime,ij}=\calIdp(M_{ij})\,,
\nonumber\\
\calH_{ij}=\calH(M_{ij},0)\,,
&\quad&
\calH_{\eta^\prime,ij}=\calHdp(M_{ij},0)\,,
\nonumber\\
\calH^\Delta_{ij}=\calH(M_{ij},\Delta)\,,
&\quad&
\calH^\Delta_{\eta^\prime,ij}=\calHdp(M_{ij},\Delta)\,,
\nonumber\\
\label{eq:Kij_def}
\calK_{ij}=\calK(M_{ij},\Delta)\,,
&\quad&
\calK_{\eta^\prime,ij}=\calKdp(M_{ij},\Delta)\,,
\end{eqnarray}
\begin{eqnarray}
\calI_{X}=\calI(M_{X})\,,
&\quad&
\calH_{X}=\calH(M_{X},0)\,,
\nonumber \\
\calH^\Delta_{X}=\calH(M_{X},\Delta)\,,
&\quad&
\calK_{X}=\calK(M_{X},\Delta)  \,,
  \label{eq:x_defs}
\end{eqnarray}
and
\begin{eqnarray}
\calH_{abc} = 2\, \calH(M_{ab},0) + \calH(M_{ac},0)\,,
&\quad\quad&
\hat\calH_{abc} =  \calH(M_{ab},0) - \calH(M_{ac},0)\,,
\nonumber \\
\calH_{abc}^\Delta = 2\, \calH(M_{ab},\Delta) + \calH(M_{ac},\Delta)\,,
&\quad\quad&
\hat\calH_{abc}^\Delta =  \calH(M_{ab},\Delta) - \calH(M_{ac},\Delta)\,,
\nonumber \\
\calI_{abc} = 2\, \calI(M_{ab}) + \calI(M_{ac})\,,
&\quad\quad&
\hat\calI_{abc} = \calI(M_{ab}) - \calI(M_{ac})\,,
\nonumber \\
\label{eq:Kujr_def}
\calK_{abc} = 2\, \calK(M_{ab},\Delta) + \calK(M_{ac},\Delta)\,,
&\quad\quad&
\hat\calK_{abc} =  \calK(M_{ab},\Delta) - \calK(M_{ac},\Delta)\,,
\end{eqnarray}
and finally
\begin{equation}
 \tilde{y} = \frac{y_{2}}{1+y_{1}} \qquad {\rm and} \qquad
 \tilde{q} = \frac{1+q_{2}}{1+q_{1}}\,.
\end{equation}

\subsection{Wave-function renormalisation}

The wave-function renormalisations for the different octet states are
%
\begin{eqnarray}
 \calW^{(p)}_{SU(6|3)} &=& \frac{i}{f^{2}}
  \bigg \{ 
   \bigg [
    2 \calH_{uu}^\Delta + \calH_{ujr}^\Delta
   \bigg ]\, C^{2} 
   +  \calH_{ujr} \,
    (5 D^{2} - 6 DF + 9 F^{2}) -4 \calH_{uu}\,D\,(D-3F)
    -3\, i\, R_{uu} (D-3F)^{2}
  \bigg \}
\hspace*{5mm}
\end{eqnarray}
for the proton,
\begin{eqnarray}
\calW^{(\Lambda)}_{SU(6|3)} &=&
\frac{i }{3 f^{2}}\,\bigg \{  3\,
     \bigg [ \calH_{us}^\Delta + \calH_{uu}^\Delta + 
       \calH_{ujr}^\Delta \bigg ] \,C^2
+ \calH_{sjr}
 (D + 3\, F)^{2}
+  2\,\calH_{ujr}
 (7 D^{2} - 12 DF + 9 F^{2})
\nonumber\\
& &\,\,\,\,\,\,\,\,\,\,\,\,\,\,\, - 
    2\,\calH_{us}\,\left( 5\,D^2 - 6\,D\,F - 9\,F^2
       \right)  - 2\,\calH_{uu}\,
     \left( D^2 - 12\,D\,F + 9\,F^2 \right)  
\nonumber\\
& &\,\,\,\,\,\,\,\,\,\,\,\,\,\,\, +i \bigg [ - 
    R_{ss} ( 3 F + D )^{2} 
    + R_{us} \left ( 8 D^2 + 12 D F - 36 F^{2} \right )
 - 
     4 \,R_{uu}\,
     {\left( 2\,D - 3\,F \right) }^2 
  \bigg ]
\bigg \} 
\end{eqnarray}
for the $\Lambda$ baryons,
\begin{eqnarray}
 \calW^{(\Sigma)}_{SU(6|3)} &=&
 \frac{i }{3 f^{2}}\,\bigg \{ \bigg [  5\,\calH_{us}^\Delta + 
       \calH_{uu}^\Delta + 2\,\calH_{sjr}^\Delta + \calH_{ujr}^\Delta 
-        2\,i  \,
        \left( T_{ss} - 2\,T_{us} + T_{uu} \right)  \bigg ]
      \,C^2 
+     3\,\bigg [  3\,\calH_{sjr}
      \,{\left (D-F \right) }^2 
\nonumber\\
& &\,\,\,\,\,\,\,\,\,\,\,\,\,\,\,\,\,\,
     + 2\,\calH_{ujr}\,\left (D^{2} + 3\, F^{2} \right )   
      -  2\,\calH_{uu}\,\left( D^2 - 3\,F^2 \right) 
          - 2\,\calH_{us}\,
     \left( D^2 - 6\,D\,F + 3\,F^2 \right) \bigg ]
\nonumber\\
& & \,\,\,\,\,\,\,\,
   - 9\,i\, \bigg [
          \,R_{ss}\,(D-F)^2 + 
        4 \,R_{us}\,F (F-D) + 
        4 \,R_{uu}\,F^2
       \bigg ]  
\bigg \}
\end{eqnarray}
for the $\Sigma$ baryons, and
\begin{eqnarray}
 \calW^{(\Xi)}_{SU(6|3)} &=& \frac{i}{3 f^{2}}
\bigg \{
 \bigg [
   5\,\calH_{us}^\Delta + \calH_{sjr}^\Delta + \calH_{ss}^\Delta + 
   2\,\calH_{ujr}^\Delta
-     2\,i \,\left( T_{ss} - 2\,T_{us} + T_{uu} \right)      
 \bigg ]\, C^{2}\nonumber\\
& &\,\,\,\,\,\,\,\,\,\,\,\,
+ 3 \bigg [
   3\,\calH_{ujr}
    (D-F)^{2}
    + 2\,\calH_{sjr}
     (D^{2} + 3 F^{2})
  -2 \calH_{us} (D^{2} - 6 DF + 3F^{2})
  -2 \calH_{ss} (D^{2} - 3 F^{2})
 \bigg ]\nonumber\\
& &\,\,\,\,\,\,\,\,\,\,\,\,
-9\, i\, \bigg [
   4\,R_{ss}\,F^{2} -4\,R_{us}\,F(D-F) + R_{uu}\,(D-F)^{2}
 \bigg ]
\bigg \} 
\end{eqnarray}
for the $\Xi$ baryons.

\subsection{Isovector unpolarised matrix elements}

\begin{eqnarray}
\label{eq:isovec_unpol_proton_su63}
 \la p | \op^{(3)}_{\mu_{0}\cdots\mu_{n}} | p \ra &=&
 \frac{\overline{U}_p v_{\mu_{0}}\ldots v_{\mu_{n}}U_p\big ( (5 + q_{1}) \alpha^{(a)}_{n}
  + 2 (1 + 2 q_{1} ) \beta^{(a)}_{n} \big )}{6}
 \times \left (1 + \calW^{(p)}_{SU(6|3)}(1-\delta_{n0}) 
  \right )
\\
 & & + \frac{\overline{U}_p v_{\mu_{0}}\ldots v_{\mu_{n}}U_p(1-\delta_{n0})}{f^{2}}
\bigg \{
  \frac{i}{9}\,
  \left( \,{{\gamma^{(a)}_{n}}} - \frac{{\sigma^{(a)}_{n}}}{3} \right)
 \bigg[ \bigg ( 3\,\calH_{uu}^\Delta + 
    \calH_{ujr}^\Delta \bigg ) \,
  \left( 5 + {q_1} \right)
+3(1+q_2) \hat\calH^{\Delta}_{ujr}\bigg] 
 C^{2}\nonumber\\
& &\,\,\,\,\,\,\,\,\,\,\,\,\,\,\,
+\alpha^{(a)}_{n}
 \bigg [\frac{1}{6}(-5 - q_{1}) 
   \, \calI_{ujr}
+(1+q_2)\hat\calI_{ujr}
  -\,i\, \calH_{ujr} 
  \bigg ( 3\,D^2 - 6\,D\,F + 5\,F^2 + 
  \left( D^2 + F^2 \right) \,{q_1}
   \bigg )\nonumber\\
& &\,\,\,\,\,\,\,\,\,\,\,\,\,\,\,\,\,\,\,\,\,\,\,\,\,\,\,\,\,\,\,
  + \,4\,i\, \calH_{uu}\,D\,(D-2F-F q_{1}) 
-i(1+q_2)(D^2+3F^2)\hat\calH_{ujr}  - \frac{1}{2} R_{uu} (D-3F)^{2} (5+q_{1})
 \bigg ]\nonumber\\
& &\,\,\,\,\,\,\,\,\,\,\,\,\,\,\,
+\beta^{(a)}_{n}
 \bigg [\frac{-1}{3}(1 + 2 q_{1}) 
   \, \calI_{ujr} 
+(1+q_2)\hat\calI_{ujr}
  -\,2\,i\, \calH_{ujr} 
  \bigg ( D^2 + F^2 (1+2{q_1})
   \bigg )\nonumber\\
& &\,\,\,\,\,\,\,\,\,\,\,\,\,\,\,\,\,\,\,\,\,\,\,\,\,\,\,\,\,\,\,
  + \,4\,i\, \calH_{uu}\,D\,(D-2F-F q_{1}) 
-3i(1+q_2)(D-F)^2\hat\calH_{ujr}  -  R_{uu} (D-3F)^{2} (1+2q_{1})
 \bigg ]
\bigg \} \,. 
\nonumber
\end{eqnarray}
\begin{eqnarray}
 \la \Lambda^{0} | \op^{(3)}_{\mu_{0}\cdots\mu_{n}} | 
 \Lambda^{0}\ra &=& 
   -\frac{\overline{U}_{\Lambda^{0}} v_{\mu_{0}}\ldots v_{\mu_{n}} U_{\Lambda^{0}} (1+q_{1})}{4} 
    \left (\alpha^{(a)}_{n} -2 \beta^{(a)}_{n} \right )\times
 \left ( 1 + \calW^{(\Lambda)}_{SU(6|3)}(1-\delta_{n0})\right )\nonumber\\
& & + \frac{\overline{U}_{\Lambda^{0}} v_{\mu_{0}}\ldots v_{\mu_{n}}
  U_{\Lambda^{0}} (1+q_{1})(1-\delta_{n0})}{f^{2}} 
\bigg \{
  -\frac{i}{6}\,
  \left( \,{{\gamma^{(a)}_{n}}} - \frac{{\sigma^{(a)}_{n}}}{3} \right)
\bigg ( 3\,\calH_{us}^\Delta + 
    \calH_{ujr}^\Delta -2\tilde{q}\hat\calH_{ujr}^\Delta\bigg ) 
  C^{2}\nonumber\\
& &\,\,\,\,\,\,\,\,\,\,\,\,\,\,\,\,\,\,\,\,\,
+\alpha^{(a)}_{n}
 \bigg [ \frac{1}{4} \bigg (-\,\calI_{ujr} + 2\,\calI_{sjr}\bigg )
+\frac{\tilde{q}}{2}\left(\hat\calI_{ujr} + \hat\calI_{sjr}\right)
\nonumber\\
& &\,\,\,\,\,\,\,\,\,\,\,\,\,\,\,\,\,\,\,\,\,
 +\frac{i}{12}  \calH_{ujr} 
 \,(7D - 9 F)\, (D-3F)
 -\frac{i}{12}  \calH_{sjr} 
 \,(D+3F)^{2}\nonumber\\
& &\,\,\,\,\,\,\,\,\,\,\,\,\,\,\,\,\,\,\,\,\,\,\,\,\,\,\,\,\,\,\,\,\,\,
 -\frac{i}{6} \calH_{uu} \left ( 7 D^{2} - 12 DF + 9 F^{2}\right )
 -\frac{i}{6} \calH_{us} (5D - 3F) (D+3F)
 \nonumber\\
& &\,\,\,\,\,\,\,\,\,\,\,\,\,\,\,\,\,\,\,\,\,\,\,\,\,\,\,\,\,\,\,\,\,\,
-\frac{i}{6} \tilde{q}(D+3F)^2 \hat\calH_{sjr}
-\frac{i}{6} \tilde{q}(5D^2-6DF+9F^2) \hat\calH_{ujr}
 \nonumber\\
& &\,\,\,\,\,\,\,\,\,\,\,\,\,\,\,\,\,\,\,\,\,\,\,\,\,\,\,\,\,\,\,\,\,\,
 +\frac{1}{12}
  \bigg (
   R_{ss}\,(D+3F)^{2} -4 R_{us}\,(2D-3F)(D+3F) 
   +4 R_{uu}\, (2D-3F)^{2}
  \bigg )
 \bigg ]\nonumber\\
& &\,\,\,\,\,\,\,\,\,\,\,\,\,\,\,\,\,\,\,\,\,
+\beta^{(a)}_{n}
 \bigg [-\frac{1}{2} \bigg (2 \calI_{uj}
 + \calI_{ur}\bigg ) +\tilde{q}\hat\calI_{ujr}
 +\frac{i}{6}  \calH_{ujr} 
 \,(7D^{2} + 6DF - 9F^{2})
 -\frac{i}{6}  \calH_{sjr} 
 \,(D+3F)^{2}\nonumber\\
& &\,\,\,\,\,\,\,\,\,\,\,\,\,\,\,\,\,\,\,\,\,\,\,\,\,\,\,\,\,\,\,\,\,\,
 -\frac{i}{3} \calH_{uu} \left ( 7 D^{2} - 12 DF + 9 F^{2}\right )
 -\frac{i}{3} \calH_{us} (13D^{2} -12 DF + 9F^{2})
-3i \tilde{q}(D-F)^2 \hat\calH_{ujr}
 \nonumber\\
& &\,\,\,\,\,\,\,\,\,\,\,\,\,\,\,\,\,\,\,\,\,\,\,\,\,\,\,\,\,\,\,\,\,\,
 -\frac{1}{6}
  \bigg (
   R_{ss}\,(D+3F)^{2} -4 R_{us}\,(2D-3F)(D+3F) 
   +4 R_{uu}\, (2D-3F)^{2}
  \bigg )
 \bigg ]
\bigg \} \,.
\label{eq:isovec_unpol_lambda_su63}
\end{eqnarray}
\begin{eqnarray}
 \la \Sigma^{+} | \op^{(3)}_{\mu_{0}\cdots\mu_{n}} | 
 \Sigma^{+}\ra &=& 
   -\frac{\overline{U}_{\Sigma^{+}} v_{\mu_{0}}\ldots v_{\mu_{n}} U_{\Sigma^{+}} \left (
    (-4+q_{1})\alpha^{(a)}_{n} + 2 (1 + 2 q_{1}) \beta^{(a)}_{n}
   \right )}{6}
   \times
 \left ( 1 + \calW^{(\Sigma)}_{SU(6|3)}(1-\delta_{n0})\right )\nonumber\\
& & + \frac{\overline{U}_{\Sigma^{+}} v_{\mu_{0}}\ldots v_{\mu_{n}} U_{\Sigma^{+}} (1-\delta_{n0})}{f^{2}}
\bigg \{
 \frac{i}{9}
   \left (\gamma^{(a)}_{n} - \frac{\sigma^{(a)}_{n}}{3} \right )\bigg (
   \calH_{uu}^\Delta + 11\,\calH_{us}^\Delta + 4\,\calH_{sjr}^\Delta\nonumber\\
& &\,\,\,\,\,\,\,\,\,\,\,\,\,\,\,\,\,\,\,\,\,\,\,\,\,\,
   \,\,\,\,\,\,\,\,\,\,\,\,\,\,\,\,\,\,\,\,\,\, 
 -   \big [ \calH_{uu}^\Delta +2\,\calH_{us}^\Delta + \calH_{ujr}^\Delta \big ] \,
   {q_1}
+(1+q_2)\left(\hat\calH^{\Delta}_{ujr} +2
  \hat\calH^{\Delta}_{sjr}\right)
\nonumber\\
& &\,\,\,\,\,\,\,\,\,\,\,\,\,\,\,\,\,\,\,\,\,\,\,\,\,\,
   \,\,\,\,\,\,\,\,\,\,\,\,\,\,\,\,\,\,\,\,\,\, 
-   2\,i\,
   \left( T_{ss} - 2\,T_{us} + T_{uu} \right)(1-q_{1})
  \bigg )
  C^{2}\nonumber\\
& &\,\,\,\,\,\,\,\,\,\,\,\,\,\,\,\,\,\,\,\,\,
+\alpha^{(a)}_{n}
 \bigg [ \frac{1}{6} \bigg (-5\,\calI_{ujr} 
 +  \calI_{sjr} (1+q_{1}) 
+(1+q_2)\,\left(5\,\hat\calI_{ujr}+\hat\calI_{sjr}\right)
  \bigg )
\nonumber\\
& &\,\,\,\,\,\,\,\,\,\,\,\,\,\,\,\,\,\,\,\,\,\,\,\,\,\,\,\,\,\,\,\,
 +\frac{i}{2}  \calH_{ujr} 
 \,\bigg ( (D-F)(D+3F)+2\,(D^{2}+F^{2})\,q_{1}\bigg )\nonumber\\
& &\,\,\,\,\,\,\,\,\,\,\,\,\,\,\,\,\,\,\,\,\,\,\,\,\,\,\,\,\,\,\,\,\,\,
 -\frac{5i}{2}  \calH_{sjr} 
 \,(D-F)^{2}
 - i \calH_{uu} \bigg (4F^{2} + (D-F)(D+F)\,q_{1} \bigg )\nonumber\\
& &\,\,\,\,\,\,\,\,\,\,\,\,\,\,\,\,\,\,\,\,\,\,\,\,\,\,\,\,\,\,\,\,\,\,
 + i \calH_{us} \bigg ( 4 (D^{2}-DF+F^{2}) + (D^{2}+4DF - F^{2})\,q_{1}\bigg )
 \nonumber\\
& &\,\,\,\,\,\,\,\,\,\,\,\,\,\,\,\,\,\,\,\,\,\,\,\,\,\,\,\,\,\,\,\,\,\,
-\frac{i}{2}(1+q_2)(D^2+2DF+5F^2)\hat\calH_{ujr}
-\frac{i}{2}(1+q_2)(D-F)^2\hat\calH_{sjr}
 \nonumber\\
& &\,\,\,\,\,\,\,\,\,\,\,\,\,\,\,\,\,\,\,\,\,\,\,\,\,\,\,\,\,\,\,\,\,\,
 +\frac{1}{2}
  \bigg (
   R_{ss}\,(D-F)^{2} -4 R_{us}\,F\,(D-F) 
   +4 R_{uu}\,F^{2}
  \bigg ) (q_{1}-4)
 \bigg ]\nonumber\\
& &\,\,\,\,\,\,\,\,\,\,\,\,\,\,\,\,\,\,\,\,\,
+\beta^{(a)}_{n}
 \bigg [
\frac{1}{3} \bigg (-\,\calI_{ujr} 
 +  2\,\calI_{sjr} (1+q_{1}) 
+(1+q_2)\,\left(\hat\calI_{ujr}+2\,\hat\calI_{sjr}\right)
   \bigg )\nonumber\\
& &\,\,\,\,\,\,\,\,\,\,\,\,\,\,\,\,\,\,\,\,\,\,\,\,\,\,\,\,\,\,\,\,\,\,
 - i  \calH_{ujr} 
 \,\bigg ( (D-F)(D+3F) - 4 F^{2}\,q_{1}\bigg )\nonumber\\
& &\,\,\,\,\,\,\,\,\,\,\,\,\,\,\,\,\,\,\,\,\,\,\,\,\,\,\,\,\,\,\,\,\,\,
 - i  \calH_{sjr} 
 \,(D-F)^{2}
 +2\,i\,\calH_{uu} \bigg (D^{2} + F^{2} + 2 F^{2}\,q_{1} \bigg )\nonumber\\
& &\,\,\,\,\,\,\,\,\,\,\,\,\,\,\,\,\,\,\,\,\,\,\,\,\,\,\,\,\,\,\,\,\,\,
 +2\,i\,\calH_{us} \bigg ( D^{2}-2DF-F^{2} + 2(D-F)F\,q_{1}\bigg )
 \nonumber\\
& &\,\,\,\,\,\,\,\,\,\,\,\,\,\,\,\,\,\,\,\,\,\,\,\,\,\,\,\,\,\,\,\,\,\,
-i(1+q_2)(D-F)^2\left(\hat\calH_{ujr}+2\,\hat\calH_{sjr}\right) \nonumber\\
& &\,\,\,\,\,\,\,\,\,\,\,\,\,\,\,\,\,\,\,\,\,\,\,\,\,\,\,\,\,\,\,\,\,\,
 +
  \bigg (
   R_{ss}\,(D-F)^{2} -4 R_{us}\,F\,(D-F) 
   +4 R_{uu}\,F^{2}
  \bigg ) (1+2q_{1})
 \bigg ]
\bigg \} \,. 
\label{eq:isovec_unpol_sigma_su63}
\end{eqnarray}
\begin{eqnarray}
 \la \Xi^{-} | \op^{(3)}_{\mu_{0}\cdots\mu_{n}} | 
 \Xi^{-}\ra &=& 
   -\frac{\overline{U}_{\Xi^{-}} v_{\mu_{0}}\ldots v_{\mu_{n}} U_{\Xi^-}\left (
    (5+4q_{1})\alpha^{(a)}_{n} - 2 (-1 + q_{1}) \beta^{(a)}_{n}
   \right )}{6}
   \times
 \left ( 1 + \calW^{(\Xi)}_{SU(6|3)}(1-\delta_{n0})\right )\nonumber\\
& & + \frac{\overline{U}_{\Xi^{-}} v_{\mu_{0}}\ldots v_{\mu_{n}} U_{\Xi^-}(1-\delta_{n0})}{f^{2}}
\bigg \{
-\frac{i}{9}
   \left (\gamma^{(a)}_{n} - \frac{\sigma^{(a)}_{n}}{3} \right )
 \bigg (
  \calH_{sjr}^\Delta 
+   \calH_{ss}^\Delta\,\left( 2 + {q_1} \right)  + 
  \calH_{us}^\Delta\,\left( 13 + 11\,{q_1} \right)\nonumber\\
& &\,\,\,\,\,\,\,\,\,\,\,\,\,\,\,\,\,\,\,\,\,\,\,\,\,\,\,\,
  + 4\, \calH_{ujr}^\Delta \left( 1 + {q_1} \right)
-(1+q_2)\left(2\,\hat\calH^\Delta_{ujr}+\hat\calH^\Delta_{sjr}\right)
 - 2i\left( T_{ss} - 2\,T_{us} + T_{uu} \right) 
      \left( 2 + {q_1} \right)
 \bigg )
 C^{2}\nonumber\\
& &\,\,\,\,\,\,\,\,\,\,\,\,\,\,\,\,\,\,\,\,\,
+\alpha^{(a)}_{n}
 \bigg [ \frac{1}{6} \bigg (-\calI_{ujr}\, q_{1}
 + 5\,\calI_{sjr} (1+q_{1}) 
+(1+q_2)\left(\hat\calI_{ujr}+5\,\hat\calI_{sjr}\right)
   \bigg )
\nonumber\\
& &\,\,\,\,\,\,\,\,\,\,\,\,\,\,\,\,\,\,\,\,\,\,\,\,\,\,\,\,\,\,\,\,\,\,
 +\frac{5 i}{2}  \calH_{ujr} 
 \, (D-F)^{2}\,(1+q_{1})
 +\frac{i}{2}  \calH_{sjr} 
 \bigg ( D^{2} - 2DF + 5 F^{2} - (D-F) (D+3F) q_{1}\bigg )\nonumber\\
& &\,\,\,\,\,\,\,\,\,\,\,\,\,\,\,\,\,\,\,\,\,\,\,\,
 - i \calH_{ss} \bigg (D^{2}-5\,F^{2}-4\,F^{2}\,q_{1} \bigg )
 - i \calH_{us} 
 \bigg ( (3D-5F)(D-F) + 4 \left (D^{2}-DF + F^{2}\right ) q_{1}\bigg )
 \nonumber\\
& &\,\,\,\,\,\,\,\,\,\,\,\,\,\,\,\,\,\,\,\,\,\,\,\,\,\,\,\,\,\,\,\,\,\,
-\frac{i}{2}(1+q_2)(D-F)^2\hat\calH_{ujr}
-\frac{i}{2}(1+q_2)(D^2+2DF+5F^2)\hat\calH_{sjr}
 \nonumber\\
& &\,\,\,\,\,\,\,\,\,\,\,\,\,\,\,\,\,\,\,\,\,\,\,\,\,\,\,\,\,\,\,\,\,\,
 +\frac{1}{2}
  \bigg (
   4 R_{ss}\,F^{2} -4 R_{us}\,F\,(D-F) 
   +R_{uu}\,(D-F)^{2}
  \bigg ) (5+4\,q_{1})
 \bigg ]\nonumber\\
& &\,\,\,\,\,\,\,\,\,\,\,\,\,\,\,\,\,\,\,\,\,
+\beta^{(a)}_{n}
 \bigg [ \frac{1}{3} \bigg (-2 \,\calI_{ujr}\, q_{1}
 +  \calI_{sjr} (1+q_{1}) 
+(1+q_2)\left(2\,\hat\calI_{ujr}+\hat\calI_{sjr}\right)
   \bigg )
\nonumber\\
& &\,\,\,\,\,\,\,\,\,\,\,\,\,\,\,\,\,\,\,\,\,\,\,\,\,\,\,\,\,\,\,\,\,\,
 +\,i\, \calH_{sjr} 
 \,\bigg ( (D+F)^{2} + (D-F) (D+3F) q_{1}\bigg )
 - 2\,i\,\calH_{ss} \bigg (D^{2}-F^{2}+(D^{2}+F^{2})\,q_{1} \bigg )\nonumber\\
& &\,\,\,\,\,\,\,\,\,\,\,\,\,\,\,\,\,\,\,\,\,\,\,\,\,\,\,\,\,\,\,\,\,\,
 - 2\,i\,\calH_{us} 
 \bigg (  D^{2} - 4DF + F^{2} + \left (D^{2}-2DF - F^{2}\right ) q_{1}\bigg )
 \nonumber\\
& &\,\,\,\,\,\,\,\,\,\,\,\,\,\,\,\,\,\,\,\,\,\,\,\,\,\,\,\,\,\,\,\,\,\,
-i(1+q_2)(D-F)^2\left(2\,\hat\calH_{ujr}+\hat\calH_{sjr}\right)
 +i\, \calH_{ujr} 
 \,\bigg ( (D-F)^{2}\,(1+q_{1})\bigg )
 \nonumber\\
& &\,\,\,\,\,\,\,\,\,\,\,\,\,\,\,\,\,\,\,\,\,\,\,\,\,\,\,\,\,\,\,\,\,\,
 +
  \bigg (
   4 R_{ss}\,F^{2} -4 R_{us}\,F\,(D-F) 
   +R_{uu}\,(D-F)^{2}
  \bigg ) (1-q_{1})
 \bigg ]
\bigg \} \,.
\label{eq:isovec_unpol_xi_su63}
\end{eqnarray}
\begin{eqnarray}
 \la \Sigma^{0} | \op^{(3)}_{\mu_{0}\cdots\mu_{n}} | 
 \Lambda^{0}\ra &=& 
   -\frac{ \overline{U}_{\Sigma^0} v_{\mu_{0}}\ldots v_{\mu_{n}} U_{\Lambda^0}\left (
    -1+q_{1}
   \right )}{4\sqrt{3}}\left (\alpha^{(a)}_{n}-2 \beta^{(a)}_{n}\right )
   \times
 \left [ 1 +
   \frac{1}{2}\left (\calW^{(\Lambda)}_{SU(6|3)}  +\calW^{(\Sigma)}_{SU(6|3)}
   \right )(1-\delta_{n0})\right ]\nonumber\\
& & + \frac{\overline{U}_{\Sigma^0} v_{\mu_{0}}\ldots v_{\mu_{n}} U_{\Lambda^0}(-1+q_{1})(1-\delta_{n0})}{\sqrt{3}f^{2}}
\bigg \{
 -\frac{i}{6}
   \left (\gamma^{(a)}_{n} - \frac{\sigma^{(a)}_{n}}{3} \right )
\bigg (
   2 \calH_{uu}^\Delta + \calH_{us}^\Delta + \calH_{ujr}^\Delta
  \bigg )
  C^{2}\nonumber\\
& &\,\,\,\,\,\,\,\,\,\,\,\,\,\,\,\,\,\,\,\,\,
+\alpha^{(a)}_{n}
 \bigg [ \frac{1}{4} \calI_{ujr} 
 +\frac{i}{4}  \calH_{ujr} 
 \,(D-3F)(3D-F)
 -\frac{i}{4}  \calH_{sjr} 
 \,(D-F)(D+3F)
\nonumber\\
& &\,\,\,\,\,\,\,\,\,\,\,\,\,\,\,\,\,\,\,\,\,\,\,\,\,\,\,\,\,\,\,\,\,\,
 - i \calH_{uu} D (D+F)
 - i \calH_{us} D (D-F)
 \nonumber\\
& &\,\,\,\,\,\,\,\,\,\,\,\,\,\,\,\,\,\,\,\,\,\,\,\,\,\,\,\,\,\,\,\,\,\,
 -\frac{1}{4}
  \bigg (
   R_{ss}\,(D-F)\,(D+3F) 
   -4 R_{us}\,(D^{2}-2DF + 3F^{2}) - 4 R_{uu}\,F\,(-2D+3F)
  \bigg )
 \bigg ]\nonumber\\
& &\,\,\,\,\,\,\,\,\,\,\,\,\,\,\,\,\,\,\,\,\,
+\beta^{(a)}_{n}
 \bigg [
-\frac{1}{2} \calI_{ujr} 
 +\frac{i}{2}  \calH_{ujr} 
 \,(D-3F)(D+F)
 +\frac{i}{2}  \calH_{sjr} 
 \,(D-F)(D+3F)\nonumber\\
& &\,\,\,\,\,\,\,\,\,\,\,\,\,\,\,\,\,\,\,\,\,\,\,\,\,\,\,\,\,\,\,\,\,\,
 - 2 i \calH_{uu} D (D - F)
 - 2 i \calH_{us} D F
 \nonumber\\
& &\,\,\,\,\,\,\,\,\,\,\,\,\,\,\,\,\,\,\,\,\,\,\,\,\,\,\,\,\,\,\,\,\,\,
 +\frac{1}{2}
  \bigg (
   R_{ss}\,(D-F)\,(D+3F) 
   -4 R_{us}\,(D^{2}-2DF + 3F^{2})-4 R_{uu}\,F\,(-2D+3F)
  \bigg )
 \bigg ]
\bigg \} \,.\nonumber \\
\label{eq:isovec_unpol_lambda_sigma_su63}
\end{eqnarray}

\subsection{Isovector helicity matrix elements}

\begin{eqnarray}
\label{eq:isovec_heli_proton_su63}
 \la p | \tilde{\op}^{(3)}_{\mu_{0}\cdots\mu_{n}} | p \ra &=&
 \frac{\overline{U}_p v_{\{\mu_{0}}\ldots v_{\mu_{n-1}} S_{\mu_{n}\}}U_p\big ( (5 + q_{1})
   \Delta\alpha^{(a)}_{n}
  + 2 (1 + 2 q_{1} ) \Delta\beta^{(a)}_{n} \big )}{6}
 \times \left (1 + \calW^{(p)}_{SU(6|3)} 
  \right )
\\
 & & + \frac{\overline{U}_p v_{\{\mu_{0}}\ldots v_{\mu_{n-1}} S_{\mu_{n}\}}U_p}{f^{2}}
\bigg \{
  \frac{5 i}{81}\,
  \left( \,{{\Delta\gamma^{(a)}_{n}}} 
 - \frac{{\Delta\sigma^{(a)}_{n}}}{5} \right)
\bigg[
\bigg ( 3\,\calH_{uu}^\Delta + 
    \calH_{ujr}^\Delta \bigg ) \,
  \left( 5 + {q_1} \right) 
+3(1+q_2)\hat\calH^\Delta_{ujr}
\bigg]
 C^{2}\nonumber\\
& &\,\,\,\,\,\,\,\,\,\,\,\,\,\,\,\,\,\,\,\,\,\,\,\,\,\,
 +\frac{8i}{9}\sqrt{\frac{2}{3}}
 \bigg [
  \calK_{uu} (D+3F) (-1+q_{1})   -  \calK_{ujr} 
    (3D - F - 2 F q_{1})
+3(1+q_2)(D-F)\hat\calK_{ujr}
 \bigg ] C \Delta c_{n}^{(a)} 
\nonumber\\
& &\,\,\,\,\,\,\,\,\,\,\,\,\,\,\,
+\Delta\alpha^{(a)}_{n}
 \bigg [-\frac{1}{6}(5 + q_{1}) 
   \, \calI_{ujr} +(1+q_2)\hat\calI_{ujr}
  +\frac{i}{3}\, \calH_{ujr} 
  \bigg ( 3\,D^2 - 6\,D\,F + 5\,F^2 + 
  \left( D^2 + F^2 \right) \,{q_1}
   \bigg )\nonumber\\
& &\,\,\,\,\,\,\,\,\,\,\,\,\,\,\,\,\,\,\,\,\,\,\,\,\,\,\,\,
  - \frac{4i}{3}\, \calH_{uu}\,D\,(D-2F-F q_{1}) 
 +\frac{i}{3}(1+q_2)(D^2+3F^2)\hat\calH_{ujr}
  + \frac{1}{6} R_{uu} (D-3F)^{2} (5+q_{1})
 \bigg ]\nonumber\\
& &\,\,\,\,\,\,\,\,\,\,\,\,\,\,\,
+\Delta\beta^{(a)}_{n}
 \bigg [-\frac{1}{3}(1 + 2 q_{1}) 
   \, \calI_{ujr}  +(1+q_2)\hat\calI_{ujr}
  +\frac{2i}{3}\, \calH_{ujr} 
  \bigg ( D^2 + F^2 (1+2{q_{1}})
   \bigg )\nonumber\\
& &\,\,\,\,\,\,\,\,\,\,\,\,\,\,\,\,\,\,\,\,\,\,\,\,\,\,\,\,
  -\frac{4i}{3}\, \calH_{uu}\,D\,(D-2F-F q_{1}) 
 +i(1+q_2)(D-F)^2\hat\calH_{ujr}
  +\frac{1}{3}  R_{uu} (D-3F)^{2} (1+2q_{1})
 \bigg ]
\bigg \} \,. 
\nonumber
\end{eqnarray}
\begin{eqnarray}
 \la \Lambda^{0} | \tilde{\op}^{(3)}_{\mu_{0}\cdots\mu_{n}} | 
 \Lambda^{0}\ra &=& 
   -\frac{\overline{U}_{\Lambda^{0}} v_{\{\mu_{0}}\ldots v_{\mu_{n-1}}
     S_{\mu_{n}\}} U_{\Lambda^{0}} (1+q_{1})}{4} 
    \left (\Delta\alpha^{(a)}_{n} -2 \Delta\beta^{(a)}_{n} \right )\times
 \left ( 1 + \calW^{(\Lambda)}_{SU(6|3)}\right )\nonumber\\
& & + \frac{\overline{U}_{\Lambda^{0}} v_{\{\mu_{0}}\ldots
  v_{\mu_{n-1}} S_{\mu_{n}\}} U_{\Lambda^{0}} (1+q_{1})}{f^{2}}
\bigg \{
  -\frac{5i}{54}\,
  \left( \,{{\Delta\gamma^{(a)}_{n}}} 
 - \frac{{\Delta\sigma^{(a)}_{n}}}{5} \right)\bigg ( 3\,\calH_{us}^\Delta + 
    \calH_{ujr}^\Delta 
-2\tilde{q}\hat\calH^\Delta_{ujr}\bigg ) 
 C^{2}
\nonumber\\
& &\,\,\,\,\,\,\,\,\,\,\,\,\,\,\,\,\,\,\,\,\,\,\,\,\,\,
 +\frac{4i}{3}\sqrt{\frac{2}{3}}
 \bigg [ -\calK_{us} (D-3F)
  +2\calK_{uu}\,D    +  \calK_{ujr} 
    (D+F)
+2\tilde{q}(D-F)\hat\calK_{ujr} \bigg ] C \Delta c_{n}^{(a)} 
\nonumber\\
& &\,\,\,\,\,\,\,\,\,\,\,\,\,\,\,
+\Delta\alpha^{(a)}_{n}
 \bigg [ \frac{1}{4} \bigg (-\,\calI_{ujr} + 2\,\calI_{sjr}
+2\tilde{q}\left(\hat\calI_{ujr} + \hat\calI_{sjr}\right) 
\bigg )
 -\frac{i}{36}  \calH_{ujr} 
 \,(7D - 9 F)\, (D-3F)
\nonumber\\
& &\,\,\,\,\,\,\,\,\,\,\,\,\,\,\,\,\,\,\,\,\,\,\,\,\,\,\,\,\,\,\,\,\,\,
 +\frac{i}{36}  \calH_{sjr} 
 \,(D+3F)^{2} +\frac{i}{18} \calH_{uu} \left ( 7 D^{2} - 12 DF + 9 F^{2}\right )
 +\frac{i}{18} \calH_{us} (5D - 3F) (D+3F)
 \nonumber\\
& &\,\,\,\,\,\,\,\,\,\,\,\,\,\,\,\,\,\,\,\,\,\,\,\,\,\,\,\,\,\,\,\,\,\,
+\frac{i}{18}\tilde{q}(5D^2-6DF+9F^2)\hat\calH_{ujr} 
+\frac{i}{18}\tilde{q}(D+3F)^2\hat\calH_{sjr} 
\nonumber\\
& &\,\,\,\,\,\,\,\,\,\,\,\,\,\,\,\,\,\,\,\,\,\,\,\,\,\,\,\,\,\,\,\,\,\,
 -\frac{1}{36}
  \bigg (
   R_{ss}\,(D+3F)^{2} -4 R_{us}\,(2D-3F)(D+3F) 
   +4 R_{uu}\, (2D-3F)^{2}
  \bigg )
 \bigg ]\nonumber\\
& &\,\,\,\,\,\,\,\,\,\,\,\,\,\,\,\,\,\,\,\,\,
+\Delta\beta^{(a)}_{n}
 \bigg [-\frac{1}{2}  \calI_{ujr} +\tilde{q}\hat\calI_{ujr} 
 -\frac{i}{18}  \calH_{ujr} 
 \,(7D^{2} + 6DF - 9F^{2})
 +\frac{i}{18}  \calH_{sjr} 
 \,(D+3F)^{2}\nonumber\\
& &\,\,\,\,\,\,\,\,\,\,\,\,\,\,\,\,\,\,\,\,\,\,\,\,\,\,\,\,\,\,\,\,\,\,
 -\frac{i}{9} \calH_{uu} \left ( 7 D^{2} - 12 DF + 9 F^{2}\right )
 +\frac{i}{9} \calH_{us} (13D^{2} -12 DF + 9F^{2})
+i\tilde{q}(D-F)^2\hat\calH_{ujr}
 \nonumber\\
& &\,\,\,\,\,\,\,\,\,\,\,\,\,\,\,\,\,\,\,\,\,\,\,\,\,\,\,\,\,\,\,\,\,\,
 +\frac{1}{18}
  \bigg (
   R_{ss}\,(D+3F)^{2} -4 R_{us}\,(2D-3F)(D+3F) 
   +4 R_{uu}\, (2D-3F)^{2}
  \bigg )
 \bigg ]
\bigg \} \,. \nonumber\\
\label{eq:isovec_heli_lambda_su63}&&
\end{eqnarray}
\begin{eqnarray}
 \la \Sigma^{+} | \tilde{\op}^{(3)}_{\mu_{0}\cdots\mu_{n}} | 
 \Sigma^{+}\ra &=& 
   -\frac{\overline{U}_{\Sigma^{+}} v_{\{\mu_{0}}\ldots v_{\mu_{n-1}}
     S_{\mu_{n}\}} U_{\Sigma^{+}} \left (
    (-4+q_{1})\Delta\alpha^{(a)}_{n} + 2 (1 + 2 q_{1}) \Delta\beta^{(a)}_{n}
   \right )}{6}
   \times
 \left ( 1 + \calW^{(\Sigma)}_{SU(6|3)}\right )\nonumber\\
& & + \frac{\overline{U}_{\Sigma^{+}} v_{\{\mu_{0}}\ldots
  v_{\mu_{n-1}} S_{\mu_{n}\}} U_{\Sigma^{+}} }{f^{2}}
\bigg \{
\frac{5i}{81}
   \left (\Delta\gamma^{(a)}_{n} - \frac{\Delta\sigma^{(a)}_{n}}{5} \right )
\bigg (
   \calH_{uu}^\Delta + 11\,\calH_{us}^\Delta + 4\,\calH_{sjr}^\Delta\nonumber\\
& &\,\,\,\,\,\,\,\,\,\,\,\,\,\,\,\,\,\,\,\,\,\,\,\,\,\,
   \,\,\,\,\,\,\,\,\,\,\,\,\,\,\,\,\,\,\,\,\,\,\,\,\,\,\,\,
  -   \big [ \calH_{uu}^\Delta +2\,\calH_{us}^\Delta 
       + \calH_{ujr}^\Delta \big ] \,  {q_1} 
+(1+q_2)\left(\hat\calH^\Delta_{ujr}+2\,\hat\calH^\Delta_{sjr}\right)
\nonumber\\
& &\,\,\,\,\,\,\,\,\,\,\,\,\,\,\,\,\,\,\,\,\,\,\,\,\,\,
   \,\,\,\,\,\,\,\,\,\,\,\,\,\,\,\,\,\,\,\,\,\,\,\,\,\,\,\,
-  2\,i\,
   \left( T_{ss} - 2\,T_{us} + T_{uu} \right)(1-q_{1})
  \bigg )
  C^{2}
\nonumber\\
& &\,\,\,\,\,\,\,\,\,\,\,\,\,\,\,\,\,\,\,\,\,\,\,\,\,\,
 -\frac{8i}{9}\sqrt{\frac{2}{3}}\,
 \bigg [ \calK_{ujr} 
 (D+3F+2F q_{1}) 
+ 2  \calK_{sjr}   (D-F)
 + 2 \calK_{uu}\,F\,(2 + q_{1}) 
\nonumber\\
& &\,\,\,\,\,\,\,\,\,\,\,\,\,\,\,\,\,\,\,\,\,\,\,\,\,\,\,\,\,\,\,\, 
\,\,\,\,\,\,\,\,\,\,\,\,\,\,\,\,\,\,\,\,\,\,\,\,\,\,\,\,\,\,\,\,
  + \calK_{us} (D+F) (2+q_{1})
+(1+q_2)(F-D)\left(\hat\calK_{ujr}+2\,\hat\calK_{sjr}\right)
\nonumber\\
& &\,\,\,\,\,\,\,\,\,\,\,\,\,\,\,\,\,\,\,\,\,\,\,\,\,\,\,\,\,\,\,\, 
\,\,\,\,\,\,\,\,\,\,\,\,\,\,\,\,\,\,\,\,\,\,\,\,\,\,\,\,\,\,\,\,
  -2\,i \,
  \bigg ( S_{ss}\,\left( D - F \right)  + 2\,S_{uu}\,F - 
    S_{us}\,\left( D + F \right)  \bigg ) \,
  \left( 2 + {q_1} \right)
 \bigg ] C \Delta c_{n}^{(a)} 
\nonumber\\
& &\,\,\,\,\,\,\,\,\,\,\,\,\,\,\,
+\Delta\alpha^{(a)}_{n}
 \bigg [ \frac{1}{6} \bigg (-5\,\calI_{ujr} 
 +  \calI_{sjr} (1+q_{1}) 
+(1+q_2)\left(5\,\hat\calI_{ujr}+\hat\calI_{sjr}\right)   \bigg )
\nonumber\\
& &\,\,\,\,\,\,\,\,\,\,\,\,\,\,\,\,\,\,\,\,\,\,\,\,\,\,\,\,\,\,\,\,\,\,
 -\frac{i}{6}  \calH_{ujr} 
 \,\bigg ( (D-F)(D+3F)+2\,(D^{2}+F^{2})\,q_{1}\bigg )
\nonumber\\
& &\,\,\,\,\,\,\,\,\,\,\,\,\,\,\,\,\,\,\,\,\,\,\,\,\,\,\,\,\,\,\,\,\,\,
 +\frac{5i}{6}  \calH_{sjr} 
 \,(D-F)^{2} +\frac{i}{3} 
 \calH_{uu} \bigg (4F^{2} + (D-F)(D+F)\,q_{1} \bigg )\nonumber\\
& &\,\,\,\,\,\,\,\,\,\,\,\,\,\,\,\,\,\,\,\,\,\,\,\,\,\,\,\,\,\,\,\,\,\,
 - \frac{i}{3} 
 \calH_{us} \bigg ( 4 (D^{2}-DF+F^{2}) + (D^{2}+4DF - F^{2})\,q_{1}\bigg )
\nonumber\\
& &\,\,\,\,\,\,\,\,\,\,\,\,\,\,\,\,\,\,\,\,\,\,\,\,\,\,\,\,\,\,\,\,\,\,
+\frac{i}{6}(1+q_2)(D^2+2DF+5F^2)\hat\calH_{ujr}
+\frac{i}{6}(1+q_2)(D-F)^2\hat\calH_{sjr}
 \nonumber\\
& &\,\,\,\,\,\,\,\,\,\,\,\,\,\,\,\,\,\,\,\,\,\,\,\,\,\,\,\,\,\,\,\,\,\,
 -\frac{1}{6}
  \bigg (
   R_{ss}\,(D-F)^{2} -4 R_{us}\,F\,(D-F) 
   +4 R_{uu}\,F^{2}
  \bigg ) (q_{1}-4)
 \bigg ]\nonumber\\
& &\,\,\,\,\,\,\,\,\,\,\,\,\,\,\,\,\,\,\,\,\,
+\Delta\beta^{(a)}_{n}
 \bigg [
\frac{1}{3} \bigg (-\,\calI_{ujr} 
 + 2\,\calI_{sjr}  (1+q_{1}) 
+(1+q_2)\left(\hat\calI_{ujr}+2\,\hat\calI_{sjr}\right)   \bigg )
   \bigg )
\nonumber\\
& &\,\,\,\,\,\,\,\,\,\,\,\,\,\,\,\,\,\,\,\,\,\,\,\,\,\,\,\,\,\,\,\,\,\,
 + \frac{i}{3}  \calH_{ujr} 
 \,\bigg ( (D-F)(D+3F) - 4 F^{2}\,q_{1}\bigg )\nonumber\\
& &\,\,\,\,\,\,\,\,\,\,\,\,\,\,\,\,\,\,\,\,\,\,\,\,\,\,\,\,\,\,\,\,\,\,
 +\frac{i}{3}  \calH_{sjr} 
 \,(D-F)^{2}
 -\frac{2i}{3}\,
 \calH_{uu} \bigg (D^{2} + F^{2} + 2 F^{2}\,q_{1} \bigg )\nonumber\\
& &\,\,\,\,\,\,\,\,\,\,\,\,\,\,\,\,\,\,\,\,\,\,\,\,\,\,\,\,\,\,\,\,\,\,
 -\frac{2i}{3}\,
 \calH_{us} \bigg ( D^{2}-2DF-F^{2} + 2(D-F)F\,q_{1}\bigg )
+\frac{i}{3}(1+q_2)(D-F)^2\left(\hat\calH_{ujr}+2\,\hat\calH_{sjr}\right)
 \nonumber\\
& &\,\,\,\,\,\,\,\,\,\,\,\,\,\,\,\,\,\,\,\,\,\,\,\,\,\,\,\,\,\,\,\,\,\,
 -\frac{1}{3}
  \bigg (
   R_{ss}\,(D-F)^{2} -4 R_{us}\,F\,(D-F) 
   +4 R_{uu}\,F^{2}
  \bigg ) (1+2q_{1})
 \bigg ]
\bigg \} \,.
\label{eq:isovec_heli_sigma_su63}
\end{eqnarray}
\begin{eqnarray}
 \la \Xi^{-} | \tilde{\op}^{(3)}_{\mu_{0}\cdots\mu_{n}} | 
 \Xi^{-}\ra &=& 
   -\frac{\overline{U}_{\Xi^{-}} v_{\{\mu_{0}}\ldots v_{\mu_{n-1}}
     S_{\mu_{n}\}} U_{\Xi^{-}} \left (
    (5+4q_{1})\Delta\alpha^{(a)}_{n} - 2 (-1 + q_{1}) \Delta\beta^{(a)}_{n}
   \right )}{6}
   \times
 \left ( 1 + \calW^{(\Xi)}_{SU(6|3)}\right )\nonumber\\
& & + \frac{\overline{U}_{\Xi^{-}} v_{\{\mu_{0}}\ldots v_{\mu_{n-1}}
  S_{\mu_{n}\}} U_{\Xi^{-}} }{f^{2}}
\bigg \{
 -\frac{5i}{81}
   \left (\Delta\gamma^{(a)}_{n} - \frac{\Delta\sigma^{(a)}_{n}}{5} \right )
 \bigg (
  \calH_{sjr}^\Delta + 
  \calH_{ss}^\Delta\,\left( 2 + {q_1} \right)  + 
  \calH_{us}^\Delta\,\left( 13 + 11\,{q_1} \right)\nonumber\\
& &\,\,\,\,\,\,\,\,\,\,\,\,\,\,\,\,\,\,\,\,\,\,\,\,\,\,\,\,
 +   4\,\calH_{ujr}^\Delta \big ] \left( 1 + {q_1} \right)
-(1+q_2)\left(2\,\hat\calH^\Delta_{ujr}+\hat\calH^\Delta_{sjr}\right)
 - 2i\left( T_{ss} - 2\,T_{us} + T_{uu} \right) 
      \left( 2 + {q_1} \right)
 \bigg )
  C^{2}
\nonumber\\
& &\,\,\,\,\,\,\,\,\,\,\,\,\,\,\,\,\,\,\,\,\,\,\,\,\,\,
 +\frac{8i}{9}\sqrt{\frac{2}{3}}\,
 \bigg [ 2  \calK_{ujr}  
 (D-F) (1+q_{1})
+  \calK_{sjr}  
  ( D+F+(D+3F) q_{1} )
 + 2 \calK_{ss} F (1 + 2 q_{1})
\nonumber\\
& &\,\,\,\,\,\,\,\,\,\,\,\,\,\,\,\,\,\,\,\,\,\,\,\,\,\,\,\,\,\,\,\, 
\,\,\,\,\,\,\,\,\,\,\,\,\,\,\,\,\,\,\,\,\,
  + \calK_{us} (D+F) (1+ 2 q_{1})
+(1+q_2)(D-F)\left(2\,\hat\calK_{ujr}+\hat\calK_{sjr}\right)
\nonumber\\
& &\,\,\,\,\,\,\,\,\,\,\,\,\,\,\,\,\,\,\,\,\,\,\,\,\,\,\,\,\,\,\,\, 
\,\,\,\,\,\,\,\,\,\,\,\,\,\,\,\,\,\,\,\,\,
  -2 i \,
  \bigg ( S_{uu}\,\left( D - F \right)  + 2\,S_{ss}\,F - 
    S_{us}\,\left( D + F \right)  \bigg ) \,
  \left( 1 + 2\,{q_1} \right)
 \bigg ] C \Delta c_{n}^{(a)} 
\nonumber\\
& &\,\,\,\,\,\,\,\,\,\,\,\,\,\,\,
+\Delta\alpha^{(a)}_{n}
 \bigg [ \frac{1}{6} \bigg (- \calI_{ujr}  q_{1}
 + 5\,\calI_{sjr}  (1+q_{1}) 
+(1+q_2)\left(\hat\calI_{ujr}+5\,\hat\calI_{sjr}\right)    \bigg )
\nonumber\\
& &\,\,\,\,\,\,\,\,\,\,\,\,\,\,\,\,\,\,\,\,\,\,\,\,\,\,\,\,\,\,\,\,\,\,
 -\frac{5 i}{6}  \calH_{ujr} 
 \, (D-F)^{2}\,(1+q_{1}) -\frac{i}{6}  \calH_{sjr} 
\bigg ( D^{2} - 2DF + 5 F^{2} - (D-F) (D+3F) q_{1}\bigg )
\nonumber\\
& &\,\,\,\,\,\,\,\,\,\,\,\,\,\,\,\,\,\,\,\,\,\,\,\,\,\,
 + \frac{i}{3} 
 \calH_{ss} \bigg (D^{2}-5\,F^{2}-4\,F^{2}\,q_{1} \bigg )
 + \frac{i}{3} \calH_{us} 
 \bigg ( (3D-5F)(D-F) + 4 \left (D^{2}-DF + F^{2}\right ) q_{1}\bigg )
\nonumber\\
& &\,\,\,\,\,\,\,\,\,\,\,\,\,\,\,\,\,\,\,\,\,\,\,\,\,\,\,\,\,\,\,\,\,\,
+\frac{i}{6}(1+q_2)(D-F)^2\hat\calH_{ujr}
+\frac{i}{6}(1+q_2)(D^2+2DF+5F^2)\hat\calH_{sjr}
 \nonumber\\
& &\,\,\,\,\,\,\,\,\,\,\,\,\,\,\,\,\,\,\,\,\,\,\,\,\,\,\,\,\,\,\,\,\,\,
 -\frac{1}{6}
  \bigg (
   4 R_{ss}\,F^{2} -4 R_{us}\,F\,(D-F) 
   +R_{uu}\,(D-F)^{2}
  \bigg ) (5+4\,q_{1})
 \bigg ]\nonumber\\
& &\,\,\,\,\,\,\,\,\,\,\,\,\,\,\,\,\,\,\,\,\,
+\Delta\beta^{(a)}_{n}
 \bigg [ \frac{1}{3} \bigg (-2  \calI_{ujr}  q_{1}
 +  \calI_{sjr} (1+q_{1}) 
+(1+q_2)\left(2\,\hat\calI_{ujr}+\hat\calI_{sjr}\right)
   \bigg )\nonumber\\
& &\,\,\,\,\,\,\,\,\,\,\,\,\,\,\,\,\,\,\,\,\,\,\,\,\,\,\,\,\,\,\,\,\,\,
 -\frac{i}{3}\, \calH_{ujr} 
 \, (D-F)^{2}\,(1+q_{1}) -\frac{i}{3}\, \calH_{sjr} 
 \,\bigg ( (D+F)^{2} + (D-F) (D+3F) q_{1}\bigg )
\nonumber\\
& &\,\,\,\,\,\,\,\,\,\,\,\,\,\,\,\,\,\,\,\,\,\,\,\,\,\,\,\,\,\,\,\,\,\,
 +\frac{2i}{3}\,
  \calH_{ss} \bigg (D^{2}-F^{2}+(D^{2}+F^{2})\,q_{1} \bigg )
+\frac{i}{3}(1+q_2)(D-F)^2\left(2\,\hat\calH_{ujr}+\hat\calH_{sjr}\right)
 \nonumber\\
& &\,\,\,\,\,\,\,\,\,\,\,\,\,\,\,\,\,\,\,\,\,\,\,\,\,\,\,\,\,\,\,\,\,\,
 +\frac{2i}{3}\,\calH_{us} 
 \bigg (  D^{2} - 4DF + F^{2} + \left (D^{2}-2DF - F^{2}\right ) q_{1}\bigg )
 \nonumber\\
& &\,\,\,\,\,\,\,\,\,\,\,\,\,\,\,\,\,\,\,\,\,\,\,\,\,\,\,\,\,\,\,\,\,\,
 -\frac{1}{3}
  \bigg (
   4 R_{ss}\,F^{2} -4 R_{us}\,F\,(D-F) 
   +R_{uu}\,(D-F)^{2}
  \bigg ) (1-q_{1})
 \bigg ]
\bigg \} \,. 
\label{eq:isovec_heli_xi_su63}
\end{eqnarray}
\begin{eqnarray}
 \la \Sigma^{0} | \tilde{\op}^{(3)}_{\mu_{0}\cdots\mu_{n}} | 
 \Lambda^{0}\ra &=& 
   -\frac{\overline{U}_{\Sigma^{0}} v_{\{\mu_{0}}\ldots v_{\mu_{n-1}}
     S_{\mu_{n}\}} U_{\Lambda^{0}} \left (
    -1+q_{1}
   \right )}{4\sqrt{3}}\left (\Delta\alpha^{(a)}_{n}
  -2 \Delta\beta^{(a)}_{n}\right )
   \times
 \left [ 1 +
   \frac{1}{2}\left (\calW^{(\Lambda)}_{SU(6|3)}  +\calW^{(\Sigma)}_{SU(6|3)}
   \right )\right ]\nonumber\\
& & + \frac{\overline{U}_{\Sigma^{0}} v_{\{\mu_{0}}\ldots
  v_{\mu_{n-1}} S_{\mu_{n}\}} U_{\Lambda^{0}} (-1+q_{1})}{\sqrt{3}f^{2}}
\bigg \{
 -\frac{5i}{54}
   \left (\Delta\gamma^{(a)}_{n} - 
 \frac{\Delta\sigma^{(a)}_{n}}{5} \right )\bigg (
   2 \calH_{uu}^\Delta + \calH_{us}^\Delta + \calH_{ujr}^\Delta
  \bigg )
  C^{2}
\nonumber\\
& &\,\,\,\,\,\,\,\,\,\,\,\,\,\,\,\,\,\,\,\,\,\,\,\,\,\,
 +\frac{4i}{9}\sqrt{\frac{2}{3}}
 \bigg [ 2\calK_{us} (2D+3F)
  - \calK_{uu}\,(D-3F) 
   +2  \calK_{ujr} 
    \,D   + \calK_{sjr}
    \,(D+3F)
  \nonumber\\
& &\,\,\,\,\,\,\,\,\,\,\,\,\,\,\,\,\,\,\,\,\,\,\,\,\,\,\,\,\,\,\,\, 
\,\,\,\,\,\,\,\,\,\,\,\,\,\,\,\,\,\,\,\,\,\,
   -i \,\bigg ( 2\,S_{uu}\,\left( 2\,D - 3\,F \right)  + 
    S_{us}\,\left( -5\,D + 3\,F \right)  + 
    S_{ss}\,\left( D + 3\,F \right)  \bigg )
 \bigg ] C \Delta c_{n}^{(a)}
\nonumber\\
& &\,\,\,\,\,\,\,\,\,\,\,\,\,\,\,
+\Delta\alpha^{(a)}_{n}
 \bigg [ \frac{1}{4} \calI_{ujr}
 -\frac{i}{12}  \calH_{ujr} 
 \,(D-3F)(3D-F)
 +\frac{i}{12}  \calH_{sjr} 
 \,(D-F)(D+3F)\nonumber\\
& &\,\,\,\,\,\,\,\,\,\,\,\,\,\,\,\,\,\,\,\,\,\,\,\,\,\,\,\,\,\,\,\,\,\,
 +\frac{i}{3} \calH_{uu} D (D+F)
 +\frac{i}{3} \calH_{us} D (D-F)
 +\frac{1}{12}
  \bigg (
   R_{ss}\,(D-F)\,(D+3F) \nonumber\\
& &\,\,\,\,\,\,\,\,\,\,\,\,\,\,\,\,\,\,\,\,\,\,\,\,\,\,\,\,\,\,\,\,\,\,
 \,\,\,\,\,\,\,\,\,\,\,\,\,\,\,\,\,\,\,\,\,\,\,\,\,\,\,\,\,\,\,\,\,
   -4 R_{us}\,(D^{2}-2DF + 3F^{2}) 
   -4 R_{uu}\,F\,(-2D+3F)
  \bigg )
 \bigg ]\nonumber\\
& &\,\,\,\,\,\,\,\,\,\,\,\,\,\,\,\,\,\,\,\,\,
+\Delta\beta^{(a)}_{n}
 \bigg [
-\frac{1}{2} \calI_{ujr} 
 -\frac{i}{6}  \calH_{ujr} 
 \,(D-3F)(D+F)
 -\frac{i}{6}  \calH_{sjr} 
 \,(D-F)(D+3F)\nonumber\\
& &\,\,\,\,\,\,\,\,\,\,\,\,\,\,\,\,\,\,\,\,\,\,\,\,\,\,\,\,\,\,\,\,\,\,
 +\frac{2i}{3} \calH_{uu} D (D - F)
 +\frac{2i}{3} \calH_{us} D F
 -\frac{1}{6}
  \bigg (
   R_{ss}\,(D-F)\,(D+3F) \nonumber\\
& &\,\,\,\,\,\,\,\,\,\,\,\,\,\,\,\,\,\,\,\,\,\,\,\,\,\,\,\,\,\,\,\,\,\,
\,\,\,\,\,\,\,\,\,\,\,\,\,\,\,\,\,\,\,\,\,\,\,\,\,\,\,\,\,\,\,\,\,\,
   -4 R_{us}\,(D^{2}-2DF + 3F^{2}) 
   -4 R_{uu}\,F\,(-2D+3F)
  \bigg )
 \bigg ]
\bigg \} \,. 
\label{eq:isovec_heli_lambda_sigma_su63}
\end{eqnarray}

\subsection{Isoscalar unpolarised matrix elements}

\begin{eqnarray}
\label{eq:isoscal_unpol_proton_su63}
 \la p | \op^{(0)}_{\mu_{0}\cdots\mu_{n}} | p \ra &=&
 \overline{U}_p v_{\mu_{0}}\ldots v_{\mu_{n}}U_p \big (
   \alpha^{(s)}_{n}
  + \beta^{(s)}_{n} \big )
 \times \left (1 + (1-\delta_{n0})\calW^{(p)}_{SU(6|3)} 
  \right )
\\
&& + \frac{\overline{U}_p v_{\mu_{0}}\ldots v_{\mu_{n}}U_p(1-\delta_{n0}) }{f^{2}}
\bigg \{
  i\,
  \left( \,{{\gamma^{(s)}_{n}}} 
 - \frac{{\sigma^{(s)}_{n}}}{3} \right)
 \bigg ( 2\,\calH_{uu}^\Delta + 
    \calH_{ujr}^\Delta \bigg ) \,
  C^{2}
\nonumber\\
& &\,\,\,\,\,\,\,\,\,\,\,\,\,\,\,
+\left ( \alpha^{(s)}_{n} + \beta^{(s)}_{n} \right )
 \bigg [
  -i\, \calH_{ujr} 
  ( 5\,D^2 - 6\,D\,F + 9\,F^2 )
  + 4\,i\, \calH_{uu}\,D\,(D-3F)
  -3  R_{uu} (D-3F)^{2}
 \bigg ]
\bigg \} \,. 
\nonumber
\end{eqnarray}
\begin{eqnarray}
 \la \Lambda^{0} | \op^{(0)}_{\mu_{0}\cdots\mu_{n}} | 
 \Lambda^{0}\ra &=& 
    \overline{U}_{\Lambda^{0}} v_{\mu_{0}}\ldots v_{\mu_{n}} U_{\Lambda^{0}} 
    \left (\alpha^{(s)}_{n} + \beta^{(s)}_{n} \right )\times
 \left ( 1 + (1-\delta_{n0})\,\calW^{(\Lambda)}_{SU(6|3)}\right )\nonumber\\
& & + \frac{\overline{U}_{\Lambda^{0}} v_{\mu_{0}}\ldots v_{\mu_{n}}
  U_{\Lambda^{0}} (1-\delta_{n0})}{f^{2}} 
\bigg \{
  i\,
  \left( \,{{\gamma^{(s)}_{n}}} - \frac{{\sigma^{(s)}_{n}}}{3} \right)
\bigg ( \calH_{us}^\Delta + \calH_{uu}^\Delta 
    + \calH_{ujr}^\Delta \bigg ) 
  C^{2}
\nonumber\\
& &\,\,\,\,\,\,\,\,\,\,\,\,\,\,\,
+\left ( \alpha^{(s)}_{n} + \beta^{(s)}_{n} \right )
 \bigg [ 
 -\frac{2i}{3}  \calH_{ujr} 
 \,(7D^{2} - 12 D F + 9 F^{2})
 -\frac{i}{3}  \calH_{sjr} 
 \,(D+3F)^{2}\nonumber\\
& &\,\,\,\,\,\,\,\,\,\,\,\,\,\,\,\,\,\,\,\,\,\,\,\,\,\,\,
 +\frac{2i}{3} \calH_{uu} \left ( D^{2} - 12 DF + 9 F^{2}\right )
 +\frac{2i}{3} \calH_{us} (5D^{2} -6DF - 9 F^{2})
 \nonumber\\
& &\,\,\,\,\,\,\,\,\,\,\,\,\,\,\,\,\,\,\,\,\,\,\,\,\,\,\,
 -\frac{1}{3}
  \bigg (
   R_{ss}\,(D+3F)^{2} -4 R_{us}\,(2D-3F)(D+3F) 
   +4 R_{uu}\, (2D-3F)^{2}
  \bigg )
 \bigg ]
\bigg \} \,. 
\label{eq:isoscal_unpol_lambda_su63}
\end{eqnarray}
\begin{eqnarray}
 \la \Sigma^{+} | \op^{(0)}_{\mu_{0}\cdots\mu_{n}} | 
 \Sigma^{+}\ra &=& 
   \overline{U}_{\Sigma^{+}} v_{\mu_{0}}\ldots v_{\mu_{n}} U_{\Sigma^{+}} 
   \left ( \alpha^{(s)}_{n} + \beta^{(s)}_{n} \right )
   \times
 \left ( 1 + (1-\delta_{n0})\,\calW^{(\Sigma)}_{SU(6|3)}\right )\nonumber\\
& & + \frac{\overline{U}_{\Sigma^{+}} v_{\mu_{0}}\ldots v_{\mu_{n}}
  U_{\Sigma^{+}}  (1-\delta_{n0})}{f^{2}} 
\bigg \{
 \frac{i}{3}
   \left (\gamma^{(s)}_{n} - \frac{\sigma^{(s)}_{n}}{3} \right )
  \bigg (
   \calH_{uu}^\Delta + 5\,\calH_{us}^\Delta
 + 2\,\calH_{sjr}^\Delta + \calH_{ujr}^\Delta
\nonumber \\
&&\hspace*{6cm} -   2\,i\,
   \left( T_{ss} - 2\,T_{us} + T_{uu} \right)
  \bigg )
  C^{2}
\nonumber\\
& &\,\,\,\,\,\,\,\,\,\,\,\,\,\,\,
+\left ( \alpha^{(s)}_{n} + \beta^{(s)}_{n} \right )
 \bigg [ 
 -2\,i\, \calH_{ujr} 
 \,(D^{2}+3 F^{2})
 -3\,i\,  \calH_{sjr} 
 \,(D-F)^{2}
 +2\,i\,
 \calH_{uu} (D^{2}-3F^{2})\nonumber\\
& &\,\,\,\,\,\,\,\,\,\,\,\,\,\,\,\,\,\,\,\,\,\,\,\,\,\,\,\,\,\,\,\,\,\,
 +2\,i\, 
 \calH_{us} (D^{2}-6DF + 3F^{2})
 -3\,
  \bigg (
   R_{ss}\,(D-F)^{2} -4 R_{us}\,F\,(D-F) 
   +4 R_{uu}\,F^{2}
  \bigg )
 \bigg ]
\bigg \} \,. \nonumber\\
\label{eq:isoscal_unpol_sigma_su63}&&
\end{eqnarray}
\begin{eqnarray}
 \la \Xi^{-} | \op^{(0)}_{\mu_{0}\cdots\mu_{n}} | 
 \Xi^{-}\ra &=& 
  \overline{U}_{\Xi^{-}} v_{\mu_{0}}\ldots v_{\mu_{n}} U_{\Xi^{-}} 
 \left ( \alpha^{(s)}_{n} + \beta^{(s)}_{n}\right )
   \times
 \left ( 1 + (1-\delta_{n0})\,\calW^{(\Xi)}_{SU(6|3)}\right )\nonumber\\
& & + \frac{\overline{U}_{\Xi^{-}} v_{\mu_{0}}\ldots v_{\mu_{n}} U_{\Xi^{-}} (1-\delta_{n0}) }{f^{2}}
\bigg \{
\frac{i}{3}
   \left (\gamma^{(s)}_{n} - \frac{\sigma^{(s)}_{n}}{3} \right )
\bigg ( 2\, \calH_{ujr}^\Delta
  + \calH_{sjr}^\Delta  
+ 
  \calH_{ss}^\Delta + 
  5\,\calH_{us}^\Delta 
\nonumber \\
&&\hspace*{6cm}  - 2i\left( T_{ss} - 2\,T_{us} + T_{uu} \right) 
 \bigg )
  C^{2}
\nonumber\\
& &\,\,\,\,\,\,\,\,\,\,\,\,\,\,\,
+\left ( \alpha^{(s)}_{n} + \beta^{(s)}_{n} \right )
 \bigg [
 -3\,i\, \calH_{ujr} 
 \,(D-F)^{2}
 -2\,i\, \calH_{sjr} 
 \,(D^{2} + 3F^{2})
 + 2\,i\,
 \calH_{ss} (D^{2} - 3 F^{2})\nonumber\\
& &\,\,\,\,\,\,\,\,\,\,\,\,\,\,\,\,\,\,\,\,\,\,\,\,\,\,\,\,\,\,\,\,\,\,
 +2\,i\, \calH_{us} 
 (D^{2} - 6 D F + 3 F^{2})
 -3
  \bigg (
   4 R_{ss}\,F^{2} -4 R_{us}\,F\,(D-F) 
   +R_{uu}\,(D-F)^{2}
  \bigg ) 
 \bigg ]
\bigg \} \,. \nonumber\\
\label{eq:isoscal_unpol_xi_su63}&&
\end{eqnarray}
\begin{eqnarray}
 \la \Sigma^{0} | \op^{(0)}_{\mu_{0}\cdots\mu_{n}} | 
 \Lambda^{0}\ra &=& 0\,.
\end{eqnarray}

\subsection{Isoscalar helicity matrix elements}

\begin{eqnarray}
 \la p | \tilde{\op}^{(0)}_{\mu_{0}\cdots\mu_{n}} | p \ra &=&
 \overline{U}_p v_{\{\mu_{0}}\ldots v_{\mu_{n-1}}S_{\mu_{n}\}} U_p\big (
   \Delta\alpha^{(s)}_{n}
  + \Delta\beta^{(s)}_{n} \big )
 \times \left (1 + \calW^{(p)}_{SU(6|3)} 
  \right )\nonumber\\
 & & + \frac{\overline{U}_p v_{\{\mu_{0}}\ldots v_{\mu_{n-1}}S_{\mu_{n}\}}U_p}{f^{2}}
\bigg \{
  \frac{5i}{9}\,
  \left( \,{{\Delta\gamma^{(s)}_{n}}} 
 - \frac{{\Delta\sigma^{(s)}_{n}}}{5} \right)
\bigg ( 2\,\calH_{uu}^\Delta + 
    \calH_{ujr}^\Delta \bigg ) \,
 C^{2}
\nonumber\\
& &\,\,\,\,\,\,\,\,\,\,\,\,\,\,\,
+\left ( \Delta\alpha^{(s)}_{n} + \Delta\beta^{(s)}_{n} \right )
 \bigg [
  \frac{i}{3}\, \calH_{ujr} 
  ( 5\,D^2 - 6\,D\,F + 9\,F^2 )
  -\frac{4i}{3} \calH_{uu}\,D\,(D-3F)
  + R_{uu} (D-3F)^{2}
 \bigg ]
\bigg \} \,. 
\nonumber\\\label{eq:isoscal_heli_proton_su63}
\end{eqnarray}
\begin{eqnarray}
 \la \Lambda^{0} | \tilde{\op}^{(0)}_{\mu_{0}\cdots\mu_{n}} | 
 \Lambda^{0}\ra &=& 
    \overline{U}_{\Lambda^{0}} v_{\{\mu_{0}}\ldots
    v_{\mu_{n-1}}S_{\mu_{n}\}} U_{\Lambda^{0}} 
    \left (\Delta\alpha^{(s)}_{n} + \Delta\beta^{(s)}_{n} \right )\times
 \left ( 1 + \calW^{(\Lambda)}_{SU(6|3)}\right )\nonumber\\
& & + \frac{\overline{U}_{\Lambda^{0}} v_{\{\mu_{0}}\ldots
  v_{\mu_{n-1}}S_{\mu_{n}\}} U_{\Lambda^{0}} }{f^{2}}
\bigg \{
  \frac{5i}{9}
  \left( \,{{\Delta\gamma^{(s)}_{n}}} - \frac{{\Delta\sigma^{(s)}_{n}}}{5} \right)
\bigg ( \calH_{us}^\Delta + \calH_{uu}^\Delta 
    + \calH_{ujr}^\Delta \bigg ) 
 C^{2}
\nonumber\\
& &\,\,\,\,\,\,\,\,\,\,\,\,\,\,\,
+\left ( \Delta\alpha^{(s)}_{n} + \Delta\beta^{(s)}_{n} \right )
 \bigg [ 
 \frac{2i}{9}  \calH_{ujr} 
 \,(7D^{2} - 12 D F + 9 F^{2})
 +\frac{i}{9}  \calH_{sjr} 
 \,(D+3F)^{2}\nonumber\\
& &\,\,\,\,\,\,\,\,\,\,\,\,\,\,\,\,\,\,\,\,\,\,\,\,\,\,\,\,\,\,\,\,\,\,
 -\frac{2i}{9} \calH_{uu} \left ( D^{2} - 12 DF + 9 F^{2}\right )
 -\frac{2i}{9} \calH_{us} (5D^{2} -6DF - 9 F^{2})
 \nonumber\\
& &\,\,\,\,\,\,\,\,\,\,\,\,\,\,\,\,\,\,\,\,\,\,\,\,\,\,\,\,\,\,\,\,\,\,
 +\frac{1}{9}
  \bigg (
   R_{ss}\,(D+3F)^{2} -4 R_{us}\,(2D-3F)(D+3F) 
   +4 R_{uu}\, (2D-3F)^{2}
  \bigg )
 \bigg ]
\bigg \} \,. \nonumber\\
\label{eq:isoscal_heli_lambda_su63}&&
\end{eqnarray}
\begin{eqnarray}
 \la \Sigma^{+} | \tilde{\op}^{(0)}_{\mu_{0}\cdots\mu_{n}} | 
 \Sigma^{+}\ra &=& 
   \overline{U}_{\Sigma^{+}} v_{\{\mu_{0}}\ldots
   v_{\mu_{n-1}}S_{\mu_{n}\}} U_{\Sigma^{+}} 
   \left ( \Delta\alpha^{(s)}_{n} + \Delta\beta^{(s)}_{n} \right )
   \times
 \left ( 1 + \calW^{(\Sigma)}_{SU(6|3)}\right )\nonumber\\
& &+ \frac{\overline{U}_{\Sigma^{+}} v_{\{\mu_{0}}\ldots
  v_{\mu_{n-1}}S_{\mu_{n}\}} U_{\Sigma^{+}}  }{f^{2}}
 \bigg \{
 \frac{5i}{27}
   \left (\Delta\gamma^{(s)}_{n} - \frac{\Delta\sigma^{(s)}_{n}}{5} \right )
\nonumber\\
&&\hspace*{3cm}\times  \bigg (
   \calH_{uu}^\Delta + 5\,\calH_{us}^\Delta
 + 2\,\calH_{sjr}^\Delta + \calH_{ujr}^\Delta
  -   2\,i\,
   \left( T_{ss} - 2\,T_{us} + T_{uu} \right)
  \bigg )
  C^{2}
\nonumber\\
& &\,\,\,\,\,\,\,\,\,\,\,\,\,\,\,
+\left ( \Delta\alpha^{(s)}_{n} + \Delta\beta^{(s)}_{n} \right )
 \bigg [ 
 \frac{2i}{3}\, \calH_{ujr} 
 \,(D^{2}+3 F^{2})
 +\,i\,  \calH_{sjr} 
 \,(D-F)^{2}
 -\frac{2i}{3}\,
 \calH_{uu} (D^{2}-3F^{2})\nonumber\\
& &\,\,\,\,\,\,\,\,\,\,\,\,\,\,\,\,\,\,\,\,\,\,\,\,\,\,\,\,\,\,\,\,\,\,
 -\frac{2i}{3}\,
 \calH_{us} (D^{2}-6DF + 3F^{2})
 +
  \bigg (
   R_{ss}\,(D-F)^{2} -4 R_{us}\,F\,(D-F) 
   +4 R_{uu}\,F^{2}
  \bigg )
 \bigg ]
\bigg \} \,. \nonumber\\
\label{eq:isoscal_heli_sigma_su63}&&
\end{eqnarray}
\begin{eqnarray}
 \la \Xi^{-} | \tilde{\op}^{(0)}_{\mu_{0}\cdots\mu_{n}} | 
 \Xi^{-}\ra &=& 
  \overline{U}_{\Xi^{-}} v_{\{\mu_{0}}\ldots
  v_{\mu_{n-1}}S_{\mu_{n}\}} U_{\Xi^{-}} 
 \left ( \Delta\alpha^{(s)}_{n} + \Delta\beta^{(s)}_{n}\right )
   \times
 \left ( 1 + \calW^{(\Xi)}_{SU(6|3)}\right )\nonumber\\
& & + \frac{\overline{U}_{\Xi^{-}} v_{\{\mu_{0}}\ldots
  v_{\mu_{n-1}}S_{\mu_{n}\}} U_{\Xi^{-}} }{f^{2}}
\bigg \{
 \frac{5i}{27}
   \left (\Delta\gamma^{(s)}_{n} - \frac{\Delta\sigma^{(s)}_{n}}{5} \right )
\nonumber\\
& &\,\,\,\,\,\,\,\,\,\,\,\,\,\,\,\,\,\,\,\,\,\,\,\,\,\,\,\,\,\,\,
   \,\,\,\,\,\,\,\,\,\,\,\,\,\,\,\,\,\,\,\,\,\,\,\,\,\,\,\,\,\,\,
 \times\bigg ( 2\,\calH_{ujr}^\Delta
  + \calH_{sjr}^\Delta   + 
  \calH_{ss}^\Delta + 
  5\,\calH_{us}^\Delta 
 - 2i\left( T_{ss} - 2\,T_{us} + T_{uu} \right) 
 \bigg )
  C^{2}
\nonumber\\
& &\,\,\,\,\,\,\,\,\,\,\,\,\,\,\,
+\left ( \Delta\alpha^{(s)}_{n} + \Delta\beta^{(s)}_{n} \right )
 \bigg [
 i\, \calH_{ujr} 
 \,(D-F)^{2}
 +\frac{2i}{3}\, \calH_{sjr} 
 \,(D^{2} + 3F^{2})
 - \frac{2i}{3}\,
 \calH_{ss} (D^{2} - 3 F^{2})\nonumber\\
& &\,\,\,\,\,\,\,\,\,\,\,\,\,\,\,\,\,\,\,\,\,\,\,\,\,\,\,\,\,\,\,\,\,\,
 - \frac{2i}{3}\, \calH_{us} 
 (D^{2} - 6 D F + 3 F^{2}) +
  \bigg (
   4 R_{ss}\,F^{2} -4 R_{us}\,F\,(D-F) 
   +R_{uu}\,(D-F)^{2}
  \bigg ) 
 \bigg ]
\bigg \} \,. \nonumber\\
\label{eq:isoscal_heli_xi_su63}&&
\end{eqnarray}
\begin{eqnarray}
 \la \Sigma^{0} | \tilde{\op}^{(0)}_{\mu_{0}\cdots\mu_{n}} | 
 \Lambda^{0} \ra &=& 0 \,.
\end{eqnarray}

\subsection{Transversity matrix elements}

\begin{eqnarray}
 \la p | \tilde{\op}^{T}_{\mu_{0}\cdots\mu_{n}\alpha} | p \ra &=&
 \frac{\overline{U}_p v_{\{\mu_{0}}\ldots v_{[\mu_{n}\}} S_{\alpha ]}U_p\big ( (5 + y_{1})
   \delta\alpha_{n}
  + 2 (1 + 2 y_{1} ) \delta\beta_{n} \big )}{6}
 \times \left (1 + \calW^{(p)}_{SU(6|3)} 
  \right )\nonumber\\
 & & + \frac{\overline{U}_p v_{\{\mu_{0}}\ldots v_{[\mu_{n}\}} S_{\alpha ]}U_p}{f^{2}}
\bigg \{
  \frac{5 i}{81}\,
  \left( \,{{\delta\gamma_{n}}} 
 - \frac{{3\,\delta\sigma_{n}}}{5} \right)
\bigg ( 3\,\calH_{uu}^\Delta + 
    \calH_{ujr}^\Delta \bigg ) \,
  \left( 5 + {y_{1}} \right) 
  C^{2}\nonumber\\
& &\,\,\,\,\,\,\,\,\,\,\,\,\,\,\,\,\,\,\,\,\,\,\,\,\,\,
 +\frac{8i}{9}\sqrt{\frac{2}{3}}
 \bigg [
  \calK_{uu} (D+3F) (-1+y_{1})   -  \calK_{ujr} 
    (3D - F - 2 F y_{1})
 \bigg ] C \delta c_{n} 
\nonumber\\
& &\,\,\,\,\,\,\,\,\,\,\,\,\,\,\,
+\delta\alpha_{n}
 \bigg [-\frac{1}{6}(5 + y_{1}) 
   \,\bigg (\calI_{ujr} -  2\,i\,\quu \bigg )
  +\frac{i}{3}\, \calH_{ujr} 
  \bigg ( 3\,D^2 - 6\,D\,F + 5\,F^2 + 
  \left( D^2 + F^2 \right) \,{y_{1}}
   \bigg )\nonumber\\
& &\,\,\,\,\,\,\,\,\,\,\,\,\,\,\,\,\,\,\,\,\,\,\,\,\,\,\,\,
  - \frac{4i}{3}\, \calH_{uu}\,D\,(D-2F-F y_{1}) 
  + \frac{1}{6} R_{uu} (D-3F)^{2} (5+y_{1})
 \bigg ]\nonumber\\
& &\,\,\,\,\,\,\,\,\,\,\,\,\,\,\,
+\delta\beta_{n}
 \bigg [-\frac{1}{3}(1 + 2 y_{1}) 
   \,\bigg (\calI_{ujr} - 2\,i\,\quu\bigg )
  +\frac{2i}{3}\, \calH_{ujr} 
  \bigg ( D^2 + F^2 (1+2 y_{1})
   \bigg )\nonumber\\
\label{eq:trans_proton_su63}
& &\,\,\,\,\,\,\,\,\,\,\,\,\,\,\,\,\,\,\,\,\,\,\,\,\,\,\,\,
  -\frac{4i}{3}\, \calH_{uu}\,D\,(D-2F-F y_{1}) 
  +\frac{1}{3}  R_{uu} (D-3F)^{2} (1+2y_{1})
 \bigg ]
\bigg \} \,. 
\end{eqnarray}
\begin{eqnarray}
 \la \Lambda^{0} | \tilde{\op}^{T}_{\mu_{0}\cdots\mu_{n}\alpha} | 
 \Lambda^{0}\ra &=& 
   \frac{\overline{U}_{\Lambda^{0}} v_{\{\mu_{0}}\ldots v_{[\mu_{n}\}}
     S_{\alpha ]} U_{\Lambda^{0}} (1+y_{1})}{4} 
    \big ( (1+2\tilde{y})\delta\alpha_{n} +2 \delta\beta_{n} \big )\times
 \left ( 1 + \calW^{(\Lambda)}_{SU(6|3)}\right )\nonumber\\
& & + \frac{\overline{U}_{\Lambda^{0}} v_{\{\mu_{0}}\ldots
  v_{[\mu_{n}\}} S_{\alpha ]} U_{\Lambda^{0}} (1+y_{1})}{f^{2}}
\bigg \{
   \frac{5i}{54}\,
  \left( \,{{\delta\gamma_{n}}} - \frac{{3\,\delta\sigma_{n}}}{5} \right)
\nonumber\\
&&\hspace*{3cm}\times\bigg ( 2\calH_{uu}^\Delta\,(1+\tilde{y}) + 
   \calH_{us}^\Delta\,(1+4\tilde{y}) 
 +\calH_{ujr}^\Delta \, (1+2\tilde{y}) \bigg ) 
 C^{2}
\nonumber\\
& &\,\,\,\,\,\,\,\,\,\,\,\,\,\,\,\,\,\,\,\,\,\,\,\,\,\,
 +\frac{4i}{9}\sqrt{\frac{2}{3}}
 \bigg [ \bigg (-\calK_{us}\,(D-3F)
  +2\calK_{uu}\,D \bigg )\,(1-2\tilde{y})
   -  \calK_{ujr} 
    (D - 3F + 4D\tilde{y})
 \bigg ] C \delta c_{n} 
\nonumber\\
& &\,\,\,\,\,\,\,\,\,\,\,\,\,\,\,
+\delta\alpha_{n}
 \bigg [ -\frac{1}{4} \bigg (\calI_{ujr} + 2\,\calI_{sjr}\,\tilde{y}
 - 2\,i\,\quu - 4\,i\,\qss \tilde{y} \bigg )\nonumber\\
& &\,\,\,\,\,\,\,\,\,\,\,\,\,\,\,\,\,\,\,\,\,\,\,\,\,\,\,\,\,\,\,\,\,\,
 +\frac{i}{36}  \calH_{ujr} 
\bigg ( 13 D^{2} - 18 DF + 9 F^{2} 
     + 4 (5 D^{2} -12 DF + 9 F^{2})\tilde{y}\bigg )
 +\frac{i}{36}  \calH_{sjr} 
 \,(D+3F)^{2}\nonumber\\
& &\,\,\,\,\,\,\,\,\,\,\,\,\,\,\,\,\,\,\,\,\,\,\,\,\,\,\,\,\,\,\,\,\,\,
 +\frac{i}{18} \calH_{uu}\,
 \bigg ( D^{2} + 12 DF - 9 F^{2} - 6 (D-3F)(D-F)\tilde{y}\bigg )\nonumber\\
& &\,\,\,\,\,\,\,\,\,\,\,\,\,\,\,\,\,\,\,\,\,\,\,\,\,\,\,\,\,\,\,\,\,\,
 -\frac{i}{18} \calH_{us}\,
  \bigg ( (D-3F) (5D+3F) + 2(5D^{2}-9F^{2})\tilde{y}\bigg )
 \nonumber\\
& &\,\,\,\,\,\,\,\,\,\,\,\,\,\,\,\,\,\,\,\,\,\,\,\,\,\,\,\,\,\,\,\,\,\,
 +\frac{1}{36}
  \bigg (
   R_{ss}\,(D+3F)^{2} 
   -4 R_{us}\,(2D-3F)(D+3F) 
   +4 R_{uu}\, (2D-3F)^{2}
  \bigg )\,(1+2\tilde{y})
 \bigg ]\nonumber\\
& &\,\,\,\,\,\,\,\,\,\,\,\,\,\,\,\,\,\,\,\,\,
+\delta\beta_{n}
 \bigg [-\frac{1}{2} \bigg (\calI_{ujr} - 2\,i\,\quu\bigg )
 +\frac{i}{18}  \calH_{ujr} 
 \,\bigg ( (D-3F)^{2} + 8D^{2} \tilde{y}\bigg )
 +\frac{i}{18}  \calH_{sjr} 
 \,(D+3F)^{2}\nonumber\\
& &\,\,\,\,\,\,\,\,\,\,\,\,\,\,\,\,\,\,\,\,\,\,\,\,\,\,\,\,\,\,\,\,\,\,
 -\frac{i}{9} \calH_{uu} 
 \bigg ( 3(D-3F)(D-F) - 4D^{2}\tilde{y} \bigg )
 +\frac{i}{9} \calH_{us} 
 \bigg ( D^{2} + 9F^{2}-12D(D-F)\tilde{y}\bigg )
 \nonumber\\
& &\,\,\,\,\,\,\,\,\,\,\,\,\,\,\,\,\,\,\,\,\,\,\,\,\,\,\,\,\,\,\,\,\,\,
 +\frac{1}{18}
  \bigg (
   R_{ss}\,(D+3F)^{2} 
-4 R_{us}\,(2D-3F)(D+3F) 
   +4 R_{uu}\, (2D-3F)^{2}
  \bigg )
 \bigg ]
\bigg \} \,. 
\label{eq:trans_lambda_su63}
\end{eqnarray}
\begin{eqnarray}
 \la \Sigma^{+} | \tilde{\op}^{T}_{\mu_{0}\cdots\mu_{n}\alpha} | 
 \Sigma^{+}\ra &=& 
   \frac{\overline{U}_{\Sigma^{+}} v_{\{\mu_{0}}\ldots v_{[\mu_{n}\}}
     S_{\alpha ]} U_{\Sigma^{+}} \big (
    (5+y_{2})\delta\alpha_{n} 
    + 2 (1 + 2 y_{2}) \delta\beta_{n}
   \big )}{6}
   \times
 \left ( 1 + \calW^{(\Sigma)}_{SU(6|3)}\right )\nonumber\\
& & + \frac{\overline{U}_{\Sigma^{+}} v_{\{\mu_{0}}\ldots
  v_{[\mu_{n}\}} S_{\alpha ]} U_{\Sigma^{+}} }{f^{2}}
\bigg \{
 \frac{5i}{81}
   \left (\delta\gamma_{n} - \frac{3\delta\sigma_{n}}{5} \right )
  \bigg (
   \calH_{uu}^\Delta\,(2+y_{2}) + \calH_{us}^\Delta\,(13+2y_{2})
\nonumber\\
& &\,\,\,\,\,\,\,\,\,\,\,\,\,\,\,\,\,\,\,\,\,\,\,\,\,\,
   \,\,\,\,\,\,\,\,\,\,\,\,\,\,\,\,\,\,\,\,\,\,\,\,\,\,
   \,\,\,\,\,\,\,\,\,\,\,\,\,\,\,\,\, \,\,\,\,\,\,\,\,\,  
 + 4\,\calH_{sjr}^\Delta
     + \calH_{ujr}^\Delta \,(1+y_{2})
  -2\,i\,
   \left( T_{ss} - 2\,T_{us} + T_{uu} \right)(2+y_{2})
  \bigg )
 C^{2}
\nonumber\\
& &\,\,\,\,\,\,\,\,\,\,\,\,\,\,\,\,\,\,\,\,
 -\frac{8i}{9}\sqrt{\frac{2}{3}}\,
 \bigg [ \calK_{ujr} 
 (D+F-2F y_{2}) 
+ 2  \calK_{sjr}  
 (D-F)
 +\bigg (  2 \calK_{uu}\,F
  + \calK_{us}\,(D+F) \bigg )\,(1-y_{2})\nonumber\\
& &\,\,\,\,\,\,\,\,\,\,\,\,\,\,\,\,\,\,\,\,\,\,\,\,\,\,\,\,\,\,\,\, 
\,\,\,\,\,\,\,\,\,\,\,\,\,\,\,\,\,\,\,\,\,\,\,\,\,\,\,\,\,\,\,\,
  -2\,i \,
  \bigg ( S_{ss}\,\left( D - F \right)  + 2\,S_{uu}\,F - 
    S_{us}\,\left( D + F \right)  \bigg ) \,
  \left(1-{y_{2}} \right)
 \bigg ] C \delta c_{n} 
\nonumber\\
& &\,\,\,\,\,\,\,\,\,\,\,\,\,\,\,
+\delta\alpha_{n}
 \bigg [ -\frac{1}{6} \bigg (5\,\calI_{ujr} - 10\,i\,\quu
 +\big [ \calI_{sjr} -2\,i\,\qss\big ]\,y_{2}
   \bigg )
\nonumber\\
& &\,\,\,\,\,\,\,\,\,\,\,\,\,\,\,\,\,\,\,\,\,\,\,\,\,\,\,\,\,\,\,\,\,\,
 +\frac{i}{6}  \calH_{ujr} 
 \,\bigg ( D^{2} - 2DF + 5F^{2} + 2 (D^{2}+F^{2}) y_{2}\bigg )
 +\frac{5i}{6}  \calH_{sjr} 
 \,(D-F)^{2}\nonumber\\
& &\,\,\,\,\,\,\,\,\,\,\,\,\,\,\,\,\,\,\,\,\,\,\,\,\,\,\,\,\,\,\,\,\,\,
 -\frac{i}{3} 
 \calH_{uu} 
  \bigg ( D^{2} - 5F^{2} + (D-F)(D+F)y_{2}\bigg )\nonumber\\
& &\,\,\,\,\,\,\,\,\,\,\,\,\,\,\,\,\,\,\,\,\,\,\,\,\,\,\,\,\,\,\,\,\,\,
 - \frac{i}{3} 
 \calH_{us} 
  \bigg (3D^{2}-8DF+5F^{2}-(D^{2}+4DF-F^{2})y_{2} \bigg )
 \nonumber\\
& &\,\,\,\,\,\,\,\,\,\,\,\,\,\,\,\,\,\,\,\,\,\,\,\,\,\,\,\,\,\,\,\,\,\,
 +\frac{1}{6}
  \bigg (
   R_{ss}\,(D-F)^{2} -4 R_{us}\,F\,(D-F) 
   +4 R_{uu}\,F^{2}
  \bigg ) (5+y_{2})
 \bigg ]\nonumber\\
& &\,\,\,\,\,\,\,\,\,\,\,\,\,\,\,\,\,\,\,\,\,
+\delta\beta_{n}
 \bigg [
- \frac{1}{3} \bigg (\calI_{ujr} - 2\,i\,\quu
 + 2\,\big [ \calI_{sjr} -2\,i\,\qss\big ] y_{2} 
   \bigg )
 + \frac{i}{3}  \calH_{ujr} 
 \,\bigg ((D+F)^{2} + 4F^{2} y_{2}\bigg )\nonumber\\
& &\,\,\,\,\,\,\,\,\,\,\,\,\,\,\,\,\,\,\,\,\,\,\,\,\,\,\,\,\,\,\,\,\,\,
 -\frac{2i}{3}\,
 \calH_{uu} \bigg (D^{2}-F^{2}-2F^{2}y_{2} \bigg )
 -\frac{2i}{3}\,
 \calH_{us} \bigg ( D^{2}-4DF+F^{2}-2F(D-F)y_{2}\bigg )
 \nonumber\\
& &\,\,\,\,\,\,\,\,\,\,\,\,\,\,\,\,\,\,\,\,\,\,\,\,\,\,\,\,\,\,\,\,\,\,
+\frac{i}{3}  \calH_{sjr} 
 \,(D-F)^{2}
 +\frac{1}{3}
  \bigg (
   R_{ss}\,(D-F)^{2} -4 R_{us}\,F\,(D-F) 
   +4 R_{uu}\,F^{2}
  \bigg ) (1+2y_{2})
 \bigg ]
\bigg \} \,. \nonumber\\
\label{eq:trans_sigma_su63}&&
\end{eqnarray}
\begin{eqnarray}
 \la \Xi^{-} | \tilde{\op}^{T}_{\mu_{0}\cdots\mu_{n}\alpha} | 
 \Xi^{-}\ra &=& 
   -\frac{\overline{U}_{\Xi^{-}} v_{\{\mu_{0}}\ldots v_{[\mu_{n}\}}
     S_{\alpha ]} U_{\Xi^{-}} \big (
    (y_{1}+5 y_{2})\delta\alpha_{n}+(4y_{1}+2y_{2}) \delta\beta_{n}
   \big )}{6}
   \times
 \left ( 1 + \calW^{(\Xi)}_{SU(6|3)}\right )\nonumber\\
& & + \frac{\overline{U}_{\Xi^{-}} v_{\{\mu_{0}}\ldots v_{[\mu_{n}\}}
  S_{\alpha ]} U_{\Xi^{-}} }{f^{2}}
\bigg \{
 \frac{5i}{81}
   \left (\delta\gamma_{n} - \frac{3\delta\sigma_{n}}{5} \right )
\bigg (
   \calH_{sjr}^\Delta
  (y_{1} + y_{2} )  + 
  4\, \calH_{ujr}^\Delta \, y_{2} + 
  \calH_{ss}^\Delta\,\left(y_{1}+2y_{2} \right)\nonumber\\
& &\,\,\,\,\,\,\,\,\,\,\,\,\,\,\,\,\,\,\,\,\,\,\,\,\,\,\,\,
   \,\,\,\,\,\,\,\,\,\,\,\,\,\,\,\,\,\,\,\,\,\,\,\,\,  
+   \calH_{us}^\Delta\,\left( 2y_{1}+13y_{2} \right)
 - 2i\left( T_{ss} - 2\,T_{us} + T_{uu} \right) 
      \left(y_{1}+2y_{2} \right)
 \bigg ) \bigg ] C^{2}
\nonumber\\
& & \hspace*{5mm}
 -\frac{8i}{9}\sqrt{\frac{2}{3}}\,
 \bigg [ 2  \calK_{ujr}  
 (D-F) y_{2}
-  \calK_{sjr}  
  \bigg ( 2F y_{1} -(D+F)y_{2} \bigg )
  - \calK_{us} (D+F) (y_{1}-y_{2})\nonumber\\
& & \hspace*{15mm}
 - 2 \calK_{ss} F (y_{1}-y_{2})
  +2 i \,
  \bigg ( S_{uu}\,\left( D - F \right)  + 2\,S_{ss}\,F - 
    S_{us}\,\left( D + F \right)  \bigg ) \,
  \left({y_{1}}-y_{2} \right)
 \bigg ] C \delta c_{n} 
\nonumber\\
& &\,\,\,\,\,\,\,\,\,\,\,\,\,\,\,
+\delta\alpha_{n}
 \bigg [-\frac{1}{6} \bigg (\big [\calI_{ujr} -2\,i\,\quu\big ] y_{1}
 + 5\,\big [ \calI_{sjr} -2 \,i\,\qss\big ]y_{2}
   \bigg )\nonumber\\
& &\,\,\,\,\,\,\,\,\,\,\,\,\,\,\,\,\,\,\,\,\,\,\,\,\,\,\,\,\,\,\,\,\,\,
 +\frac{5 i}{6}  \calH_{ujr} 
 \, (D-F)^{2}\,y_{2}
 +\frac{i}{6}  \calH_{sjr} 
 \,\bigg ( 2(D^{2}+F^{2})y_{1}+(D^{2}-2DF+5F^{2})y_{2}\bigg )\nonumber\\
& &\,\,\,\,\,\,\,\,\,\,\,\,\,\,\,\,\,\,\,\,\,\,\,\,\,\,\,\,\,\,\,\,\,\,
 - \frac{i}{3} 
 \calH_{ss} 
 \bigg ( (D-F)(D+F) y_{1} + (D^{2}-5F^{2})y_{2}\bigg )\nonumber\\
& &\,\,\,\,\,\,\,\,\,\,\,\,\,\,\,\,\,\,\,\,\,\,\,\,\,\,\,\,\,\,\,\,\,\,
 + \frac{i}{3} \calH_{us} 
 \bigg ( (D^{2}+4DF-F^{2})y_{1}- (3D-5F)(D-F)y_{2}\bigg )
 \nonumber\\
& &\,\,\,\,\,\,\,\,\,\,\,\,\,\,\,\,\,\,\,\,\,\,\,\,\,\,\,\,\,\,\,\,\,\,
 +\frac{1}{6}
  \bigg (
   4 R_{ss}\,F^{2} -4 R_{us}\,F\,(D-F) 
   +R_{uu}\,(D-F)^{2}
  \bigg ) (y_{1}+5y_{2})
 \bigg ]\nonumber\\
& &\,\,\,\,\,\,\,\,\,\,\,\,\,\,
+\delta\beta_{n}
 \bigg [-\frac{1}{3} \bigg (2 \big [\calI_{ujr} -2\,i\,\quu\big ] y_{1}
 + \big [ \calI_{sjr} -2\,i\,\qss\big ] y_{2} 
   \bigg )
 +\frac{i}{3}\, \calH_{ujr} 
 \, (D-F)^{2}\,y_{2}\nonumber\\
& &\,\,\,\,\,\,\,\,\,\,\,\,\,\,\,\,\,\,\,\,\,\,\,\,\,\,\,\,\,\,\,\,\,\,
 +\frac{i}{3}\, \calH_{sjr} 
 \,\bigg ( 4 F^{2} y_{1}+ (D+F)^{2} y_{2}\bigg )
 +\frac{2i}{3}\,
  \calH_{ss} \bigg (2F^{2}y_{1}-(D^{2}-F^{2})y_{2}\bigg )\nonumber\\
& &\,\,\,\,\,\,\,\,\,\,\,\,\,\,\,\,\,\,\,\,\,\,\,\,\,\,\,\,\,\,\,\,\,\,
 +\frac{2i}{3}\,\calH_{us} 
 \bigg ( 2F(D-F)y_{1}-(D^{2}-4DF+F^{2})y_{2}\bigg )
 \nonumber\\
& &\,\,\,\,\,\,\,\,\,\,\,\,\,\,\,\,\,\,\,\,\,\,\,\,\,\,\,\,\,\,\,\,\,\,
 +\frac{1}{3}
  \bigg (
   4 R_{ss}\,F^{2} -4 R_{us}\,F\,(D-F) 
   +R_{uu}\,(D-F)^{2}
  \bigg ) (2y_{1}+y_{2})
 \bigg ]
\bigg \} \,.
\label{eq:trans_xi_su63}
\end{eqnarray}
\begin{eqnarray}
 \la \Sigma^{0} | \tilde{\op}^{T}_{\mu_{0}\cdots\mu_{n}\alpha} | 
 \Lambda^{0}\ra &=& 
   -\frac{\overline{U}_{\Sigma^{0}} v_{\{\mu_{0}}\ldots v_{[\mu_{n}\}}
     S_{\alpha ]} U_{\Lambda^{0}} \left (
    -1+y_{1}
   \right )}{4\sqrt{3}}\left (\delta\alpha_{n}
  -2 \delta\beta_{n}\right )
   \times
 \left [ 1 +
   \frac{1}{2}\left (\calW^{(\Lambda)}_{SU(6|3)}  +\calW^{(\Sigma)}_{SU(6|3)}
   \right )\right ]\nonumber\\
& & + \frac{\overline{U}_{\Sigma^{0}} v_{\{\mu_{0}}\ldots
  v_{[\mu_{n}\}} S_{\alpha ]}U_{\Lambda^{0}}
  (-1+y_{1})}{\sqrt{3}f^{2}} 
\bigg \{
 -\frac{5i}{54}
   \left (\delta\gamma_{n} - 
 \frac{3\,\delta\sigma_{n}}{5} \right )\bigg (
   2 \calH_{uu}^\Delta + \calH_{us}^\Delta + \calH_{ujr}^\Delta
  \bigg )
  C^{2}
\nonumber\\
& &\,\,\,\,\,\,\,\,\,\,\,\,\,\,\,\,\,\,\,\,\,\,\,\,\,\,
 +\frac{4i}{9}\sqrt{\frac{2}{3}}
 \bigg [ 2\calK_{us} (2D+3F)
  - \calK_{uu}\,(D-3F) 
   +2  \calK_{ujr} 
    \,D
   + \calK_{sjr}
    \,(D+3F)
  \nonumber\\
& &\,\,\,\,\,\,\,\,\,\,\,\,\,\,\,\,\,\,\,\,\,\,\,\,\,\,\,\,\,\,\,\, 
\,\,\,\,\,\,\,\,\,\,\,\,\,\,\,\,\,\,\,\,
   -i \,\bigg ( 2\,S_{uu}\,\left( 2\,D - 3\,F \right)  + 
    S_{us}\,\left( -5\,D + 3\,F \right)  + 
    S_{ss}\,\left( D + 3\,F \right)  \bigg )
 \bigg ] C \delta c_{n} 
\nonumber\\
& &\,\,\,\,\,\,\,\,\,\,\,\,\,\,\,
+\delta\alpha_{n}
 \bigg [ \frac{1}{4} \bigg (\calI_{ujr} -2\,i\,\quu
   \bigg ) 
 -\frac{i}{12}  \calH_{ujr} 
 \,(D-3F)(3D-F)
 +\frac{i}{12}  \calH_{sjr} 
 \,(D-F)(D+3F)\nonumber\\
& &\,\,\,\,\,\,\,\,\,\,\,\,\,\,\,\,\,\,\,\,\,\,\,\,\,\,\,\,\,\,\,\,\,\,
 +\frac{i}{3} \calH_{uu} D (D+F)
 +\frac{i}{3} \calH_{us} D (D-F)
 +\frac{1}{12}
  \bigg (
   R_{ss}\,(D-F)\,(D+3F) 
\nonumber\\
& &\,\,\,\,\,\,\,\,\,\,\,\,\,\,\,\,\,\,\,\,\,\,\,\,\,\,\,\,\,\,\,\, 
\,\,\,\,\,\,\,\,\,\,\,\,\,\,\,\,\,\,\,\,
   -4 R_{us}\,(D^{2}-2DF + 3F^{2}) 
   -4 R_{uu}\,F\,(-2D+3F)
  \bigg )
 \bigg ]\nonumber\\
& &\,\,\,\,\,\,\,\,\,\,\,\,\,\,\,\,\,\,\,\,\,
+\delta\beta_{n}
 \bigg [
-\frac{1}{2} \bigg (\calI_{ujr} -2\,i\,\quu
   \bigg ) 
 -\frac{i}{6}  \calH_{ujr} 
 \,(D-3F)(D+F)
 -\frac{i}{6}  \calH_{sjr} 
 \,(D-F)(D+3F)\nonumber\\
& &\,\,\,\,\,\,\,\,\,\,\,\,\,\,\,\,\,\,\,\,\,\,\,\,\,\,\,\,\,\,\,\,\,\,
 +\frac{2i}{3} \calH_{uu} D (D - F)
 +\frac{2i}{3} \calH_{us} D F
 -\frac{1}{6}
  \bigg (
   R_{ss}\,(D-F)\,(D+3F) 
\nonumber\\
& &\,\,\,\,\,\,\,\,\,\,\,\,\,\,\,\,\,\,\,\,\,\,\,\,\,\,\,\,\,\,\,\, 
\,\,\,\,\,\,\,\,\,\,\,\,\,\,\,\,\,\,\,\,
   -4 R_{us}\,(D^{2}-2DF + 3F^{2}) 
   -4 R_{uu}\,F\,(-2D+3F)
  \bigg )
 \bigg ]
\bigg \} \,.
\label{eq:trans_lambda_sigma_su63}
\end{eqnarray}
%


\section{Results in SU(2$|$2) \qxpt}
\label{A4}

In SU(2$|$2) quenched \xpt, the Lagrangian
Eq.~(\ref{eq:free_lagrangian}) receives additional contributions since
the theory has no axial anomaly and the singlet meson field remains
light. Thus,
\begin{eqnarray}
{\cal L}_B^{Q} & = & 
i\left(\overline{\cal B} v\cdot {\cal D} {\cal B}\right)
 - i \left(\overline{\cal T}^\mu v\cdot {\cal D} {\cal T}_\mu\right)
+ \Delta\ \left(\overline{\cal T}^\mu {\cal T}_\mu\right)
\nonumber\\
 && + 2\alpha \left(\overline{\cal B} S^\mu {\cal B} A_\mu\right)
 +  2\beta \left(\overline{\cal B} S^\mu A_\mu {\cal B} \right)
 +  2{\cal H} \left(\overline{\cal T}^\nu S^\mu A_\mu {\cal T}_\nu \right)
+ \sqrt{3\over 2}{\cal C} 
\left[
\left( \overline{\cal T}^\nu A_\nu {\cal B}\right) + 
\left(\overline{\cal B} A_\nu {\cal T}^\nu\right) \right]
\nonumber\\
 && + 2\gamma \left(\overline{\cal B} S^\mu {\cal B}\right) \str\left(A_\mu\right)
 +  2\gamma^\prime \left(\overline{\cal T}^\nu S^\mu {\cal T}_\nu
 \right) \str\left(A_\mu\right)
\,,
\label{eq:quenched_free_lagrangian}
\end{eqnarray}
with two additional couplings $\gamma$ and $\gamma^\prime$.  There is
no relation between the other couplings in Eq.~
(\ref{eq:quenched_free_lagrangian}) and those in (PQ)\xpt\ (though we
use the same notation for convenience).

Defining
\begin{equation}
  \label{eq:221}
  \hat\tau_3={\rm diag}\left( 1, -1, \hat q, -\hat q\right)\,,
\end{equation}
\begin{equation}
  \label{eq:222}
  \hat\tau_0={\rm diag}\left( 1, 1, 1, 1 \right)\,,
\end{equation}
and
\begin{equation}
  \label{eq:223}
  \hat\tau_T={\rm diag}\left( 1,\hat y_1,\hat y_2,\hat y_3\right)\,,
\end{equation}
the quenched \xpt\ twist-two operators correspond (for the most part)
to those given in Eqs.~(\ref{eq:hadron_op})--(\ref{eq:hadron_sigmaop})
with the replacement $\tau \to \hat \tau$ everywhere.  Again, the LECs
($\alpha_n^{(a)}$ etc) occurring in the quenched theory are different
from those in SU(4$|$2) but with this caution, we use the same
notation.  For the transversity operators one needs additional
operators proportional to ${\rm str}\left(\hat\tau^{\xi_T^+}\right)$:
\begin{equation}
\left[\delta\alpha^\prime_n  v_{\{\mu_0}\ldots v_{[\mu_{n}\}} 
\left(\bar{\cal B}S_{\alpha]}{\cal B}\right)
+
\delta\gamma^\prime_n  v_{\{\mu_0}\ldots v_{[\mu_{n}\}} 
\left(\bar{\cal T}^\nu S_{\alpha]}{\cal T}_\nu\right)
+ 
\delta\sigma^\prime_n v_{\{\mu_0}\ldots v_{\mu_{n-2}} 
\left(\overline{\cal T}_{\mu_{n-1}} S_{[\alpha}\bar{\cal T}_{\mu_n]\}} \right) 
\right]
{\rm str}\left(\hat\tau^{\xi_T^+}\right)\,.
\end{equation}

With these definitions it is then easy to calculate the quenched
matrix elements in the isospin limit. The wave-function
renormalisation is
\begin{equation}
\label{eq:waverenorm_su22}
 \calW_{SU(2|2)} = \frac{i}{f^{2}}\,\bigg (
  (g_{1}+g_{A})\,(6\gamma -g_{1} + 2 g_{A})\,\calH_{uu}
  + 2\,g_{\Delta N}^{2}\,\calH_{uu}^\Delta
  + (g_{1}+g_{A})^{2}\,m^{2}_{0}\,\calH_{\eta^\prime,uu}
  \bigg ) \,.
\end{equation}

The isovector, unpolarised matrix element is
\begin{eqnarray}
\la N | \op^{(3)}_{\mu_{0}\cdots\mu_{n}} |N\ra
 &=& \frac{1}{3} \overline{U}_{N} v_{\mu_{0}}\ldots v_{\mu_{n}} U_N
  \bigg ( 2 \alpha^{(a)}_{n} - \beta^{(a)}_{n} \bigg )
  \times \bigg ( 1+ (1-\delta_{n0})\,\calW_{SU(2|2)} \bigg )
\label{eq:result_isovec_unpol_su22}
\\
 &&+ \frac{1}{3f^{2}} \overline{U}_{N} v_{\mu_{0}}\ldots v_{\mu_{n}}
   U_N (1-\delta_{n0})
 \bigg \{
  4\,i\,g_{\Delta N}^{2} \left ( \gamma^{(a)}_{n} - \frac{\sigma^{(a)}_{n}}{3} \right )
    \calH_{uu}^\Delta
\nonumber\\
& &\,\,\,\,\,\,\,\,\,\,\,\,\,\,\,\,\,\,\,\,\,\,\,\,
 -i\, \bigg ( 2 \alpha^{(a)}_{n} - \beta^{(a)}_{n} \bigg )
 \times
 \bigg [ 6g_{A}\gamma \calH_{uu} 
   + (g_{1}+g_{a})^{2} m_{0}^{2} \calH_{\eta^\prime,uu} \bigg ]
\nonumber\\
& &\,\,\,\,\,\,\,\,\,\,\,\,\,\,\,\,\,\,\,\,\,\,\,
  + \alpha^{(a)}_{n} 
  \bigg [ \frac{3i}{2} g_{1} 
  \bigg ( g_{1} - 2 (4\gamma + g_{A}) \bigg ) \calH_{uu} \bigg ]
  + \beta^{(a)}_{n} 
  \bigg [ \frac{3i}{2} g_{1} 
  \bigg ( g_{1} + 2 (2\gamma - g_{A})\bigg ) \calH_{uu} \bigg ]
 \bigg \} \,,
\nonumber
\end{eqnarray}

and the isovector, helicity matrix element is
\begin{eqnarray}
\la N | \tilde{\op}^{(3)}_{\mu_{0}\cdots\mu_{n}} |N\ra
 &=& \frac{1}{3} \overline{U}_{N} v_{\{\mu_{0}}\ldots v_{\mu_{n-1}} S_{\mu_{n}\}} U_N
  \bigg ( 2 \Delta\alpha^{(a)}_{n} - \Delta\beta^{(a)}_{n} \bigg )
  \bigg (1 + \calW_{SU(2|2)}\bigg ) 
\label{eq:result_isovec_heli_su22}
\\
 &&+ \frac{1}{3f^{2}} \overline{U}_{N} v_{\{\mu_{0}}\ldots
 v_{\mu_{n-1}} S_{\mu_{n}\}} U_N
 \bigg \{
\frac{8i}{3}\sqrt{\frac{2}{3}} (g_{1}+4g_{A}) \calK_{uu}
  g_{\Delta N} \Delta c_{n}
  + \frac{20}{9}\,i\,g_{\Delta N}^{2} \left ( \Delta\gamma^{(a)}_{n} -
    \frac{\Delta\sigma^{(a)}_{n}}{5} \right ) \calH_{uu}^\Delta
\nonumber\\
& &\,\,\,\,\,\,\,\,\,\,\,\,\,\,\,\,\,\,\,\,\,\,\,\,\,\,\,\,\,\,\,\,\,
+ \frac{i}{3}\,\bigg ( 2 \Delta\alpha^{(a)}_{n} - \Delta\beta^{(a)}_{n} \bigg )
 \times
 \bigg [  6g_{A}\gamma \calH_{uu} 
   + (g_{1}+g_{A})^{2} m_{0}^{2} \calH_{\eta^\prime,uu} \bigg ]
\nonumber\\
& &\,\,\,\,\,\,\,\,\,\,\,\,\,\,\,\,\,\,\,\,\,\,\,\,\,\,\,\,\,\,\,\,\,
  -\frac{i}{2}  \Delta\alpha^{(a)}_{n}
   g_{1} 
  \bigg [ g_{1} - 2 (4\gamma + g_{A}) \bigg ] \calH_{uu}
  -\frac{i}{2}  \Delta\beta^{(a)}_{n}
   g_{1} 
  \bigg [ g_{1} + 2 (2\gamma - g_{A})\bigg ] \calH_{uu} 
 \bigg \} .
\nonumber \,.
\end{eqnarray}

The isoscalar, unpolarised matrix element is
\begin{eqnarray}
\la N | \op^{(0)}_{\mu_{0}\cdots\mu_{n}} |N\ra
 &=&  \overline{U}_{N} v_{\mu_{0}}\ldots v_{\mu_{n}} U_N
  \bigg ( \alpha^{(s)}_{n} + \beta^{(s)}_{n} \bigg )
  \times \bigg ( 1+ (1-\delta_{n0})\,\calW_{SU(2|2)} \bigg )
\label{eq:result_isoscal_unpol_su22}
\\
 &&+  \frac{\overline{U}_{N} v_{\mu_{0}}\ldots v_{\mu_{n}}U_N(1-\delta_{n0})}{f^{2}}
 \bigg \{
  \left ( \gamma^{(s)}_{n} - \frac{\sigma^{(s)}_{n}}{3} \right )
  \times \bigg [ 2\,i\,g_{\Delta N}^{2} \calH_{uu}^\Delta\bigg ]
\nonumber\\
& &\,\,\,\,\,\,\,\,\,\,\,\,\,\,\,\,\,\,\,\,\,\,\,\,\,\,\,\,\,\,\,\,\,
-\,i\, \bigg ( \alpha^{(s)}_{n} + \beta^{(a)}_{n} \bigg )
(g_{1}+g_{A})\,
  \bigg ( 
   (6\gamma-g_{1}+2g_{A}) \calH_{uu} 
    + (g_{1}+g_{A}) m_{0}^{2}\calH_{\eta^\prime,uu} 
  \bigg )
 \bigg \} \,.
\nonumber
\end{eqnarray}

The isoscalar, helicity matrix element is
\begin{eqnarray}
\la N | \tilde{\op}^{(0)}_{\mu_{0}\cdots\mu_{n}} |N\ra
 &=&  \overline{U}_{N} v_{\{\mu_{0}}\ldots v_{\mu_{n-1}}S_{\mu_{n}\}} U_N
  \bigg ( \Delta\alpha^{(s)}_{n} + \Delta\beta^{(s)}_{n} \bigg )
  \times \bigg ( 1+ \calW_{SU(2|2)} \bigg )
\label{eq:result_isoscal_heli_su22}
\\
 &&+  \frac{\overline{U}_{N} v_{\{\mu_{0}}\ldots v_{\mu_{n-1}}S_{\mu_{n}\}}U_N}{f^{2}}
 \bigg \{
  \left ( \Delta\gamma^{(s)}_{n} - \frac{\Delta\sigma^{(s)}_{n}}{5} \right )
  \times \bigg [ \frac{10\,i}{9}\,g_{\Delta N}^{2} \calH_{uu}^\Delta\bigg ]
\nonumber\\
& &\,\,\,\,\,\,\,\,\,\,\,\,\,\,\,\,\,\,\,\,\,\,\,\,\,\,\,\,\,\,\,\,\,
+ \frac{i}{3}\,\bigg ( \Delta\alpha^{(s)}_{n} + \Delta\beta^{(a)}_{n} \bigg )
 (g_{1}+g_{A})\,
  \bigg ( 
   (6\gamma-g_{1}+2g_{A}) \calH_{uu} 
    + (g_{1}+g_{A}) m_{0}^{2}\calH_{\eta^\prime,uu} 
  \bigg )
 \bigg \} \,.\nonumber
\end{eqnarray}

Finally, the transversity matrix elements are
\begin{eqnarray}
 \la N | \tilde{\op}^{T}_{\mu_{0}\cdots\mu_{n}\alpha} |N\ra
 &=&  \frac{\overline{U}_{N} v_{\{\mu_{0}}\ldots v_{[\mu_{n}\}}S_{\alpha ]}U_N}{6}
  \bigg ( (5+\hqone ) \delta\alpha_{n} 
    +  (2 + 4\hqone) \delta\beta_{n} + 6 (1 + \hqone -\hqtwo - \hqthree) 
   \delta\alpha^{\prime}_{n}\bigg )
 \bigg( 1+ \calW_{SU(2|2)} \bigg)\nonumber\\
 && + \frac{\overline{U}_{N} v_{\{\mu_{0}}\ldots v_{[\mu_{n}\}}S_{\alpha ]} U_N}{f^{2}}
 \bigg \{
  \frac{i}{27}\,\calH_{uu}^\Delta \bigg [
    \left( 6 + 2\,{\hqone} - {\hqthree} - 
     {\hqtwo} \right) \,\left(5\,\delta\gamma_{n} -3\,\delta\sigma_{n}\right)
\nonumber\\
& &\,\,\,\,\,\,\,\,\,\,\,\,\,\,\,\,\,\,\,\,\,\,\,\,\,
\,\,\,\,\,\,\,\,\,\,\,\,\,\,\,\,\,\,\,\,\,\,\,\,\,
\,\,\,\,\,\,\,\,\,\,\,\,\,\,\,\,\,\,\,\,\,\,\,\,\,
+ 6\,\left( 1 + {\hqone} - {\hqthree} - 
        {\hqtwo} \right) \,
      \left( 5\,{\delta\gamma^{\prime}_{n}} - 
        3\,{\delta\sigma^{\prime}_{n}} \right) 
  \bigg ] g_{\Delta N}^{2}
\nonumber\\
& &\,\,\,\,\,\,\,\,\,\,\,\,\,\,\,\,\,\,\,\,\,\,\,\,\,
+  \frac{4i}{9} \sqrt{\frac{2}{3}} \,\calK_{uu}
 \bigg [
  4 g_{A} (1-\hqone) + g_{1} (4 + 2 \hqone - 3 \hqtwo - 3 \hqthree)
 \bigg ] g_{\Delta N} \,\delta c_{n}
\nonumber\\
& &\,\,\,\,\,\,\,\,\,\,\,\,\,\,\,\,\,\,\,\,\,\,\,\,\,
+ \delta\alpha_{n}
 \bigg [(-1-\hqone+\hqtwo+\hqthree) \calI_{uu}
  - \frac{1}{9} m^{2}_{0} (5+\hqone) \calI_{\eta^\prime,uu}
\nonumber\\
& &\,\,\,\,\,\,\,\,\,\,\,\,\,\,\,\,\,\,\,\,\,\,\,\,\,
   \,\,\,\,\,\,\,\,\,\,\,\,\,
  + \frac{i}{12}\bigg ( 
    - {{g_{1}^{2}}}\,
     \left( 2 + {\hqthree} + {\hqtwo} \right)
          + 4\,{g_A}\,
   \big [ {g_A}\,
      \left( 2 + 2\,{\hqone} - {\hqthree} - 
        {\hqtwo} \right)  + 
     \left( 5 + {\hqone} \right) \,\gamma  \big ]  
\nonumber\\
& &\,\,\,\,\,\,\,\,\,\,\,\,\,\,\,\,\,\,\,\,\,\,\,\,\,
   \,\,\,\,\,\,\,\,\,\,\,\,\,\,\,\,\,\,\,\,
+ 
  2\,{g_1}\,\big [ {g_A}\,
      \left( 3 + {\hqone} - {\hqthree} - 
        {\hqtwo} \right)  + 
     2\,\left( 5 + {\hqone} \right) \,\gamma  \big ]
   \bigg )\,\calH_{uu}\nonumber\\
& &\,\,\,\,\,\,\,\,\,\,\,\,\,\,\,\,\,\,\,\,\,\,\,\,\,
\,\,\,\,\,\,\,\,\,\,\,\,\,
 +\frac{i}{18} (g_{1}+g_{A})^{2} m^{2}_{0} (5 + \hqone)
 \calH_{\eta^\prime,uu}
 \bigg ]
\nonumber\\
& &\,\,\,\,\,\,\,\,\,\,\,\,\,\,\,\,\,\,\,\,\,\,\,\,\,
+ \delta\beta_{n}
 \bigg [(-1-\hqone+\hqtwo+\hqthree) \calI_{uu}
  - \frac{2}{9} m^{2}_{0} (1+2\hqone) \calI_{\eta^\prime,uu}
\nonumber\\
& &\,\,\,\,\,\,\,\,\,\,\,\,\,\,\,\,\,\,\,\,\,\,\,\,\,
   \,\,\,\,\,\,\,\,\,\,\,\,\,
  + \frac{i}{12}\bigg (
   4 g_{1} (2\gamma+g_{A}+4\gamma\hqone) 
   + 4 g_{A} \big [ 2\gamma + g_{A} + (4\gamma+g_{A})\hqone\big ]
\nonumber\\
& &\,\,\,\,\,\,\,\,\,\,\,\,\,\,\,\,\,\,\,\,\,\,\,\,\,
   \,\,\,\,\,\,\,\,\,\,\,\,\,\,\,\,\,\,\,\,\,
    + g^{2}_{1} \big [ 2\hqone - 3 (\hqtwo + \hqthree)\big ]
  \bigg ) \,\calH_{uu}
+ \frac{i}{9}(g_{1}+g_{A})^{2} m^{2}_{0}
   (1+2\hqone) \,\calH_{\eta^\prime,uu}
 \bigg ]
\nonumber\\
& &\,\,\,\,\,\,\,\,\,\,\,\,\,\,\,\,\,\,\,\,\,\,\,\,\,
+ \delta\alpha^{\prime}_{n}
 \bigg [ 4(\hqtwo + \hqthree) \calI_{uu}
   -\frac{2}{3} m^{2}_{0} (1 + \hqone -\hqtwo-\hqthree)\calI_{\eta^\prime,uu}
\nonumber\\
& &\,\,\,\,\,\,\,\,\,\,\,\,\,\,\,\,\,\,\,\,\,\,\,\,\,
   \,\,\,\,\,\,\,\,\,\,\,\,\,
-\frac{i}{3} (g_{1}+g_{A}) \bigg ( g_{1} -2 (3\gamma + g_{A})\bigg )
 \, (1 + \hqone-\hqtwo-\hqthree)\,\calH_{uu}
\nonumber\\
& &\,\,\,\,\,\,\,\,\,\,\,\,\,\,\,\,\,\,\,\,\,\,\,\,\,
   \,\,\,\,\,\,\,\,\,\,\,\,\,
+\frac{i}{3}(g_{1}+g_{A})^{2} m^{2}_{0} (1+\hqone-\hqtwo-\hqthree)
 \,\calH_{\eta^\prime,uu}
 \bigg ]
 \bigg \} \,.
\label{eq:result_trans_su22}
\end{eqnarray}
%


\end{document}